\documentclass[english,aps,prb,reprint,superscriptaddress, nofootinbib]{revtex4-2}
\usepackage[T1]{fontenc}
\usepackage[latin9]{inputenc}
\usepackage{verbatim}
\usepackage{amsmath}
\usepackage{amsthm}
\usepackage{amssymb}
\usepackage{graphicx}
\usepackage{esint}
\usepackage{textcomp}

\makeatletter

\renewcommand{\fnum@figure}{FIG. \thefigure}

\makeatother

\usepackage{babel}

\begin{document}
	
\author{Yonatan Messica}
\affiliation{
	Department of Physics, Bar-Ilan University, Ramat Gan, 52900, Israel
}
\author{Dmitri B. Gutman}
\affiliation{
	Department of Physics, Bar-Ilan University, Ramat Gan, 52900, Israel
}
\date{\today}

\begin{abstract}
We study the influence of pseudospin-orbit coupling on electron-electron
scattering in the Coulomb drag setup. We study a setup made of a time-reversal-symmetry-broken Weyl semimetal (WSM) layer and a normal metal layer. The interlayer
drag force consists of two components. The first one is conventional
and is parallel to the relative electronic boost velocity between
the layers. This part of the drag tends to equilibrate the momentum
distribution in the two layers, analogous to shear viscosity in hydrodynamics.
In the WSM layer, the shift of the Fermi surface is not parallel to
the electric field, due to skew scattering in the WSM. This induces
a Hall current in the normal metal via the conventional component
of the drag force. The second component of the drag force is perpendicular
to the boost velocity in the Weyl semimetal and arises from interlayer
e-e skew scattering, which results from two types of processes. The
first process is an interference between electron-electron and electron-disorder
scattering. The second process is due to the side jumps in electron-electron
collisions in an external electric field. Both the parallel and perpendicular
components of the drag are important for the anomalous Hall drag conductivity.
On the other hand, for the Hall drag resistivity, the contribution
from the parallel friction is partially cancelled in a broad temperature
regime. This work provides insight into the microscopic mechanisms
of Hall-like friction in electronic fluids.
\end{abstract}

\title{Hall Coulomb drag induced by electron-electron skew scattering}

\maketitle

\section{Introduction}

The Coulomb drag experiment is an efficient tool for probing properties
of two-dimensional conductors. It provides information which is not
directly accessible from single-layer measurements \cite{Narozhny2016}.
In a drag experiment, two layers are placed parallel in close proximity
but are electrically isolated from each other. An electric current
is driven through one layer (active layer), dragging a current
in the second layer (passive layer) through the interlayer electron-electron
interaction. The Coulomb drag has been thoroughly studied in a broad
variety of systems, both experimentally and theoretically. In particular,
it was studied in the quantum Hall regime \cite{Shimshoni1994,Bonsager1996,Lilly1998,Oppen2001,Muraki2004,Gornyi2004,Brener2005},
systems of bilayer excitons \cite{Nandi2012}, graphene-based materials
\cite{Tse2007,Kim2011,Gorbachev2012,Narozhny2012,Titov2013,Li2016,Zhu2020,Zhu2023}
and in the hydrodynamic regime \cite{Apostolov2014,Chen2015,Apostolov2019,Holder2019,Hasdeo2023}.
The Coulomb drag in normal metals is well understood and has been
analyzed by several theoretical methods, including the Boltzmann equation
\cite{Jauho1993}, memory-matrix formalism \cite{Zheng1993} and diagrammatics
\cite{Kamenev1995}. For normal metals, among other properties, it
allows us to quantify the electron-electron (e-e) scattering rates
\cite{Wierkowski1995,Kellogg2002,Asgari2008}.

In recent years, the topological properties of materials have introduced
a fresh perspective on electronic transport \cite{Xiao2010,Cayssol2021}.
Band topology emerges from the geometry of the Bloch wavefunctions
in the momentum space. The band geometry influences the evolution
of an electron wave packet in time, consequently affecting transport
\cite{Sundaram1999}. One of the most recognizable manifestations of
this is the Berry curvature, which induces anomalous processes that
do not have a simple semiclassical interpretation, and changes the
way the physical observables are expressed. For example, the velocity
of electrons is no longer given by the derivative of the single-particle
spectrum but acquires an additional term, known as the anomalous velocity
\cite{Xiao2010}. Band geometry plays a role in various effects, such
as the anomalous Hall effect (AHE) \cite{Haldane2004,Nagaosa2010,Culcer2024},
the spin Hall effect \cite{Sinova2004}, and the photogalvanic effect
\cite{Hosur2011,Morimoto2016,Juan2017,Konig2017,Holder2020,Orenstein2021,Ma2021}.
The interplay between band geometry and electron interactions is a
rich and interesting problem. The Coulomb drag setup is a convenient
platform for studying such effects.

In this work, we propose a simple setup which exemplifies Coulomb
drag in topological metals. This setup consists of a
Weyl semimetal (WSM) layer with broken time-reversal symmetry (TRS) and a normal metal
layer. Having only one layer with non-trivial band topology enables
us to trace all anomalous processes to the WSM. Non-interacting WSMs
exhibit anomalous processes known as side jumps and skew scattering,
occurring in the scattering of electrons off static disorder \cite{Sinitsyn2006,Sinitsyn2007}.
If TRS is broken, these processes contribute to the AHE \cite{Sinitsyn2007}.
It is natural to expect that anomalous processes arise not only in
electron-disorder scattering but in all other collision processes,
such as electron-phonon and electron-electron scattering. Indeed,
it was shown that in e-e scattering, an electron wavepacket acquires
a coordinate shift, which can be interpreted as a side jump \cite{Pesin2018}.
Furthermore, a skew-scattering contribution to the momentum-conserving
e-e collision integral that arises from e-e scattering via an intermediate
state was recently computed \cite{Glazov2022}. 

Despite recent progress, the exploration of band-geometry effects
in e-e scattering remains largely uncharted territory. Specifically,
one may ask: How do these effects manifest in transport properties?
The Coulomb drag setup provides a natural platform to address
this question, as e-e scattering serves as the primary driver of Coulomb
drag, rather than being a secondary process as is typically the case.
One of our main findings is that e-e skew scattering gives rise to
a Hall-like drag force, similar to a Hall viscous response in electron
hydrodynamics.

It is worth mentioning that e-e skew-scattering processes were anticipated
and phenomenologically postulated in the context of spin Hall drag
conductivity \cite{Badalyan2009}. A related problem of the anomalous
Hall drag between two layers of 2D massive Dirac fermions was studied
in Refs. \cite{Liu2017,Liu2019}. These works studied a setup where
both layers are made of topological materials, giving rise to numerous
anomalous processes. Our work focuses on a simpler system, with only
one layer being topological. This enables us to acquire a relatively
transparent physical picture, and to interpret our results in terms
of parallel and perpendicular friction forces.

This paper is organized as follows.

In Sec. \ref{sec:warmup} we briefly review the Hall drag between
two normal metals and contrast it with the WSM-normal metal system.
We describe the phenomena on the level of a qualitative picture.

In Sec. \ref{sec:The-model} we define the model and outline the key
steps of the microscopic calculation. We identify the parts of the
drag response originating from parallel and perpendicular friction.

In Sec. \ref{sec:Results:-drag-conductivity} we compute the Hall
drag conductivity and resistivity.

In Sec. \ref{sec:Summary-and-outlook} we summarize the results and
give a brief outlook for future directions. 

The technical details are delegated to the Appendices.

\section{Hall drag from the kinetic equation: a qualitative discussion\label{sec:warmup}}

In this section, we present a qualitative picture of the Coulomb drag
from the point of view of kinetic theory. Before considering a setup
involving a WSM, we start with the standard
setup of the Coulomb drag between two normal metals.

\subsection{Hall drag between two normal metals in a magnetic field\label{subsec:Hall-drag-between simple picture Hall drag}}

We briefly review the
Hall Coulomb drag in a normal metal-metal bilayer system in a magnetic
field. We consider the simplest case of a parabolic dispersion $\epsilon_{\boldsymbol{k}}=k^{2}/\left(2m\right)$
and constant relaxation times for both layers (for convenience, we
set $\hbar=k_{B}=1$). In this case, while the Hall drag conductivity
$\sigma_{xy}^{D}$ is non-zero, a cancellation leads to a vanishing
Hall drag resistivity $\rho_{xy}^{D}=0$. To show this, we start with
the Boltzmann equation for the electron distribution functions in
constant and spatially uniform fields. In this case, the steady-state
distribution function satisfies

\begin{equation}
e\left(\boldsymbol{E}^{l}+\boldsymbol{v}_{\boldsymbol{k}}^{l}\times\boldsymbol{B}^{l}\right)\cdot\frac{\partial f^{l}}{\partial\boldsymbol{k}}=I^{\textrm{e-e }(l,\bar{l})}\left[f^{l},f^{\bar{l}}\right]+I^{\textrm{dis.}(l)}\left[f^{l}\right].\label{eq:Boltzmann eq with fields}
\end{equation}
Here, the index $l\in\left(\textrm{a},\textrm{p}\right)$ denotes
the active and passive layers, $e$ is the electron charge, $v_{\boldsymbol{k}}^{l}=\partial\epsilon_{\boldsymbol{k}}^{l}/\partial\boldsymbol{k}$
is the electron velocity, $\boldsymbol{E}^{l},\boldsymbol{B}^{l}$
are the electric and magnetic fields in each layer, and $I^{\textrm{e-e }(l,\bar{l})}$,$I^{\textrm{dis.}(l)}$
are the collision integrals corresponding to interlayer electron-electron
and electron-disorder scattering, respectively. We assume the disorder
to be the dominant relaxation mechanism in both layers, so that intralayer
electron-electron scattering can be neglected (the opposite case of
dominant intralayer e-e scattering corresponds to the hydrodynamic
regime, studied in the context of Coulomb drag in a magnetic field
in Refs. \cite{Apostolov2019,Hasdeo2023}). In the relaxation time
approximation, the disorder collision integral reads $I^{\textrm{dis.}(l)}$$\left[\delta f_{\boldsymbol{k}}\right]=-\delta f_{\boldsymbol{k}}^{l}/\tau^{l}$,
with $\delta f_{\boldsymbol{k}}^{l}\equiv f_{\boldsymbol{k}}^{l}-f_{0}(\epsilon_{\boldsymbol{k}}^{l})$
being the non-equilibrium part of the distribution, ($f_{0}(\epsilon)$
being the Fermi-Dirac distribution) and $\tau^{l}$ describing the
momentum relaxation time in the layer.

The non-equilibrium parts of the distribution functions can be parametrized
as boosted velocity distributions

\begin{equation}
\delta f_{\boldsymbol{k}}^{l}=-\frac{\partial f_{0}\left(\epsilon_{\boldsymbol{k}}^{l}\right)}{\partial\epsilon_{\boldsymbol{k}}^{l}}\boldsymbol{k}\cdot\boldsymbol{u}^{l},\label{eq:boost velocity dists}
\end{equation}
with $\boldsymbol{u}^{l}$ being the boost velocity. In the active
layer, the boost velocity is related to the external fields by (assuming the
disorder scattering rate to be much faster than interlayer e-e scattering)
\begin{equation}
\boldsymbol{u}^{\textrm{a}}=\frac{e\tau^{\textrm{a}}}{m}\frac{\boldsymbol{E}^{\textrm{a}}+\frac{e\tau^{\textrm{a}}}{m}\boldsymbol{E}^{\textrm{a}}\times\boldsymbol{B}^{\textrm{a}}}{1+\left(\frac{eB^{\textrm{a}}\tau^{\textrm{a}}}{m}\right)^{2}}.\label{eq:boost velocity active layer}
\end{equation}
The boosted velocity distribution {[}Eq. (\ref{eq:boost velocity dists}){]}
corresponds to an electric current

\begin{equation}
\boldsymbol{j}^{l}=en^{l}\boldsymbol{u}^{l},\label{eq:current and boost velocity}
\end{equation}
where $n^{l}$ is the carrier density of the layer. To analyze the
Coulomb drag, it is useful to consider the force balance acting on
the electrons in each layer. To do so, we multiply the Boltzmann equation
{[}Eq. (\ref{eq:Boltzmann eq with fields}){]} by the momentum $\boldsymbol{k}$
and integrate over\textbf{ $\boldsymbol{k}$} \cite{Song2013}. This
yields

\begin{equation}
-en^{l}\boldsymbol{E}^{l}-\boldsymbol{j}^{l}\times\boldsymbol{B}^{l}=\boldsymbol{F}^{l,\bar{l}}-\frac{m}{e\tau^{l}}\boldsymbol{j}^{l},\label{eq:steady-state Euler eq}
\end{equation}
where
\begin{equation}
\boldsymbol{F}^{l,\bar{l}}\equiv\intop\left(d\boldsymbol{k}\right)\boldsymbol{k}I_{\boldsymbol{k}}^{\textrm{e-e }(l,\bar{l})}\label{eq:momentum transfer rate def}
\end{equation}
is the momentum transfer rate between the layers due to the interlayer
collisions (the drag force). To linear order in the boost velocities,
$\boldsymbol{F}^{l,\bar{l}}$ is given by

\begin{equation}
\boldsymbol{F}^{l,\bar{l}}=\frac{\eta^{D}}{d}\left(\boldsymbol{u}^{\bar{l}}-\boldsymbol{u}^{l}\right)=\frac{\eta^{D}}{d}\frac{1}{e}\left(\frac{\boldsymbol{j}^{\bar{l}}}{n^{\bar{l}}}-\frac{\boldsymbol{j}^{l}}{n^{l}}\right),\label{eq:simplified drag sym scattering}
\end{equation}
where $d$ is the interlayer distance and $\eta^{D}$ is a scalar
coefficient with dimensions of viscosity\footnote{We note that Ref. \cite{Hasdeo2023} defines a different drag viscosity
constant $\nu_{D}$, which quantifies an interlayer drag force response
to the velocity gradients in a single layer. We express the conventional
drag force with the coefficient $\eta^{D}$ with dimensions of viscosity,
conceptualizing drag as a response to the velocity difference along
the axis perpendicular to the bilayer system (the direction normal
to the layers).}. The drag force $\boldsymbol{F}^{l,\bar{l}}$ can be interpreted
as a friction force arising from the relative boost velocity between
the layers.

For the drag \textbf{resistivity} $\rho_{\alpha\beta}^{D}\equiv -E_{\alpha}^{\textrm{p}}/j_{\beta}^{\textrm{a}}$ (the minus sign is conventional),
one sets $\boldsymbol{j}^{\textrm{p}}=0$ and computes $\boldsymbol{E}^{\textrm{p}}$,
finding $e\boldsymbol{E}^{\textrm{p}}=-\boldsymbol{F}^{\textrm{p,a}}=-\eta^{D}\boldsymbol{j}^{\textrm{a}}/\left(en^{\textrm{a}}d\right)$.
Thus, the resulting voltage in the passive layer is parallel to $\boldsymbol{j}^{\textrm{a}}$,
and the drag resistivity is purely longitudinal, i.e.,

\begin{equation}
\rho_{xy}^{D}=0.
\end{equation}
However, for the drag \textbf{conductivity} $\sigma_{\alpha\beta}^{D}\equiv j_{\alpha}^{\textrm{p}}/E_{\beta}^{\textrm{a}}$,
one sets $\boldsymbol{E}^{\textrm{p}}=0$ and computes $\boldsymbol{j}^{\textrm{p}}$.
In the absence of a magnetic field in the passive layer, $\boldsymbol{j}^{\textrm{p}}$
aligns with $\boldsymbol{j}^{\textrm{a}}$ {[}Eqs. (\ref{eq:steady-state Euler eq})
and (\ref{eq:simplified drag sym scattering}){]} and thus a transverse
component in $\boldsymbol{j}^{\textrm{a}}$ creates a corresponding
one in $\boldsymbol{j}^{\textrm{p}}$, leading to a finite Hall drag
conductivity $\sigma_{xy}^{D}$. The fact that $\boldsymbol{j}^{\textrm{p}}\parallel\boldsymbol{j}^{\textrm{a}}$
implies that the ratio $\sigma_{xy}^{D}/\sigma_{xx}^{D}$ is equal
to the Hall ratio of the conductivities of the active layer, $\sigma_{xy}^{\textrm{a}}/\sigma_{xx}^{\textrm{a}}$.
A non-zero magnetic field in the passive layer rotates $\boldsymbol{j}^{\textrm{p}}$
relative to $\boldsymbol{j}^{\textrm{a}}$ {[}Eq. (\ref{eq:steady-state Euler eq}){]},
and the general result is \cite{Kamenev1995}
\begin{equation}
\sigma_{xy}^{D}=\sigma_{xx}^{D}\sum_{l=\textrm{p,a}}\sigma_{xy}^{l}/\sigma_{xx}^{l}.
\end{equation}

In the case of energy-dependent relaxation times $\tau^{l}$ or non-parabolic
dispersion, electrons at different energies are boosted with different
velocities. Therefore, the momentum-relaxing force due to disorder
scattering is no longer given by the rightmost term in Eq. (\ref{eq:steady-state Euler eq}),
and may exist even in the absence of a current \cite{Hu1997,Narozhny2016}.
Additionally, Eq. (\ref{eq:simplified drag sym scattering}) for the
drag force is no longer valid. In that case, there is a weak $\rho_{xy}^{D}\sim T^{4}$
signal in the regime of low temperatures ($T\ll v_{F}/d$), which
is usually considered. However, we note that for high temperatures
($T\gg v_{F}/d$), energy-dependent lifetimes or non-parabolic dispersion
lead to $\rho_{xy}^{D}\sim T$, which is the same temperature dependence as the one of
$\rho_{xx}^{D}$ in this regime.

\subsection{Anomalous Hall drag between a WSM and a normal metal\label{subsec:Anomalous-Hall-drag}}

We now proceed to the case which is the focus of our work, Coulomb
drag between a TRS-broken WSM and a normal metal. Due to Onsager's
symmetry relations \cite{Onsager1931,Onsager1931b,Solomon1991,Gornyi2004}, the tensor of
the kinetic coefficients is symmetric up to a reversal of the magnetic field. This implies that the drag conductivity and resistivity tensors
are the same regardless of which layer is chosen as the active (passive) layer\footnote{In more detail, the kinetic tensor in this case is the generalized conductivity $\sigma^{l,l'}_{\alpha \beta}$
satisfying $j^l_\alpha=\sigma^{l,l'}_{\alpha \beta} E_\beta^{l'}$. The Onsager symmetry relations dictate $\sigma^{l,l'}_{\alpha \beta}(\mathbf{B}) = \sigma^{l',l}_{\beta \alpha }(-\mathbf{B})$, with $\mathbf{B}$ being a TRS-breaking field (in our case, it corresponds to the mean pseudospin in the WSM). We note that the Hall drag conductivity turns out to be symmetric with respect to the choice of the layers for systems with rotational symmetry around the z-axis, due to the following: since the Hall components are proportional to the TRS-breaking field, Onsager's relations imply $\sigma_{xy}^{\textrm{p},\textrm{a}} = -\sigma_{yx}^{\textrm{a},\textrm{p}}$. In the presence of rotational symmetry, $\sigma_{yx}^{l,l'} = -\sigma_{xy}^{l,l'}$, hence $\sigma_{xy}^{\textrm{a},\textrm{p}}=\sigma_{xy}^{\textrm{p},\textrm{a}}\equiv\sigma_{xy}^{D}$.} (we assume rotational symmetry around the z-axis).
From now on, we focus on the case where the WSM is chosen as the active
layer and the normal metal is the passive layer.

First, we discuss the Coulomb drag in this system on a qualitative
level. When an electric field $\boldsymbol{E}^{\textrm{a}}$ acts
on the WSM layer, it induces Hall drag current in the passive layer
through two mechanisms. The first is due to the transverse component
of the Fermi surface shift in the active layer (the component of the
boost velocity $\boldsymbol{u}^{\textrm{a}}$ perpendicular to $\boldsymbol{E}^{\textrm{a}}$).
It appears due to disorder skew scattering in the WSM, enabled by
the broken TRS. The transverse part of $\boldsymbol{u}^{\textrm{a}}$
induces a corresponding transverse Fermi surface shift in the passive
layer. This part of the drag is intuitively clear. The leading interlayer
e-e scattering processes drive the two layers to equilibrate their
momenta, aligning their boost velocities. We thus call this mechanism
\textit{parallel friction}. This mechanism contributes to the Hall
drag conductivity, but not to the Hall drag resistivity, due to the
same reasoning given in the previous section (under the condition
of low temperatures, since the WSM spectrum is non-parabolic). The
second mechanism of Hall drag arises due to a skew-like term in the
interlayer electron-electron scattering rate, originating from the
pseudospin-orbit coupling in the WSM. The interlayer skew scattering gives
rise to a transverse momentum exchange between the two layers, which
we refer to as \textit{Hall friction}.

\begin{widetext}

\begin{figure}[h]
\begin{centering}
\includegraphics[scale=0.95]{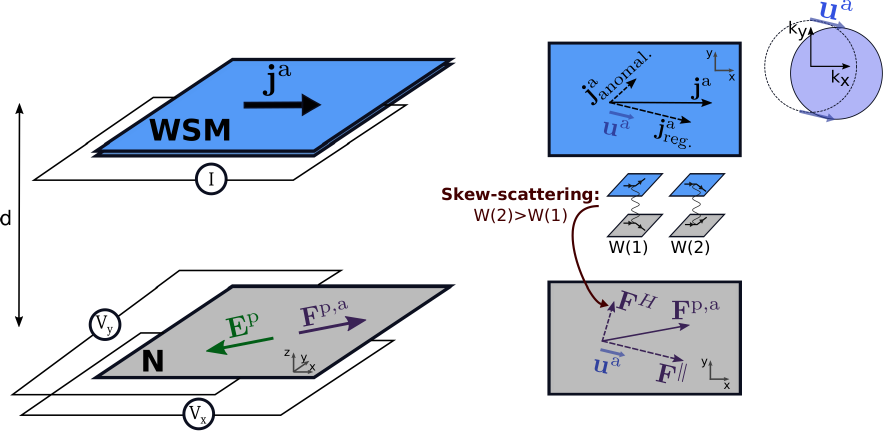}
\par\end{centering}
\caption{Scheme of Coulomb drag between an active WSM layer and a passive normal
metal layer (N). A current is driven through the active (top) layer
and the resulting drag voltage is measured in the passive (bottom)
layer. The electric current in the WSM is made of a regular and an
anomalous part. The regular part corresponds to the shift of the Fermi surface
in the WSM, parametrized by the boost velocity $\boldsymbol{u}^{\textrm{a}}$.
In the regime of low temperatures ($T\ll\min(v_{F}^{\textrm{a}},v_{F}^{\textrm{p}})/d$),
symmetric (in the scattering angle) interlayer e-e collisions result
in a drag force parallel to $\boldsymbol{u}^{a}$, while skew-scattering
collisions result in a Hall drag force $\boldsymbol{F}^{H}$ which
is perpendicular to $\boldsymbol{u}^{\textrm{a}}$. Hall voltage is
induced in the passive layer due to the angles between $\boldsymbol{j}^{\textrm{a}}$
and $\boldsymbol{u}^{\textrm{a}}$ and between $\boldsymbol{F}^{\textrm{p,a}}$
and $\boldsymbol{u}^{\textrm{a}}$. \label{fig:Scheme Hall drag}}
\end{figure}

\end{widetext}

An important difference between the drag between normal metals and
the WSM-normal metal system comes from the anomalous current in the
WSM layer. The electric current in the WSM consists of both a normal
and an anomalous part, resulting in an angle between the current and
the boost velocity in the WSM. Both the Hall friction and the anomalous
current contribute to a non-zero Hall drag resistivity $\rho_{xy}^{D}$.
The summary of this qualitative picture is depicted in Fig. \ref{fig:Scheme Hall drag}.

In a more technical sense, we can express the points mentioned above
as follows. Due to the pseudospin-orbit coupling in the WSM, the interlayer e-e scattering
rate acquires a skew-scattering part. Schematically, a representative
component of the skew-scattering rate is given by
\begin{equation}
W_{\boldsymbol{k},\boldsymbol{k}_{1}\rightarrow\boldsymbol{k}',\boldsymbol{k}_{1'}}^{\textrm{e-e [skew] (p,a)}}\sim\left(\boldsymbol{k}_{1}\times\boldsymbol{k}_{1'}\right)\cdot\boldsymbol{M},\label{eq:asym scat rate phenom.}
\end{equation}
where $\boldsymbol{M}$ is a vector parametrizing the TRS breaking
in the WSM, $\boldsymbol{k},\boldsymbol{k}'$ are the momenta of the
normal metal electrons, and $\boldsymbol{k}_{1},\boldsymbol{k}_{1'}$
are the momenta of the WSM electrons. This scattering process results
in a drag force perpendicular to the boost velocity in the WSM
layer,

\begin{equation}
\sum_{\boldsymbol{k}}\boldsymbol{k}I^{\textrm{e-e [skew] (p,a)}}\propto\eta_{H}^{D}\boldsymbol{u}^{\textrm{a}}\times\boldsymbol{M},\label{eq:transverse friction phenom.}
\end{equation}
with $\eta_{H}^{D}$ being a drag coefficient in the transverse direction.
This drag response from skew-scattering collisions resembles Hall
viscosity \cite{Avron1995,Bradlyn2012}, with $\eta_{H}^{D}$ being
an anti-symmetric part of a response tensor\footnote{In more detail, the viscosity tensor $\eta$ is a four-index tensor
satisfying $\sigma_{\alpha\beta}=\eta_{\alpha\beta\gamma\delta}\partial v_{\delta}/\partial x_{\gamma}$,
with $\sigma_{\alpha\beta}$ being the stress tensor and \textbf{$\boldsymbol{v}$}
being the fluid velocity. The Hall viscosity is anti-symmetric with
respect to the exchange of the pairs $\alpha,\beta\leftrightarrow\gamma,\delta$.
The momentum transfer in Coulomb drag is analogous to a viscous stress
response with $\alpha=\gamma=z$, $z$ being the axis perpendicular
to the layers. The coefficient we denote as $\eta_{H}^{D}$ is thus
analogous to the anti-symmetric component $\left(\eta_{zxzy}-\eta_{zyzx}\right)/2$
of the viscosity tensor.}. We will thereby refer to this part of the friction as Hall friction.

In addition, the current in the WSM consists of both a regular and
an anomalous part, 
\begin{equation}
\boldsymbol{j}^{\textrm{a}}=\boldsymbol{j}_{\textrm{reg.}}^{\textrm{a}}+\boldsymbol{j}_{\textrm{anomal.}}^{\textrm{a}}.
\end{equation}
The part $\boldsymbol{j}_{\textrm{reg.}}^{\textrm{a}}$ corresponds
to the regular part of the velocity operator,
\begin{equation}
\boldsymbol{j}_{\textrm{reg.}}^{\textrm{a}}\equiv e\sum_{n,\boldsymbol{k}}f_{n\boldsymbol{k}}^{\textrm{a}}\boldsymbol{v}_{n\boldsymbol{k}}^{\textrm{a}}=e\sum_{n,\boldsymbol{k}}f_{n\boldsymbol{k}}^{\textrm{a}}\frac{\partial\epsilon_{n\boldsymbol{k}}^{\textrm{a}}}{\partial\boldsymbol{k}},\label{eq:regular current}
\end{equation}
where $n$ denotes the band index. For a boosted velocity distribution,
$\boldsymbol{j}_{\textrm{reg.}}^{\textrm{a}}$ is given by Eq. (\ref{eq:current and boost velocity}).
The anomalous part of the current can be attributed to the off-diagonal
(in band space) elements of the velocity operator. In the semiclassical
language, these can be taken into account as corrections to the velocity
operator from the Berry curvature and side jumps (known as intrinsic
and extrinsic velocities, respectively \cite{Atencia2022}), such
that (see Appendix \ref{subsec:Appendix-C.-Non-interacting} for more
details)

\begin{equation}
\boldsymbol{j}_{\textrm{anomal.}}^{\textrm{a}}\equiv e\sum_{n,\boldsymbol{k}}f_{n\boldsymbol{k}}^{\textrm{a}}\left(\boldsymbol{v}_{n\boldsymbol{k}}^{\textrm{int.}}+\boldsymbol{v}_{n\boldsymbol{k}}^{\textrm{ext.}}\right).\label{eq:anomalous current}
\end{equation}
Importantly, the intrinsic part of the current (corresponding to $\boldsymbol{v}_{n\boldsymbol{k}}^{\textrm{int.}}$)
is a thermodynamic contribution which remains finite even in the absence
of a Fermi surface, as in the case of a Chern insulator \cite{Xiao2010}.
On the other hand, Coulomb drag arises from real transitions on the
Fermi surface due to e-e scattering, and thus vanishes in the presence
of a gap \cite{Liu2017}.

We now proceed to the specific model and microscopic calculations.

\section{Model and outline of the calculation\label{sec:The-model}}

\subsection{The model}

The non-interacting Hamiltonians of the layers are given by

\begin{align}
H^{\textrm{a}} & =\sum_{\xi=\pm1}v_{F}\left(\xi\boldsymbol{\sigma}+C_{\xi}\boldsymbol{\hat{t}}\right)\cdot\left(\boldsymbol{k}-\xi\frac{\boldsymbol{\Delta}_{k}}{2}\right)+V_{\textrm{imp}}^{\textrm{a}},\label{eq:Weyl node Hamiltonian}\\
H^{\textrm{p}} & =\frac{k^{2}}{2m}+V_{\textrm{imp}}^{\textrm{p}}.
\end{align}
The Weyl Hamiltonian consists of two tilted Weyl nodes with opposite
chiralities $\xi=\pm1$ separated in momentum space by $\boldsymbol{\Delta}_{k}$. 
The vector of the Pauli matrices  $\boldsymbol{\sigma}$
represents the pseudospin orbital degrees of freedom.
The constants $C_{\xi}$ describe the tilt in the Weyl nodes. We emphasize that in the absence of a tilt, the low-energy Hamiltonian
describing each node has an emergent TRS (defined within a single node). Since the Coulomb drag is a Fermi-surface effect,
there is no Hall drag conductivity in such case
(we confirm the absence of a Fermi-sea contribution in our calculations, see Appendix \ref{subsec:Appendix-A.-Electron-electron}). Alternatively,
including a finite curvature in the dispersion relation of the WSM is also sufficient to break the emergent TRS.
We note that the breaking of the (ordinary) TRS is necessary for any anomalous Hall effect, including the Hall drag;
two Weyl nodes that are time-reversal partners give opposite contributions to the Hall drag conductivity.

We consider $C_{\pm}=\pm C$ with $\left|C\right|<1$,
for which the Weyl nodes are known as type-I \cite{Armitage2018}.
For simplicity, we take $\boldsymbol{\hat{t}}=\hat{\boldsymbol{\Delta}}_{k}=\boldsymbol{\hat{z}}$ with z
being the axis perpendicular to the layers, so that the AHE in the
WSM is in the x-y plane. Electrons in both layers are coupled via
the Coulomb interaction, $H^{\textrm{e-e}}=\sum V_{\alpha\beta,\alpha'\beta'}^{\textrm{e-e},ll'}c_{\alpha,l}^{\dagger}c_{\beta,l'}^{\dagger}c_{\beta'l'}c_{\alpha'l}$,
where $l,l'$ are layer indices and $\alpha,\alpha',\beta,\beta'$
represent the electron states. The disorder potential $V_{\textrm{imp}}^{l}$
in each layer is characterized by a scattering time $\tau^{l}$. In
our analysis, we make the following assumptions:
\begin{itemize}
\item The interlayer distance and the momentum distance between the Weyl
nodes satisfy $d\gg1/\Delta_{k}$. In this regime, interlayer scattering
involving internode transitions is negligible, and the total drag
is the sum of the contributions from the two independent Weyl nodes.
\item The disorder potential in both layers is characterized by a Gaussian
white-noise potential, with no correlation between the layers.
\item The interlayer e-e scattering time is much longer than the momentum
relaxation time due to the disorder.
\item The interlayer distance is much smaller than the disorder mean free
path of both layers, $d\ll v_{F}^{l}\tau^{l}$ with $v_{F}^{l}$ being
the Fermi velocity of layer $l$ (the ballistic limit of the Coulomb
drag).
\item The thickness of the two layers is much smaller than the interlayer
distance. In this limit, the Coulomb interaction simplifies to the
Coulomb interaction between 2D layers {[}Eq. (\ref{eq:U_R_RPA}) in
Appendix A{]}.
\item The thickness of the WSM is much larger than the interatomic distance,
so that momentum sums in the z-axis can be approximated as integrals.
\item Both layers are weakly interacting Fermi gases at the low-temperature
limit, i.e., $T\ll\epsilon_{F}^{l}$ ($\epsilon_{F}^{l}$ being the
Fermi energy of layer $l$).
\end{itemize}

Let us give an estimate of the scales mentioned above, considering a thin film of the TRS-breaking WSM $\textrm{Co}_{\textrm{3}}\textrm{Sn}_{\textrm{2}}\textrm{S}_{\textrm{2}}$ \cite{Liu2018, Fujiwara2019, Li2020, Tanaka2020} and a high-mobility two-dimensional electron gas in a GaAs/AlGaAs interface \cite{Gramila1991} as the two layers. WSM films are estimated to preserve bulk WSM properties for film thickness as thin as $W\approx10\textrm{nm}$ \cite{Tanaka2020}. The inverse of the internode distance in momentum space is of the order of $1/\Delta_{k}\approx5\textrm{nm}$ \cite{Liu2018, Papaj2021}. Estimating the scattering time in the WSM by $\tau\approx2\textrm{ps}$ and the Fermi velocity by $v_{F}\approx2\times10^{5}\textrm{m}/\textrm{s}$ corresponds to a mean free path of $v_{F}^{\textrm{}}\tau^{\textrm{}}\approx400\textrm{nm}$ in the WSM layer. The mobility of the two-dimensional electron gas in Ref. \cite{Gramila1991} corresponds to a mean free path in the order of 10\textmu{}m. Under these estimates, setting the interlayer distance around $d\approx100\textrm{nm}$ reasonably satisfies our assumptions.

We now proceed to outline the computation of the drag conductivity.

\subsection{Outline of the microscopic calculation}

First, we find the distribution function of the active WSM layer in
the presence of an electric field $\boldsymbol{E}^{\textrm{a}}$,
disregarding the interlayer e-e collisions. The solution of this problem
(non-interacting AHE in the model of a tilted WSM) is known \cite{Papaj2021,Zhang2023,Zhang2023b}.
Here we compute the matrix (in band space) distribution function \cite{Konig2021}
via the Keldysh formalism (see Appendix \ref{subsec:Appendix-A.-Electron-electron}).
The off-diagonal elements of the Keldysh distribution function of
the WSM are small in the dimensionless parameter $1/\left(\epsilon_{F}^{\textrm{a}}\tau^{\textrm{a}}\right)$.
This enables us to express the off-diagonal matrix elements in terms
of the diagonal ones, and consequently to write the e-e collision
integral as a functional of only the semiclassical distribution function
$f_{n\boldsymbol{k}}^{\textrm{a}}$. Due to disorder skew scattering
in the WSM, an electric field $\boldsymbol{E}^{\textrm{a}}$ shifts
its Fermi surface in a direction rotated relative to the field. The
correction to the distribution function, $\delta f_{n\boldsymbol{k}}^{\textrm{a}}\equiv f_{n\boldsymbol{k}}^{\textrm{a}}-f_{0}(\epsilon_{n\boldsymbol{k}}^{\textrm{a}})$,
is given by

\begin{equation}
\delta f_{n\boldsymbol{k}}^{\textrm{a}}=-\frac{\partial f_{0}}{\partial\epsilon_{n\boldsymbol{k}}^{\textrm{a}}}\boldsymbol{v}_{n\boldsymbol{k}}^{\textrm{a}}\cdot\left(e\boldsymbol{E}^{\textrm{a}}+\frac{\tau_{n\boldsymbol{k},\parallel}^{\textrm{a}}}{\tau_{n\boldsymbol{k},\perp}^{\textrm{a}}}e\boldsymbol{E}^{\textrm{a}}\times\hat{z}\right)\tau_{n\boldsymbol{k},\parallel}^{\textrm{a}},\label{eq:decomposition of delta_n}
\end{equation}
where $\tau_{\parallel}^{\textrm{a}}$ and $\tau_{\perp}^{\textrm{a}}$
are the momentum relaxation times in the directions parallel and perpendicular
to the electric field (in our model $\tau_{\parallel}^{\textrm{a}}/\tau_{\perp}^{\textrm{a}}\ll1$,
see Appendix \ref{subsec:Appendix-C.-Non-interacting} for details).
We note that $\delta f_{n\boldsymbol{k}}^{\textrm{a}}$ accounts for
the entire non-equilibrium part of the distribution function, including
the part known as the anomalous distribution arising from the side-jump
correction to the disorder collision integral\footnote{Skew scattering from two adjacent impurities, known as the contribution
from crossing diagrams, should also be included in $\delta\boldsymbol{f}^{\textrm{a}}$
\cite{Ado2015,Ado2016,Zhang2023b}. We neglect the crossing diagrams
in this work to simplify the derivation. The inclusion of these diagrams
amounts to the renormalization of the skew-scattering rate $\tau_{n\boldsymbol{k}}^{\textrm{a},\perp}$.} \cite{Sinitsyn2007}. 

Having solved the distribution function in the active layer, we can
now solve the Boltzmann equation for the passive layer {[}Eq. (\ref{eq:Boltzmann eq with fields}){]}
by substituting the active distribution function in the interlayer
collision integral. Setting $\boldsymbol{E}^{\textrm{p}}=0$ in Eq.
(\ref{eq:Boltzmann eq with fields}) for the calculation of the drag
conductivity, the Boltzmann equation for the passive layer reads

\begin{equation}
0=I^{\textrm{e-e (p,a)}}\left[f^{\textrm{p}},f^{\textrm{a}}\right]+I^{\textrm{dis. (p)}}\left[f^{\textrm{p}}\right],\label{eq:Boltzmann eq passive layer}
\end{equation}
where we recall that $I^{\textrm{e-e (p,a)}}$ and $I^{\textrm{dis. (p)}}$
are the collision integrals due to interlayer e-e scattering and disorder
scattering, respectively. Within the relaxation-time approximation,
the disorder collision integral in the passive layer is given by

\begin{equation}
I^{\textrm{dis. (p)}}\left[f^{\textrm{p}}\right]=-\frac{\delta f_{\boldsymbol{k}}^{\textrm{p}}}{\tau^{\textrm{p}}},\label{eq:rta passive}
\end{equation}
where $\delta f_{\boldsymbol{k}}^{\textrm{p}}\equiv f_{\boldsymbol{k}}^{\textrm{p}}-f_{0}(\epsilon_{\boldsymbol{k}}^{\textrm{p}})$
is the non-equilibrium part of the distribution function in the passive
layer. The interlayer e-e collision integral is given by 

\begin{widetext}

\begin{align}
I_{\boldsymbol{k}}^{\textrm{e-e (p,a)}}\left[f^{\textrm{p}},f^{\textrm{a}}\right] & =-W\sum_{\xi=\pm1}\sum_{n_{1},n_{1'}}\intop_{\boldsymbol{k}',\boldsymbol{k}_{1},\boldsymbol{k}_{1'}}[w_{\boldsymbol{k},n_{1}\boldsymbol{k}_{1}\rightarrow\boldsymbol{k}',n_{1'}\boldsymbol{k}_{1'}}^{\textrm{e-e}}f_{\boldsymbol{k}}^{\textrm{p}}f_{n_{1}\boldsymbol{k}_{1}}^{\textrm{a}}\left(1-f_{\boldsymbol{k}'}^{\textrm{p}}\right)\left(1-f_{n_{1'}\boldsymbol{k}_{1'}}^{\textrm{a}}\right)\nonumber \\
 & -w_{\boldsymbol{k}',n_{1'}\boldsymbol{k}_{1'}\rightarrow\boldsymbol{k},n_{1}\boldsymbol{k}_{1}}^{\textrm{e-e}}f_{\boldsymbol{k}'}^{\textrm{p}}f_{n_{1'}\boldsymbol{k}_{1'}}^{\textrm{a}}\left(1-f_{\boldsymbol{k}}^{\textrm{p}}\right)\left(1-f_{n_{1}\boldsymbol{k}_{1}}^{\textrm{a}}\right)].\label{eq:general coll integral}
\end{align}

\end{widetext}Here, we denote $\intop_{\boldsymbol{k}}\equiv\intop d\boldsymbol{k}/\left(2\pi\right)^{d}$
for the momentum integrations ($d=2,3$ for the metal and the WSM
layers, respectively). We omit the Weyl node index $\xi$ for objects
in the integrand, recalling that we neglect internode scattering.
Since disorder scattering is the dominant momentum relaxation mechanism
in the passive layer, one may replace $\ensuremath{f_{\boldsymbol{k}}^{\textrm{p}}\rightarrow f_{0}(\epsilon_{\boldsymbol{k}}^{\textrm{p}})}$
in the interlayer collision integral {[}Eq. (\ref{eq:general coll integral}){]},
making it a functional of only the active distribution function, $I_{\boldsymbol{k}}^{\textrm{e-e (p,a)}}\left[f^{\textrm{a}}\right]\equiv I_{\boldsymbol{k}}^{\textrm{e-e (p,a)}}\left[f_{\boldsymbol{k}}^{\textrm{p}}=f_{0}(\epsilon_{\boldsymbol{k}}^{\textrm{p}}),f^{\textrm{a}}\right]$.
The factor $W$ (the WSM layer's thickness) in Eq. (\ref{eq:general coll integral})
is due to the quasi-2D nature of the interlayer scattering (see Appendix
\ref{subsec:Appendix-A.-Electron-electron}).

Note that although the collision integral in Eq. (\ref{eq:general coll integral})
looks like a standard e-e collision integral, this is not the case.
The complexity is hidden in the interlayer scattering rate $w_{\boldsymbol{k},n_{1}\boldsymbol{k}_{1}\rightarrow\boldsymbol{k}',n_{1'}\boldsymbol{k}_{1'}}^{\textrm{e-e}}$,
which is computed taking into account virtual transitions in the WSM
(see Appendix \ref{subsec:Appendix-A.-Electron-electron} for more
details). Such processes are crucial for the Hall drag. Among
such processes is the interference between electron-electron
and electron-disorder scattering, breaking the momentum conservation
of the incoming and outgoing electrons. Therefore, one cannot assume
$\boldsymbol{k}_{1}+\boldsymbol{k}=\boldsymbol{k}'+\boldsymbol{k}_{1'}$
in the integrand of Eq. (\ref{eq:general coll integral}) as is usually
the case.

Substituting Eq. (\ref{eq:rta passive}) into Eq. (\ref{eq:Boltzmann eq passive layer}),
one finds

\begin{equation}
\delta f_{\boldsymbol{k}}^{\textrm{p}}=\tau^{\textrm{p}}I_{\boldsymbol{k}}^{\textrm{e-e (p,a)}}\left[f^{\textrm{a}}\right].\label{eq:delta_f_p}
\end{equation}
Employing Eq. (\ref{eq:delta_f_p}), one finds the electric current in
the passive layer

\begin{equation}
\boldsymbol{j}^{\textrm{p}}=e\intop_{\boldsymbol{k}}\boldsymbol{v}_{\boldsymbol{k}}^{\textrm{p}}\delta f_{\boldsymbol{k}}^{\textrm{p}}.\label{eq:j_p}
\end{equation}
The drag conductivity is given by
\begin{equation}
\sigma_{\alpha\beta}^{D}\equiv\frac{j_{\alpha}^{\textrm{p}}}{E_{\beta}^{\textrm{a}}}.\label{eq:drag conductivity def}
\end{equation}

For a passive layer with a parabolic spectrum and within the relaxation-time approximation,
one can relate the drag current to the drag force
(or, momentum transfer rate between the layers) $\boldsymbol{F}^{\textrm{p,a}}\equiv\intop_{\boldsymbol{k}}\boldsymbol{k}I_{\boldsymbol{k}}^{\textrm{e-e (p,a)}}$.
Employing Eqs. (\ref{eq:delta_f_p}) and (\ref{eq:j_p}), one finds

\begin{equation}
\boldsymbol{j}^{\textrm{p}}=\frac{e\tau^{\textrm{p}}}{m}\boldsymbol{F}^{\textrm{p,a}}\left[f^{\textrm{a}}\right].\label{eq:passive current and momentum transfer}
\end{equation}
Thinking of the Coulomb drag in terms of forces gives additional insight.
In the experimentally prevalent regime of low temperatures ($T\ll T_{d}$,
with $T_{d}\equiv\min\left(v_{F}^{\textrm{a}},v_{F}^{\textrm{p}}\right)/d$), the Hall drag conductivity can be divided into two parts\footnote{The meaning of the energy scale $T_{d}$ is as follows \cite{Narozhny2016}: the typical scale of momentum transfer in an interlayer collision is determined
by the interlayer screening to be $q\sim1/d$ {[}see Eq. (\ref{eq:U_R_RPA}){]}.
For this typical momentum, $T_{d}$ is the maximal energy that allows
a particle-hole excitation in both layers.}:
\begin{enumerate}
\item \textit{Parallel friction}: Drag force $\boldsymbol{F}^{\textrm{p,a}}$
which is parallel to the Fermi surface shift in the active layer.
Because an electric field $\boldsymbol{E}^{\textrm{a}}$ in the WSM
layer creates a perpendicular component in the Fermi surface shift
due to intralayer skew scattering {[}second term in Eq. (\ref{eq:decomposition of delta_n}){]},
parallel friction creates a corresponding component in the passive
layer current which is perpendicular to $\boldsymbol{E}^{\textrm{a}}$,
i.e., Hall drag current.
\item \textit{Hall friction}: Drag force $\boldsymbol{F}^{\textrm{p,a}}$
which is perpendicular to the Fermi surface shift in the active layer.
This part of the drag arises due to the many-body skew-scattering
part of the interlayer collision integral.
\end{enumerate}
In the opposite regime of high temperatures ($T\gg T_{d}$), the picture
is complicated by the energy dependence of the Fermi surface shift
in the active layer {[}Eq. (\ref{eq:decomposition of delta_n}){]}.
In this case, the drag force can be decomposed into three components:
parallel and perpendicular to the Fermi surface shift as in the previous
case, as well as a component related to the energy dependence of the
Fermi surface shift. In this case, even on the level of a simple interlayer
collision integral (disregarding the interlayer skew-scattering part),
the drag force $\boldsymbol{F}^{\textrm{p,a}}$ is generally not parallel
to the Fermi surface shift. 

Having outlined the main steps of the calculation, we now turn to
the computation of the drag conductivity and resistivity.

\section{Results: drag conductivity and resistivity\label{sec:Results:-drag-conductivity}}

First we present the interlayer e-e collision integral in more detail,
introducing its skew-scattering part.

\subsection{Interlayer e-e collision integral}

The interlayer collision integral can be written in the general form of Eq. (\ref{eq:general coll integral}),
with the e-e scattering rate separated into contributions
from three different processes,

\begin{widetext}

\begin{equation}
w_{\boldsymbol{k},n_{1}\boldsymbol{k}_{1}\rightarrow\boldsymbol{k}',n_{1'}\boldsymbol{k}_{1'}}^{\textrm{e-e}}=w_{\boldsymbol{k},n_{1}\boldsymbol{k}_{1}\rightarrow\boldsymbol{k}',n_{1'}\boldsymbol{k}_{1'}}^{\textrm{Born}}+w_{\boldsymbol{k},n_{1}\boldsymbol{k}_{1}\rightarrow\boldsymbol{k}',n_{1'}\boldsymbol{k}_{1'}}^{\textrm{s.j.}}+w_{\boldsymbol{k},n_{1}\boldsymbol{k}_{1}\rightarrow\boldsymbol{k}',n_{1'}\boldsymbol{k}_{1'}}^{\textrm{\textrm{e-e-imp}}}.\label{eq:scattering rate detail}
\end{equation}

\end{widetext}The term $w^{\textrm{Born}}$ refers to the part calculated
on the level of the Born approximation within the RPA (random-phase
approximation) approach, resulting in a scattering rate proportional to the square
of the screened Coulomb potential {[}Eq. (\ref{eq:symmetric scat rate})
in the Appendix{]}. It is an even function of the angle between the
momenta of the scattering electrons in the WSM. Although it is the
largest part of the scattering rate, the two other scattering processes are
of equal importance to Hall drag, since they give rise to interlayer
skew scattering. The term $w^{\textrm{s.j.}}$ corresponds to a correction
due to side jumps of the WSM electrons; $w^{\textrm{\textrm{e-e-imp}}}$
corresponds to the interference between interlayer e-e scattering and e-impurity
scattering in the WSM. These scattering rates are calculated using
the Keldysh formalism, accounting for processes involving virtual
transitions in the WSM layer. The virtual transitions correspond to the
interband elements of the matrix Green's functions. The full expressions
for these rates are presented in Appendix \ref{subsec:Appendix-A.-Electron-electron}
{[}Eq. (\ref{eq:W_k_k sj}) for $w^{\textrm{s.j.}}$ and Eq. (\ref{eq:W_skew full})
for $w^{\textrm{\textrm{e-e-imp}}}${]}. We now briefly describe the
physical processes giving rise to the skew-scattering terms.

The side-jump process modifies the interlayer e-e collision integral
in an analogous way to the way it modifies the electron-disorder collision
integral \cite{Sinitsyn2006}. In the context of interlayer e-e scattering,
a WSM electron acquires a coordinate shift when it scatters from the
incoming into the outgoing state (thus, ``side jump''). In the presence
of an external electric field, this coordinate shift changes the electric potential
energy of the electron. Therefore, the energy conservation condition
for the e-e scattering process is modified. Consequently, the side-jump
scattering rate $w^{\textrm{s.j.}}$ is proportional to the applied
electric field ${\bf E}^{\textrm{a}}$, and scales linearly with it
in the low-field limit. Therefore, on the level of the linear response,
one replaces the distribution functions in Eq. (\ref{eq:general coll integral})
with their equilibrium values, $f_{n\boldsymbol{k}}^{\textrm{a}}\rightarrow f_{0}(\epsilon_{n\boldsymbol{k}}^{\textrm{a}})$.
Even in this approximation of equilibrium distribution functions,
the side-jump scattering rate results in a finite contribution to
the collision integral.

Next we discuss the last term in Eq. ($\ref{eq:scattering rate detail}$),
$w_{\boldsymbol{k},n_{1}\boldsymbol{k}_{1}\rightarrow\boldsymbol{k}',n_{1'}\boldsymbol{k}_{1'}}^{\textrm{\textrm{e-e-imp}}}$.
It involves scattering through an intermediate state, and is proportional
to the imaginary part of the overlap between the Bloch wavefunctions involved in the
scattering. Because this term involves both e-e and disorder scattering,
it does not conserve the total electron momenta, unlike the other
scattering processes discussed above, which are proportional to the
delta function $\delta_{\boldsymbol{k}+\boldsymbol{k}_{1}-\boldsymbol{k}'-\boldsymbol{k}_{1'}}$.

We now move on to the calculation of the drag conductivities.

\subsection{Drag conductivity}

For the clarity of computation, we focus on two limiting cases: low
temperatures ($T\ll T_{d}$) and high temperatures ($T\gg T_{d}$)
[we remind the reader the definition $T_{d}\equiv\min\left(v_{F}^{\textrm{a}},v_{F}^{\textrm{p}}\right)/d$].

\subsubsection{Low temperatures \label{subsec:Low-temperatures}}

In the regime of low temperatures ($T\ll T_{d}$), the distribution
function $\delta f_{n\boldsymbol{k}}^{\textrm{a}}$  {[}Eq. ($\ref{eq:decomposition of delta_n}$){]}
can be approximated by a boosted velocity distribution {[}Eq. ($\ref{eq:boost velocity dists}$){]}.
This is done by replacing $\boldsymbol{v}_{n\boldsymbol{k}}^{\textrm{a}}\rightarrow\left(v_{F}^{\textrm{a}}\right)^{2}\boldsymbol{k}/\epsilon_{F}^{\textrm{a}}$
(this misses a term in the z-component of $\boldsymbol{v}_{n\boldsymbol{k}}^{\textrm{a}}$,
but we are interested in the components in the x-y plane) and neglecting
the energy dependence of the relaxation times. These approximations
are justified since the particle-hole scattering is predominantly
perpendicular in both layers, making the collision quasi-elastic
{[}i.e., $\boldsymbol{q}\perp\boldsymbol{v}_{\boldsymbol{k}}$
where $\boldsymbol{q}$ is the momentum exchange in the collision
and $\boldsymbol{v}_{\boldsymbol{k}}$ is the electron velocity. For a detailed discussion, see the text
following Eq. (\ref{eq:eta_1 integral}) of the Appendix{]}. This
accuracy is sufficient to account for the leading part of the drag
conductivities in the small parameter $T/T_{d}$. We introduce the
parametrization 

\begin{equation}
\delta f_{n\boldsymbol{k}}^{\textrm{a}}=-T\frac{\partial f_{0}(\epsilon_{n\boldsymbol{k}}^{\textrm{a}})}{\partial\epsilon_{n\boldsymbol{k}}^{\textrm{a}}}g_{n\boldsymbol{k}}^{\textrm{a}}.\label{eq:delta f parametrization}
\end{equation}
After substituting Eq. ($\ref{eq:delta f parametrization}$) in Eq.
($\ref{eq:decomposition of delta_n}$), one finds

\begin{equation}
g_{n\boldsymbol{k}}^{\textrm{a}}=\frac{\boldsymbol{k}\cdot\boldsymbol{u}^{\textrm{a}}}{T},\label{eq:g_nk boost velocity}
\end{equation}
where $\boldsymbol{u}^{\textrm{a}}$ is the boost velocity in the
active layer, given by

\begin{equation}
u_{\alpha}^{\textrm{a}}=\frac{\left(v_{F}^{\textrm{a}}\right)^{2}\tau_{\parallel}^{\textrm{a}}}{\epsilon_{F}^{\textrm{a}}}\left(\delta_{\alpha\beta}+\epsilon_{\alpha\beta}\frac{\tau_{\parallel}^{\textrm{a}}}{\tau_{\perp}^{\textrm{a}}}\right)eE_{\beta}^{\textrm{a}}.\label{eq:boost velocity const energy}
\end{equation}
Here, the momentum relaxation times $\tau_{\parallel}^{\textrm{a}},\tau_{\perp}^{\textrm{a}}$
are computed at the Fermi energy (see Appendix \ref{subsec:Appendix-C.-Non-interacting}).
We now substitute the distribution function in the active layer with
the non-equilibrium part given by Eq. ($\ref{eq:delta f parametrization}$)
into the interlayer e-e collision integral {[}Eq. $(\ref{eq:general coll integral}$)
with the scattering rates given in Eq. (\ref{eq:scattering rate detail}){]}
, and derive the linearized interlayer collision integral

\begin{widetext}

\begin{align}
{\cal I}_{\boldsymbol{k}}^{\textrm{e-e (p,a)}} & =-\frac{W}{T}\sum_{\xi=\pm1}\sum_{n_{1},n_{1'}}\intop_{\boldsymbol{q},\boldsymbol{k}_{1},\boldsymbol{k}_{1'}}f_{0}(\epsilon_{\boldsymbol{k}}^{\textrm{p}})f_{0}(\epsilon_{n_{1}\boldsymbol{k}_{1}}^{\textrm{a}})\left(1-f_{0}(\epsilon_{\boldsymbol{k}+\boldsymbol{q}}^{\textrm{p}})\right)\left(1-f_{0}(\epsilon_{n_{1'}\boldsymbol{k}_{1'}}^{\textrm{a}})\right)\nonumber \\
 & \times\left[w_{\boldsymbol{k},n_{1}\boldsymbol{k}_{1}\rightarrow\boldsymbol{k}+\boldsymbol{q},n_{1'}\boldsymbol{k}_{1'}}^{\textrm{Born}}\boldsymbol{q}\cdot\boldsymbol{u}^{\textrm{a}}+w_{\boldsymbol{k},n_{1}\boldsymbol{k}_{1}\rightarrow\boldsymbol{k}+\boldsymbol{q},n_{1'}\boldsymbol{k}_{1'}}^{\textrm{\textrm{e-e-imp}}}\left(\boldsymbol{k}_{1}-\boldsymbol{k}_{1'}\right)\cdot\boldsymbol{u}^{\textrm{a}}+2w_{\boldsymbol{k},n_{1}\boldsymbol{k}_{1}\rightarrow\boldsymbol{k}+\boldsymbol{q},n_{1'}\boldsymbol{k}_{1'}}^{\textrm{s.j.}}T\right].\label{eq:linearized coll integral}
\end{align}

\end{widetext}Here, we utilized the momentum conservation of the
process corresponding to $w^{\textrm{Born}}$, forcing $\boldsymbol{k}_{1'}=\boldsymbol{k}_{1}-\boldsymbol{q}$
in that term.

We now calculate the drag force $\boldsymbol{F}^{\textrm{p,a}}$ between
the layers by substituting Eq. (\ref{eq:linearized coll integral})
into Eq. (\ref{eq:momentum transfer rate def}). The resulting drag
force can be written as a linear response to the boost velocity in
the active layer. We identify the generation of diagonal and Hall-like
responses, writing

\begin{equation}
F_{\alpha}^{\textrm{p,a}}=\left(\delta_{\alpha\beta}\eta_{\parallel}^{D}+\epsilon_{\alpha\beta}\eta_{H}^{D}\right)\frac{u_{\beta}^{\textrm{a}}}{d}.\label{eq:transfered momentum response}
\end{equation}
Here, the diagonal response $\eta_{\parallel}^{D}$ is generated by
the Born-approximation part of the collision integral, and the Hall-like
response $\eta_{H}^{D}$ comes from the many-body skew-scattering
processes corresponding to e-e-impurity interference and side jumps\footnote{We note that $w^{\textrm{s.j. }}$ strictly generates a term in the
drag force that is proportional to $\boldsymbol{E}^{\textrm{a}}$
rather than to $\boldsymbol{u}^{\textrm{a}}$. We have substituted
$\boldsymbol{E}^{\textrm{a}}\approx\epsilon_{F}^{\textrm{a}}/\left(\left(v_{F}^{\textrm{a}}\right)^{2}\tau^{\textrm{a}}\right)\boldsymbol{u}^{\textrm{a}}$
in that term, neglecting a further subleading term in $1/\left(\epsilon_{F}^{\textrm{a}}\tau^{\textrm{a}}\right)$
which is beyond the accuracy of our calculations.}. Calculation of the momentum transfer with the total interlayer scattering
rate (see Appendix \ref{subsec:Appendix-B:-Momentum}) yields

\begin{align}
\eta_{\parallel}^{D} & =\frac{\pi\zeta(3)}{32}\frac{T^{2}}{v_{F}^{\textrm{a}}v_{F}^{\textrm{p}}\kappa^{\textrm{a}}\kappa^{\textrm{p}}d^{3}},\label{eq:eta_parallel_low_T}\\
\eta_{H}^{D} & =-\frac{C}{2\epsilon_{F}^{\textrm{a}}\tau^{\textrm{a}}}\eta_{\parallel}^{D}.\label{eq:eta_H_low_T}
\end{align}
Here, $\zeta(z)$ is the Riemann Zeta function, and $\kappa^{l}=2\pi e^{2}\nu_{2d}^{l}/\epsilon_{r}$
is the Thomas-Fermi wavevector of layer $l$ with 2D density of states
$\nu_{2d}^{l}$ ($\nu_{2d}^{\textrm{p}}\equiv\nu^{\textrm{p}}$ for
the metal layer and $\nu_{2d}^{\textrm{a}}\equiv\nu^{\textrm{a}}W$
for the WSM layer) and dielectric constant $\epsilon_{r}$. We have
thus found the dependence of the drag force on the active layer's boost
velocity. One can readily find the drag force $\boldsymbol{F}^{\textrm{p,a}}$
as a function of the electric field in the active layer by substituting
Eq. ($\ref{eq:boost velocity const energy}$) into Eq. (\ref{eq:transfered momentum response}).
The dragged current is proportional to the drag force {[}Eq. ($\ref{eq:passive current and momentum transfer}$){]}.
Employing Eq. ($\ref{eq:drag conductivity def}$), one finds the drag
conductivity. For the longitudinal component, one finds

\begin{equation}
\sigma_{xx}^{D}=e^{2}\frac{\ell^{\textrm{a}}\ell^{\textrm{p}}}{k_{F}^{\textrm{a}}k_{F}^{\textrm{p}}}\frac{\eta_{\parallel}^{D}}{d}=e^{2}\frac{\pi\zeta(3)}{64}\frac{T^{2}}{\epsilon_{F}^{\textrm{a}}\epsilon_{F}^{\textrm{p}}}\frac{\ell^{\textrm{a}}\ell^{\textrm{p}}}{\kappa^{\textrm{a}}\kappa^{\textrm{p}}d^{4}},
\end{equation}
where $\ell^{l}\equiv v_{F}^{l}\tau_{\parallel}^{l}$ is the mean
free path in layer $l$ (for the passive layer, $\tau_{\parallel}^{\textrm{p}}=\tau^{\textrm{p}})$.
This result differs from the one for the drag between
two 2D metals \cite{Kamenev1995} by a numerical
factor due to the dimensionality of the WSM layer.

Next we discuss the Hall drag conductivity. Since the boost velocity
in the active layer is rotated relative to the electric field, a corresponding
Hall current is induced in the passive layer by the parallel friction.
Denoting the corresponding contribution to the Hall drag conductivity
by $\sigma_{xy}^{D[\eta_{\parallel}^{D}]}$, one finds

\begin{equation}
\sigma_{xy}^{D[\eta_{\parallel}^{D}]}=\frac{\tau_{\parallel}^{\textrm{a}}}{\tau_{\perp}^{\textrm{a}}}\sigma_{xx}^{D}=\frac{3C}{2\epsilon_{F}^{\textrm{a}}\tau^{\textrm{a}}}\sigma_{xx}^{D}.\label{eq:parallel friction simple result}
\end{equation}

The Hall friction gives rise to a force $F_{\alpha}^{H}\equiv\epsilon_{\alpha\beta}\eta_{H}^{D}u_{\beta}^{\textrm{a}}/d$
perpendicular to the boost velocity. This results in an additional
contribution to the Hall drag conductivity,

\begin{equation}
\sigma_{xy}^{D[\eta_{H}^{D}]}=e^{2}\frac{\ell^{\textrm{a}}\ell^{\textrm{p}}}{k_{F}^{\textrm{a}}k_{F}^{\textrm{p}}}\frac{\eta_{H}^{D}}{d}.\label{eq:sigma_xy transverse fric.}
\end{equation}
Substituting the value of the Hall response coefficient $\eta_{H}^{D}$,
one gets

\begin{equation}
\sigma_{xy}^{D[\eta_{H}^{D}]}=-\frac{C}{2\epsilon_{F}^{\textrm{a}}\tau^{\textrm{a}}}\sigma_{xx}^{D}.
\end{equation}

The total Hall drag conductivity is additive in the contributions,

\begin{equation}
\sigma_{xy}^{D}=\sigma_{xy}^{D[\eta_{\parallel}^{D}]}+\sigma_{xy}^{D[\eta_{H}^{D}]}=\frac{C}{\epsilon_{F}^{\textrm{a}}\tau^{\textrm{a}}}\sigma_{xx}^{D}.
\end{equation}

The ratio $\sigma_{xy}^{D}/\sigma_{xx}^{D}\simeq C/\left(\epsilon_{F}^{\textrm{a}}\tau^{\textrm{a}}\right)$
is parametrically equal to the ratio between the Fermi-surface contribution
of the Hall conductivity and the longitudinal conductivity in the
non-interacting WSM \cite{Papaj2021,Zhang2023}. Note that unlike the anomalous Hall conductivity,
which has an intrinsic contribution proportional to the momentum distance
between the Weyl nodes ($\sigma_{xy}^{\textrm{int}}\sim\Delta_{k}$),
the Hall drag conductivity has no bulk contribution \cite{Liu2017,Liu2019}.

\subsubsection{High temperatures\label{subsec:High-temperatures drag conductivity}}

In the high-temperature limit ($T\gg T_{d}$), an additional complication
arises due to the deviation of the distribution function of the active
layer {[}Eq. ($\ref{eq:decomposition of delta_n}$){]} from a boosted
velocity distribution. This deviation is due to the non-parabolic
spectrum of the WSM and the energy dependence of the relaxation times.
In this case, one needs to replace Eq. (\ref{eq:g_nk boost velocity})
with an ``energy-dependent boost velocity'' ansatz,

\begin{equation}
g_{n\boldsymbol{k}}^{\textrm{a}}=\frac{\boldsymbol{k}\cdot\boldsymbol{u}^{\textrm{a}}(\epsilon_{n\boldsymbol{k}}^{\textrm{a}})}{T},\label{eq:g_nk energy-dependent boost velocity}
\end{equation}
where $\boldsymbol{u}^{\textrm{a}}(\epsilon)$ is the boost velocity
at a given energy, analogous to Eq. (\ref{eq:boost velocity const energy})
evaluated at energy $\epsilon$,

\begin{equation}
u_{\alpha}^{\textrm{a}}(\epsilon)=\frac{\left(v_{F}^{\textrm{a}}\right)^{2}\tau_{\parallel}^{\textrm{a}}(\epsilon)}{\epsilon}\left(\delta_{\alpha\beta}+\epsilon_{\alpha\beta}\frac{\tau_{\parallel}^{\textrm{a}}(\epsilon)}{\tau_{\perp}^{\textrm{a}}(\epsilon)}\right)eE_{\beta}^{\textrm{a}}.\label{eq:boost velocity energy dependent}
\end{equation}
Note that we have neglected the anisotropy in the WSM by approximating
$\boldsymbol{u}^{\textrm{a}}(\epsilon,\hat{k})\sim\boldsymbol{u}^{\textrm{a}}(\epsilon)$,
taking into account the leading order in the tilt parameter $C$. 

By calculating the momentum transfer from the interlayer e-e collision
integral as in the previous section, we now find the drag force

\begin{equation}
F_{\alpha}^{\textrm{p,a}}=\left(\delta_{\alpha\beta}\eta_{\parallel(0)}^{D}+\epsilon_{\alpha\beta}\eta_{H}^{D}\right)\frac{u_{\beta}^{\textrm{a}}}{d}+\delta_{\alpha\beta}\eta_{\parallel(1)}^{D}\frac{\epsilon_{F}^{\textrm{a}}}{d}\left.\frac{\partial u_{\beta}^{\textrm{a}}(\epsilon)}{\partial\epsilon}\right|_{\epsilon=\epsilon_{F}^{\textrm{a}}}.\label{eq:F_drag high T}
\end{equation}
The terms $\eta_{\parallel(0)}^{D}$ and $\eta_{H}^{D}$ correspond
respectively to the Born-approximation and skew-scattering rates as
in the previous case ($\eta_{\parallel(0)}^{D}$ has the same meaning
as the non-indexed $\eta_{\parallel}^{D}$ in the previous section).
In addition, there is a term proportional to the energy derivative
of the active layer's boost velocity, which was not present in the
previous case. This term arises even within the Born-approximation
part of the interlayer collision integral. Generally, the vector $\left[\partial\boldsymbol{u}^{\textrm{a}}(\epsilon)/\partial\epsilon\right]_{\epsilon=\epsilon_{F}^{\textrm{a}}}$
is not parallel to $\boldsymbol{u}^{\textrm{a}}$, and therefore even
in the Born-approximation level, the interlayer collision integral
generates momentum transfer perpendicular to $\boldsymbol{u}^{\textrm{a}}$.
The drag coefficients in the high-temperature and small-tilt ($C\ll1$)
limits are given by

\begin{align}
\eta_{\parallel(0)}^{D} & =\bar{\eta}Q_{1}(\frac{v_{F}^{\textrm{p}}}{v_{F}^{\textrm{a}}}),\label{eq:eta_parallel high T}\\
\eta_{\parallel(1)}^{D} & =\bar{\eta}Q_{2}(\frac{v_{F}^{\textrm{p}}}{v_{F}^{\textrm{a}}})\times\min\left[1,\left(\frac{v_{F}^{\textrm{p}}}{v_{F}^{\textrm{a}}}\right)^{2}\right],\label{eq:eta u deriv}\\
\eta_{H}^{D} & =-\frac{C}{2\epsilon_{F}^{\textrm{a}}\tau^{\textrm{a}}}\bar{\eta}Q_{3}(\frac{v_{F}^{\textrm{p}}}{v_{F}^{\textrm{a}}}),\label{eq:eta_H high T}
\end{align}
where we defined

\begin{equation}
\bar{\eta}^{D}\equiv\frac{\pi^{3}}{480}\frac{T_{d}T}{v_{F}^{\textrm{a}}v_{F}^{\textrm{p}}\kappa^{\textrm{a}}\kappa^{\textrm{p}}d^{3}},\label{eq:eta_bar}
\end{equation}
and $Q_{1,2,3}(z)$ are factors of order one given in the Appendix
{[}Eqs. (\ref{eq:F1_factor})-(\ref{eq:F3_factor}){]}. Note that
$\eta_{\parallel(1)}^{D}$ {[}the drag coefficient multiplying $\partial\boldsymbol{u}^{\textrm{a}}(\epsilon)/\partial\epsilon${]}
is suppressed in the limit $v_{F}^{\textrm{p}}/v_{F}^{\textrm{a}}\ll1$.
This comes from the phase-space restrictions of the interlayer e-e
scattering. The particle-hole pairs in both layers have to satisfy
$\boldsymbol{v}_{F}^{l}\cdot\boldsymbol{q}=\omega$, with $\omega$
being the energy transfer in the collision. In the case $v_{F}^{\textrm{p}}/v_{F}^{\textrm{a}}\ll1$,
forward scattering is suppressed in the WSM (i.e., scattering with
$\boldsymbol{q}$ parallel to the velocity of the WSM electron $\boldsymbol{v}_{\boldsymbol{k}_{1}}^{\textrm{a }}$).
Thus, the effect of the energy dependence of the boost velocity on
the drag is negligible in this limit.

It is reasonable to assume that the typical energy dependence of the
transport times in realistic materials is a power-law function. Consequently,
the boost velocity has a power-law energy dependence {[}see Eq. (\ref{eq:boost velocity energy dependent}){]}.
Generally, the parallel and perpendicular (relative to the electric
field) components of the boost velocity may have different scalings
with energy. We denote the scaling of these components {[}first and second terms
in Eq. (\ref{eq:boost velocity energy dependent}){]} as $u_{\parallel}^{\textrm{a}}(\epsilon)\sim\epsilon^{b_{\parallel}}$
and $u_{\perp}^{\textrm{a}}(\epsilon)\sim\epsilon^{b_{\perp}}$. The
drag conductivities are given by

\begin{align}
\sigma_{xx}^{D} & =e^{2}\frac{\ell^{\textrm{a}}\ell^{\textrm{p}}}{k_{F}^{\textrm{a}}k_{F}^{\textrm{p}}d}\left(\eta_{\parallel(0)}^{D}+b_{\parallel}\eta_{\parallel(1)}^{D}\right),\label{eq:sigma_xx_D high T}\\
\sigma_{xy}^{D[\eta_{\parallel}^{D}]} & =e^{2}\frac{\ell^{\textrm{a}}\ell^{\textrm{p}}}{k_{F}^{\textrm{a}}k_{F}^{\textrm{p}}d}\frac{\tau_{\parallel}^{\textrm{a}}}{\tau_{\perp}^{\textrm{a}}}\left(\eta_{\parallel(0)}^{D}+b_{\perp}\eta_{\parallel(1)}^{D}\right),\label{eq:sigma_xy_D_par_fric high T}
\end{align}
and $\sigma_{xy}^{D[\eta_{H}^{D}]}$ still given by Eq. (\ref{eq:sigma_xy transverse fric.}).
In our model, $b_{\parallel}=-3$ and $b_{\perp}=-2$. Thus, the drag
force computed within the Born approximation {[}the terms accounted
by $\eta_{\parallel(0)}^{D}$ and $\eta_{\parallel(1)}^{D}$ in Eq.
(\ref{eq:F_drag high T}){]} in our model is indeed not parallel to
$\boldsymbol{u}^{\textrm{a}}$.

Since $b_{\parallel}$ and $b_{\perp}$ are negative, the two terms
in both Eqs. (\ref{eq:sigma_xx_D high T}) and (\ref{eq:sigma_xy_D_par_fric high T})
are of opposite sign. Depending on the numerical prefactors, this
may result in an opposite sign for the drag conductivities in the
two limits of $T\ll T_{d}$ and $T\gg T_{d}$, and thus lead to a
non-monotonous temperature dependence. Physically, the non-monotonous
behavior can be understood as follows. When $u_{\alpha}^{\textrm{a}}(\epsilon)$
is a decreasing function of the energy, quasi-elastic and forward
(strongly inelastic) interlayer scattering processes give opposite
contributions to the drag force $F_{\alpha}^{\textrm{p,a}}$. Since
forward scattering gives a significant contribution only for temperatures
$T\gtrsim T_{d}$, a non-monotonous temperature dependence of the
drag conductivities may arise. The quasi-elastic contribution to the
drag is conventional, and its direction depends on the signs of the
curvatures of the single-particle spectrum in the layers \cite{Gornyi2004}.
The sign of the contribution due to forward scattering is controlled
by the energy dependence of $u_{\alpha}^{\textrm{a}}(\epsilon)$,
quantified by the coefficients $b_{\parallel}$ and $b_{\perp}$ {[}these
are related to the transport times, see Eq. (\ref{eq:boost velocity energy dependent}){]}.
We note that the scenario where forward and quasi-elastic interlayer
scattering contribute to the drag in opposite directions is quite
general. It is expected to occur in a generic Coulomb drag setup,
provided that the scattering time in one of the layers is a decreasing
function of energy. In the case where both layers have energy-dependent
scattering times, the behavior is more complicated, since forward
scattering can be more dominant in one layer than in the other depending
on the spectrum of the two layers. For a scattering event in which
both electrons in the two layers scatter in the forward direction,
the sign of the resulting contribution to the drag also depends on
the product of the derivatives $\partial\tau^{l}(\epsilon)/\partial\epsilon$.
For two identical layers, both scattering mechanisms contribute positively
to the drag conductivity. Thus, the two layers being different is
essential for a non-monotonous temperature dependence of the drag.

To summarize, on a qualitative level, the direction of the drag is
controlled by two independent mechanisms: (i) the curvatures of the
single-particle spectrum in two layers; (ii) the direction in which
the momentum relaxation rate $\tau(\epsilon)$ changes with energy.
The effect (ii) is pronounced only when the temperature is not too
low ($T\gtrsim T_{d}$).

\begin{widetext}

\begin{figure}[h]
\begin{centering}
\includegraphics[scale=0.33]{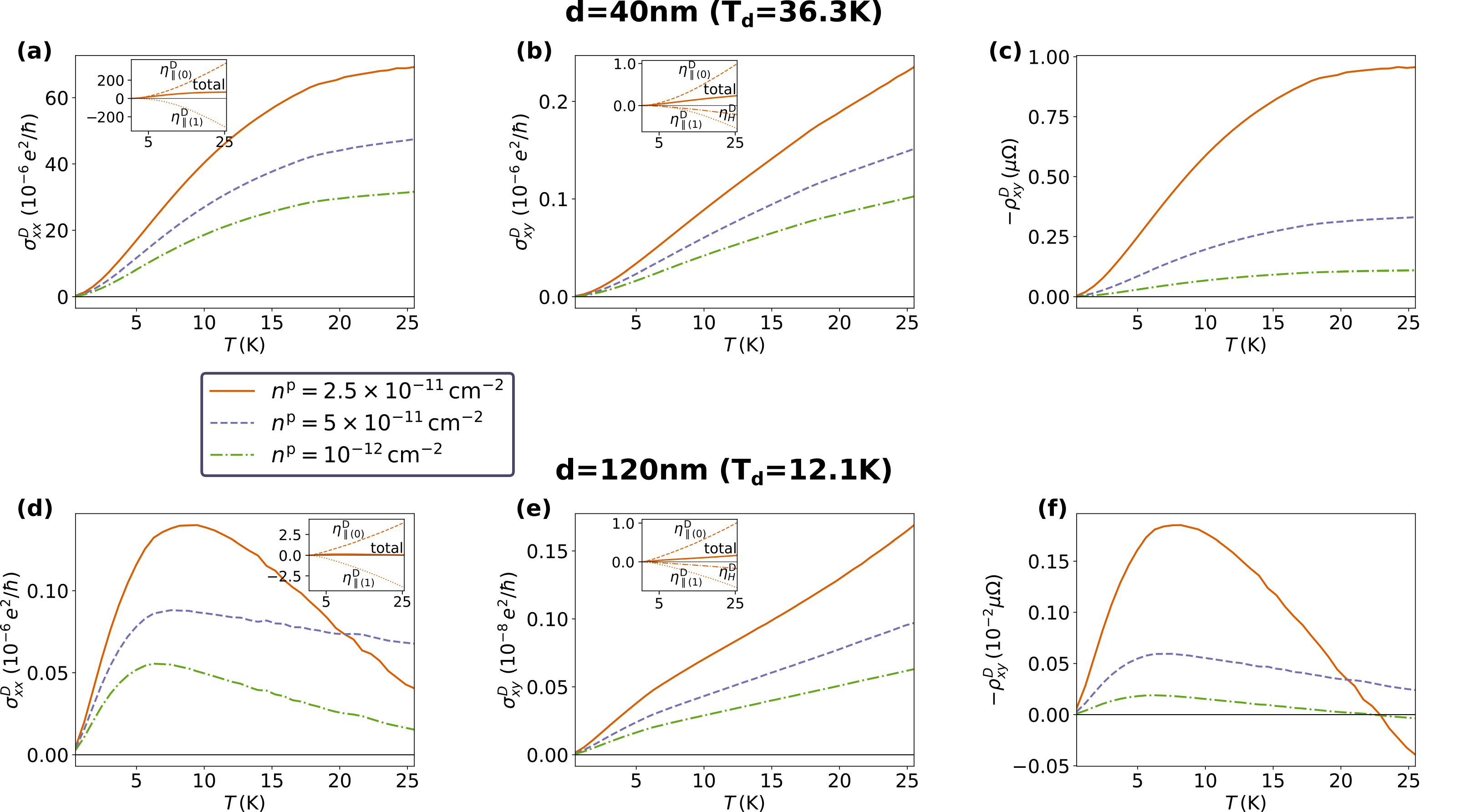}
\par\end{centering}
\caption{Drag conductivity and resistivity as a function of temperature, for interlayer distance $d=\textrm{40}\textrm{nm}$
(a-c) and $d=120\textrm{nm}$ (d-f), obtained by numerically calculating the
interlayer collision integral [Eq. (\ref{eq:general coll integral})]. (a), (d) Longitudinal drag conductivity.
(b), (e) Hall drag conductivity. (c), (f) Hall drag resistivity. The WSM is taken as a single pair of Weyl nodes, with
$\epsilon_{F}^{\textrm{a}}=60\rm{meV}$, $v_{F}^{\textrm{a}}=1.9\times10^{5}\rm{m}/\rm{s}$,
$\tau^{\textrm{a}}=2\times10^{-12}\rm{s}$, internode distance $\Delta_{k}=4.7\textrm{nm}^{-1}$,
tilt parameter $C=0.3$ and thickness $W=15\rm{nm}$. The metal consists
of one electronic band with $m=0.067m_{e}$ ($m_{e}$ being the electron
mass), $\tau^{\textrm{p}}=2\times10^{-11}\rm{s}$ and varying carrier
densities $n^\textrm{p}$ as given in the legend at the center of the figure. Dielectric constant is $\epsilon_{r}=12$. The insets show the separate contributions to the drag conductivities for the lowest value of $n^\textrm{p}$, coming from the parts of the drag force [Eq. (\ref{eq:F_drag high T})] proportional to $\eta_{\parallel(0)}^{D}$ (dashed line), $\eta_{\parallel(1)}^{D}$ (dotted line) and $\eta_{H}^{D}$ (dash-dotted line).\label{fig:numerical calc drag}}
\end{figure}

\end{widetext}

Finally, we numerically calculate the longitudinal and Hall drag conductivities
in the entire temperature range, and present the results in Figs. \ref{fig:numerical calc drag}(a-b),(d-e).
For the calculation, we restore physical units and use realistic parameters
for the TRS-breaking WSM $\textrm{Co}_{\textrm{3}}\textrm{Sn}_{\textrm{2}}\textrm{S}_{\textrm{2}}$
\cite{Liu2018,Fujiwara2019, Li2020,Tanaka2020} and two-dimensional electron gas in a GaAs/AlGaAs interface \cite{Gramila1991} as
the layers. The non-monotonous temperature dependence of $\sigma_{xx}^{D}$
can be seen from the plots, showing maxima at $T\approx T_{d}$ {[}Figs.
\ref{fig:numerical calc drag}(a,d){]}. While $\sigma_{xy}^{D}$ is always an
increasing function of temperature in our model, we stress that
it can also acquire a non-monotonous behavior in other models
with stronger energy dependence of the scattering times.

We note that within the analytic approximation for the coefficients
$\eta_{\parallel(0)}^{D}$ and $\eta_{\parallel(1)}^{D}$ at $T\gg T_{d}$
{[}Eqs. (\ref{eq:eta_parallel high T}), (\ref{eq:eta u deriv}){]},
the two terms in $\sigma_{xx}^{D}$ {[}Eq. (\ref{eq:sigma_xx_D high T}){]}
nearly cancel, but their sum is still an increasing function of $T$.
This analytic approximation takes the Coulomb interaction in the limits
$T\ll\epsilon_{F}^{\textrm{a}},\epsilon_{F}^{\textrm{p}}$, zero tilt
for the WSM, and Thomas-Fermi screening lengths ($1/\kappa^{l}$)
much shorter than the interlayer distance $d$. The small but finite
deviations from these limits in the numerical calculation enhance
the negative $\sim\eta_{\parallel(1)}^{D}$ term due to a reduction
in the screening at frequencies $\omega\approx T_{d}$, and consequently
result in $\sigma_{xx}^{D}$ being a decreasing function of temperature
at $T\gtrsim T_{d}$.

\subsection{Drag resistivity}

We now turn to the drag resistivities, defined by $\rho_{\alpha\beta}^{D}\equiv-E_{\alpha}^{\textrm{p}}/j_{\beta}^{\textrm{a}}$. We compute these by inverting the
generalized conductivity tensor $\sigma_{\alpha\beta}^{l,l'}\equiv j_{\alpha}^{l}/E_{\beta}^{l'}$
{[}for convenience, in this section we consider the sheet (2D) conductivity
and current of the WSM layer, obtained by multiplying the bulk 3D
quantities by the layer's thickness{]}. Focusing on the components $\alpha,\beta\in\left\{ x,y\right\} $,
$\sigma_{\alpha\beta}^{l,l'}$ can be viewed as a $4\times4$ tensor.
In this notation, $\sigma_{\alpha\beta}^{\textrm{p,a}}=\sigma_{\alpha\beta}^{D}$
is the drag conductivity and $\sigma_{\alpha\beta}^{l,l}$ is the
non-interacting conductivity of layer $l$. In the leading order in
the small parameter $\left(\sigma_{xx}^{D}\right)^{2}/\left(\sigma_{xx}^{\textrm{a}}\sigma_{xx}^{\textrm{p}}\right)$,
one finds

\begin{align}
\rho_{xx}^{D} & =\rho_{xx}^{\textrm{p}}\rho_{xx}^{\textrm{a}}\sigma_{xx}^{D},\label{eq:rho_xx D general}\\
\rho_{xy}^{D} & =\rho_{xx}^{D}\left(\frac{\sigma_{xy}^{D}}{\sigma_{xx}^{D}}-\frac{\sigma_{xy}^{\textrm{a}}}{\sigma_{xx}^{\textrm{a}}}\right),\label{eq:rho_xy D general}
\end{align}
where $\rho_{xx}^{l}$ is the longitudinal (2D) resistivity of layer
$l$. We now analyze these results in the low- and high-temperature
regimes.

\subsubsection{Low temperatures}

For low temperatures ($T\ll T_{d}$), the longitudinal drag resistivity
{[}Eq. (\ref{eq:rho_xx D general}){]} is given by

\begin{equation}
\rho_{xx}^{D}=\frac{\eta_{\parallel}^{D}}{e^{2}n^{\textrm{a}}n^{\textrm{p}}d},
\end{equation}
where $n^{\textrm{a}}$ and $n^{\textrm{p}}$ are the (2D) carrier
densities of the two layers. This formula represents the longitudinal
drag resistivity in terms of the parallel drag coefficient.

The analysis of the Hall drag resistivity is more delicate because
of a partial cancellation between terms in Eq. (\ref{eq:rho_xy D general}).
Let us separate the non-interacting AHE conductivity into two parts,
$\sigma_{xy}^{\textrm{a}}\equiv\sigma_{xy}^{\textrm{a},\textrm{reg.}}+\sigma_{xy}^{\textrm{a},\textrm{int.+ext.vel.}},$
corresponding to the contributions from the regular and anomalous
parts of the current {[}Eqs. (\ref{eq:regular current}) and (\ref{eq:anomalous current}){]},
respectively (see Appendix \ref{subsec:Appendix-C.-Non-interacting}
for detailed expressions). As explained qualitatively in Sec. \ref{sec:warmup}
and shown in detail in Sec. \ref{subsec:Low-temperatures}, parallel
friction drags current which is parallel to the boost velocity in
the active layer, leading to $\sigma_{xy}^{D[\eta_{\parallel}^{D}]}/\sigma_{xx}^{D}=\sigma_{xy}^{\textrm{a}\textrm{,reg}.}/\sigma_{xx}$.
Therefore, the contributions from these two terms in Eq. (\ref{eq:rho_xy D general})
cancel each other. Analogous cancellation occurs in the drag resistivity
computation for two metals placed in an external magnetic field, resulting
in zero $\rho_{xy}^{D}$ for that case, as discussed in Sec. \ref{sec:warmup}.
For our problem, drag between a WSM and a metal, the Hall drag resistivity
remains finite due to the Hall friction and the anomalous current.
It is given by

\begin{widetext}

\begin{equation}
\rho_{xy}^{D}=\rho_{xx}^{D}\left(\frac{\eta_{H}^{D}}{\eta_{\parallel}^{D}}-\frac{\sigma_{xy}^{\textrm{a},\textrm{int.+ext.vel.}}}{\sigma_{xx}^{\textrm{a}}}\right)=-\rho_{xx}^{D}\left(\frac{1}{2}\frac{v_{F}^{\textrm{a}}\Delta_{k}}{\left(\epsilon_{F}^{\textrm{a}}\right)^{2}\tau^{\textrm{a}}}+\frac{C}{\epsilon_{F}^{\textrm{a}}\tau^{\textrm{a}}}\right),\label{eq:hall resistivity T small}
\end{equation}

\end{widetext}with the last equality taken in the linear order in
$C$. Note that the intrinsic mechanism of the AHE does affect
the Hall drag resistivity, as is manifested by the term proportional
to $\Delta_{k}$ (the momentum separation between the Weyl nodes).

\subsubsection{High temperatures}

As explained in Sec. \ref{subsec:High-temperatures drag conductivity},
in the high-temperature limit ($T\gg T_{d}$), the approximation of
the active layer's distribution function by a boosted velocity distribution
is insufficient. As a result, the interlayer drag force is characterized
by a more complex response {[}Eq. ($\ref{eq:F_drag high T}$){]}.
The drag resistivity tensor in this limit can be readily obtained
from Eqs. ($\ref{eq:rho_xx D general}$), ($\ref{eq:rho_xy D general}$)
and the values of the drag conductivities computed in Sec. \ref{subsec:High-temperatures drag conductivity}.
Because the final expressions in this limit are quite cumbersome,
we do not write them in full detail here. We do emphasize that the
cancellation of the parallel friction mechanism in the Hall drag resistivity
no longer occurs, and processes rotating the boost velocity (contributing
to the regular part of the AHE conductivity, $\sigma_{xy}^{\textrm{reg}.}$)
do contribute to the Hall drag resistivity.

Qualitatively, both $\rho_{xx}^{D}$ and $\rho_{xy}^{D}$ have a linear
temperature dependence. The Hall drag resistivity can be written in
a form similar to Eq. $\eqref{eq:hall resistivity T small}$,

\begin{equation}
\rho_{xy}^{D}=-\rho_{xx}^{D}\left(\frac{1}{2}\frac{v_{F}^{\textrm{a}}\Delta_{k}}{\left(\epsilon_{F}^{\textrm{a}}\right)^{2}\tau^{\textrm{a}}}+A\frac{C}{\epsilon_{F}^{\textrm{a}}\tau^{\textrm{a}}}\right),
\end{equation}
with $A$ being a numerical coefficient of order one. Its value is
sensitive to the energy dependence of the momentum relaxation times,
and thus to the details of the disorder scattering in both layers,
see Sec. \ref{subsec:High-temperatures drag conductivity}. We present
numerical results for $\rho_{xy}^{D}$ as a function of temperature
in Figs. \ref{fig:numerical calc drag}(c,f). Note that the non-monotonous
temperature dependence of the longitudinal drag conductivity at high temperatures
leads to a similar trend in the drag resistivities (both longitudinal and Hall) due to the tensor inversion,
as can be seen in Figs. \ref{fig:numerical calc drag}(c,f).

\section{Summary and outlook\label{sec:Summary-and-outlook}}

We have studied the Coulomb drag in a setup consisting of a TRS-broken
WSM and a normal metal. The anomalous kinetics of the WSM enrich the
physics, making the problem qualitatively different from the one of
drag between normal metals. There are two ways in which the anomalous
processes affect the Coulomb drag.

The first is due to the anomalous current in the WSM layer, which
arises from the interband elements of the WSM velocity operator. Because
the anomalous current is not directly related to changes in the occupation
of the semiclassical distribution function, the relation between the
electric currents in the two layers is not straightforward. This is
in contrast to normal metals, where the drag is an equilibration process
between the distribution functions in the two layers.

Secondly, the interlayer e-e collision integral contains anomalous
terms, which originate from virtual interband transitions in the WSM.
These terms give rise to a many-body skew-scattering contribution
to the interlayer collision integral.

In our work, we computed the drag conductivity and resistivity tensors
in various temperature regimes. We now summarize the results, starting
with the experimentally common regime of low temperatures ($T\ll T_{d}$).
In this regime, the momentum transfer between the layers can be divided
into two parts:

1. $\textbf{Parallel friction}$. Drag force parallel to the relative
boost velocity between the layers, pushing the boost velocities towards
equilibration. It is analogous to shear viscosity in hydrodynamics.
This part can be computed by taking the interlayer collision integral
within the Born approximation. In the WSM layer, the boost velocity
is rotated relative to the electric field (due to disorder skew scattering).
Therefore, parallel friction gives rise to a part of the Hall drag
conductivity that is proportional to this rotation, $\sigma_{xy}^{D[\eta_{\parallel}^{D}]}=\sigma_{xx}^{D}\tau_{\parallel}^{\textrm{a}}/\tau_{\perp}^{\textrm{a}}$.

2. $\textbf{Hall friction}$. Drag force perpendicular to the WSM
boost velocity $\boldsymbol{u}^{\textrm{a}}$. It originates from
many-body skew scattering, occurring due to interference between e-e
scattering and the electric field or the disorder in the WSM. To account
for these processes, one needs to calculate the interlayer collision
integral beyond the Born approximation. Hall friction creates a second
contribution to the Hall drag conductivity. 

In the model of tilted Weyl nodes in the non-crossing approximation,
the two contributions to the Hall drag conductivity partially cancel each other, resulting in a
smaller value of $\sigma_{xy}^{D}$ than one would expect from a naive
treatment of the interlayer collision integral.

The distinction between parallel and Hall friction is more pronounced
in the Hall drag resistivity $\rho_{xy}^{D}$. This is because friction
parallel to the current does not contribute to the Hall drag resistivity.
The Hall drag resistivity is finite due to two factors: (i) the Hall
friction, which leads to momentum transfer between the layers which
is perpendicular to the WSM boost velocity; (ii) the current in the
WSM is not parallel to the boost velocity, due to the anomalous part
of the current. This leads to a term in $\rho_{xy}^{D}$ that is proportional
to the distance between Weyl nodes.

In the regime of high temperatures ($T\gg T_{d}$), one cannot attribute
a single boost velocity to the WSM layer. Instead, one considers an
energy-dependent boost velocity $\boldsymbol{u}^{\textrm{a}}(\epsilon)$.
The drag force depends on two vectors, $\boldsymbol{u}^{\textrm{a}}(\epsilon_{F}^{\textrm{a}})$
and $\partial\boldsymbol{u}^{\textrm{a}}/\partial\epsilon\vert_{\epsilon=\epsilon_{F}^{\textrm{a}}}$.
In this case, even the Born-approximation part of the interlayer collision
integral leads to a drag force that is not parallel to $\boldsymbol{u}^{\textrm{a}}(\epsilon_{F}^{\textrm{a}})$,
giving an additional contribution to $\rho_{xy}^{D}$. 

Interestingly, the two contributions due to the Born-approximation parts of the drag force are of opposite sign.
This causes a \textbf{non-monotonous} temperature dependence of the
drag conductivities at a wide range of parameter regimes (the ratio between the contributions
depends on the frequency dependence of the interlayer
screening). This behavior is quite general and arises due to the energy
dependence of the boost velocity through its dependence on the transport
times {[}Eq. (\ref{eq:boost velocity energy dependent}){]}. Thus,
we expect non-monotonous temperature behavior of the drag conductivity
in Coulomb drag setups with other materials, given that the transport
time in one layer is a sufficiently fast-decreasing function of energy.

Qualitatively, in both temperature regimes, the temperature and interlayer
distance dependences of $\sigma_{xy}^{D}$ follow the same law as
$\sigma_{xx}^{D}$. The ratio between the Hall and the longitudinal components
of the drag conductivity is given by the small parameter $\sigma_{xy}^{D}/\sigma_{xx}^{D}\simeq C/\left(\epsilon_{F}^{\textrm{a}}\tau^{\textrm{a}}\right)$.
The same parameter governs the ratio between the Fermi-surface part
of the AHE conductivity and the longitudinal conductivity of the non-interacting
WSM \cite{Papaj2021,Zhang2023}. The numerical prefactor for the
Hall drag angle (defined by $\tan\theta_{H}^{D}\equiv\sigma_{xy}^{D}/\sigma_{xx}^{D}$)
depends on the temperature.

The problem that we have studied here is closely related to the Hall viscosity
in electronic fluids. Indeed, the viscosity tensor in an electronic
fluid contains a part that is directly related to the Coulomb drag
\cite{Liao2020}. This part is due to the non-local nature of the
Coulomb collision integral, which couples layers in the fluid that
move with different velocities. Our study thus reveals a mechanism
for the Hall viscosity, stemming from electron-electron skew scattering.
We anticipate a similar term to emerge in the viscosity tensor of
the electronic fluid in a clean TRS-broken WSM. This question presents
a natural direction for future research.

\begin{acknowledgments}

We are grateful to Dimitrie Culcer and Igor Gornyi for interesting and valuable
discussions. This research was supported by ISF-China 3119/19 and
ISF 1355/20. Y. M. thanks the PhD scholarship of the Israeli Scholarship
Education Foundation (ISEF) for excellence in academic and social
leadership.

\end{acknowledgments}

\appendix

\section{Electron-electron collision integral from Keldysh formalism\label{subsec:Appendix-A.-Electron-electron}}

In this appendix, we derive the interlayer e-e collision integral
in the main text {[}Eq. (\ref{eq:general coll integral}) with the
scattering rates in Eq. (\ref{eq:scattering rate detail}){]} using
the Keldysh formalism. We follow Ref. \cite{Konig2021} in calculating
the interband elements of the WSM Keldysh Green function, which lead
to the skew-scattering part of the collision integral.

We first calculate the screened interlayer potential using the RPA
approximation \cite{Kamenev1995,Kamenev2011}. In the quasi-2D limit
(taking the thickness of the WSM layer to be small compared to the
interlayer distance, $W\ll d$), the screened interlayer potential
between the layers is given by \cite{Jauho1993,Kamenev1995} 

\begin{widetext}

\begin{equation}
U_{\textrm{RPA}}^{R}(\boldsymbol{q},\omega)=\left[\frac{4\pi e^{2}}{\epsilon_{r}q}W\Pi^{\textrm{a,}R}(\boldsymbol{q},\omega)\Pi^{\textrm{p,}R}(\boldsymbol{q},\omega)\sinh\left(qd\right)+\left(\frac{\epsilon_{r}q}{2\pi e^{2}}+W\Pi^{\textrm{a,}R}(\boldsymbol{q},\omega)+\Pi^{\textrm{p,}R}(\boldsymbol{q},\omega)\right)e^{qd}\right]^{-1},\label{eq:U_R_RPA}
\end{equation}
where $\Pi^{\textrm{a(p),}R}(\boldsymbol{q},\omega)$ is the retarded
polarization operator in the active (passive) layer {[}Eq. (\ref{eq:Polarization_R_A}){]},
and $\epsilon_{r}$ is an effective background dielectric constant,
which we assume to be uniform in the vicinity of the layers. Note
that the quasi-2D Coulomb interaction transfers only 2D momenta, $\boldsymbol{q}\equiv\left(q_{x},q_{y}\right)$.
Implicit in Eq. (\ref{eq:U_R_RPA}) is that the z- coordinate of the
Coulomb interaction is not Fourier transformed, i.e., $U_{\textrm{RPA}}^{R}(\boldsymbol{q},\omega)=U_{\textrm{RPA}}^{R}(\boldsymbol{q},\omega,z,z')$,
with $z$ ($z'$) being at the position of the 2D (3D) layer, such
that $\left|z-z'\right|\approx d$. In the ballistic limit of the
Coulomb drag ($d\gg v_{F}^{l}\tau^{l}$), and when the interlayer
distance is much larger than the inverse of the Thomas-Fermi screening
wave vectors of both layers (to be defined shortly), the squared modulus
of the interlayer interaction can be approximated by

\begin{equation}
\left|U_{\textrm{RPA}}^{R}(\boldsymbol{q},\omega)\right|^{2}=\left(\frac{\pi e^{2}q}{\epsilon_{r}\kappa^{\textrm{a}}\kappa^{\textrm{p}}\sinh\left(qd\right)}\right)^{2}\frac{1-\left(\frac{\omega}{v_{F}^{\textrm{p}}q}\right)^{2}}{\left(1+\frac{\omega}{2v_{F}^{\textrm{a}}q}\log\left|\frac{1-\omega/\left(v_{F}^{\textrm{a}}q\right)}{1+\omega/\left(v_{F}^{\textrm{a}}q\right)}\right|\right)^{2}+\frac{\pi^{2}}{4}\left(\frac{\omega}{v_{F}^{\textrm{a}}q}\right)^{2}}.\label{eq:u_R_RPA_sq}
\end{equation}
Here, $\kappa^{l}=2\pi e^{2}\nu_{2d}^{l}/\epsilon_{r}$ are the Thomas-Fermi
screening wave vectors of the layers with 2D density of states $\nu_{2d}^{l}$
($\nu_{2d}^{\textrm{p}}\equiv\nu^{\textrm{p}}$ for the metal layer
and $\nu_{2d}^{\textrm{a}}\equiv\nu^{\textrm{a}}W$ for the WSM layer).
In Eq. (\ref{eq:u_R_RPA_sq}), we have replaced the polarization operators
of the layers by their zero temperature and ballistic limits \cite{Giuliani2005}.
In these limits, the result for the polarization operator of the WSM
is identical to that of a 3D metal with a matching density of states
and Fermi velocity, up to corrections proportional to the tilt parameter
$C$.

The e-e collision integral {[}Eq. (\ref{eq:general coll integral}){]}
is given in the Keldysh formalism by (see Fig. \ref{fig:collision integral diagram} for the corresponding self-energy
diagram) \cite{Kamenev2011}

\begin{align}
I_{\boldsymbol{k}}^{\textrm{e-e (p,a)}} & =\frac{iW}{2}\intop_{\boldsymbol{q}}\left|U_{\textrm{RPA}}^{R}(\boldsymbol{q},\omega)\right|^{2}[\left(f^{\textrm{p}}(\boldsymbol{k})-f^{\textrm{p}}(\boldsymbol{k}+\boldsymbol{q})\right)\Pi_{\textrm{}}^{\textrm{a},K}(\boldsymbol{q},\omega)\nonumber \\
 & +\left(2f^{\textrm{p}}(\boldsymbol{k}+\boldsymbol{q})f^{\textrm{p}}(\boldsymbol{k})-f^{\textrm{p}}(\boldsymbol{k}+\boldsymbol{q})-f^{\textrm{p}}(\boldsymbol{k})\right)\left(\Pi^{\textrm{a},R}(\boldsymbol{q},\omega)-\Pi^{\textrm{a},A}(\boldsymbol{q},\omega)\right)].\label{eq:p-a collision integral keldysh}
\end{align}

\begin{figure}[h]
\begin{centering}
\includegraphics[scale=0.6]{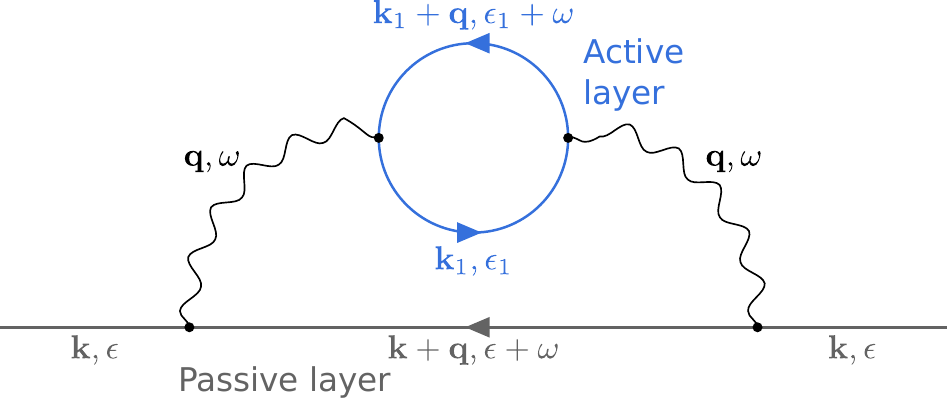}
\par\end{centering}
\caption{Diagram for the self-energy of the passive electrons due to e-e interactions
with the active layer.\label{fig:collision integral diagram}}
\end{figure}

Here, $\omega\equiv\epsilon_{\boldsymbol{k}+\boldsymbol{q}}^{\textrm{p}}-\epsilon_{\boldsymbol{k}}^{\textrm{p}}$
is the transferred energy in the collision, $U_{\textrm{RPA}}^{R}(\boldsymbol{q},\omega)$
is the retarded propagator of the screened Coulomb interlayer interaction
in the RPA approximation and $\Pi^{\textrm{a},(R,A,K)}$ are the (retarded,
advanced, Keldysh) polarization matrices in the active layer. Since
$\boldsymbol{q}$ is 2D, here $\Pi^{\textrm{a},(R,A,K)}(\boldsymbol{q},\omega)=\Pi^{\textrm{a},(R,A,K)}(q_{x},q_{y},q_{z}=0,\omega)$.
The factor of the WSM layer thickness $W$ in Eq. (\ref{eq:p-a collision integral keldysh})
is due to one free integration of the interaction $U_{\textrm{RPA}}^{R}(\boldsymbol{q},\omega,z=0,z')$
over $\intop_{d}^{d+W}dz'$ (putting the 2D layer at $z=0$ and the
WSM at $d<z<d+W$). The polarization matrices are given by (omitting
the layer index from hereon)

\begin{align}
\Pi^{R(A)}(\boldsymbol{q},\omega) & =\frac{i}{2}\textrm{Tr}\left\{ \intop_{\boldsymbol{k},\epsilon}\left[G_{\boldsymbol{k}+\boldsymbol{q},\epsilon+\omega}^{R(A)}G_{\boldsymbol{k},\epsilon}^{K}+G_{\boldsymbol{k}+\boldsymbol{q},\epsilon+\omega}^{K}G_{\boldsymbol{k},\epsilon}^{A(R)}\right]\right\} ,\label{eq:Polarization_R_A}\\
\Pi^{K}(\boldsymbol{q},\omega) & =\frac{i}{2}\textrm{Tr}\left\{ \intop_{\boldsymbol{k},\epsilon}\left[G_{\boldsymbol{k}+\boldsymbol{q},\epsilon+\omega}^{K}G_{\boldsymbol{k},\epsilon}^{K}-\left(G_{\boldsymbol{k}+\boldsymbol{q},\epsilon+\omega}^{R}-G_{\boldsymbol{k}+\boldsymbol{q},\epsilon+\omega}^{A}\right)\left(G_{\boldsymbol{k},\epsilon}^{R}-G_{\boldsymbol{k},\epsilon}^{A}\right)\right]\right\} .\label{eq:Polarization_K}
\end{align}

Note that the WSM layer Green functions are $2\times2$ matrices in
the spinor space. The objects that complicate the collision integral
{[}Eq. (\ref{eq:p-a collision integral keldysh}){]} compared to the
textbook e-e collision integral are the polarization matrices $\Pi^{\textrm{a},(R,A,K)}$,
which acquire contributions from the interband elements of the WSM
Green functions. These contributions give rise to skew-scattering
terms in the e-e collision integral.

In the next subsection, we calculate the interband elements of the
WSM Green functions perturbatively in the small parameter $1/\left(\epsilon_{F}\tau\right)$
and in the external electric field. The interband part of the Keldysh
Green function is coupled to the intraband part via the kinetic equation.
By expressing the interband elements of the Keldysh Green function
in terms of the intraband elements, we will be able to present the
collision integral as a functional of the semiclassical distribution
functions, $I_{\boldsymbol{k}}^{\textrm{e-e (p,a)}}\rightarrow I_{\boldsymbol{k}}^{\textrm{e-e (p,a)}}\left[f^{\textrm{a}},f^{\textrm{p}}\right]$.

\subsection{Kinetic equation in the Keldysh formalism and corrections to the
Green functions}

We start with briefly introducing the main objects in the Keldysh
formalism \cite{Rammer1986,Kamenev2011,Konig2021}. Consider a general
Hamiltonian

\begin{equation}
{\cal H}\equiv H+H',
\end{equation}
with $H$ being the bare part of the Hamiltonian, given by

\begin{equation}
H(\boldsymbol{x},t,\boldsymbol{x}',t')\equiv\delta(t-t')\left(H_{0}(\boldsymbol{x}-\boldsymbol{x}')+\delta(\boldsymbol{x}-\boldsymbol{x}')U_{\textrm{ext}}(\boldsymbol{x},t)\right),\label{eq:bare H}
\end{equation}
where $H_{0}$ describes the non-interacting, translation-invariant
Hamiltonian (whose Fourier transform determines the energy bands $\epsilon_{n\boldsymbol{k}}$)
and the local field $U_{\textrm{ext}}(\boldsymbol{x},t)$ describes
the external fields. The part $H'$ includes any additional complications
such as disorder and interactions. The bare retarded and advanced
Green functions are the inverse of the bare part of the Hamiltonian,

\begin{equation}
\left[G_{0}^{R}\right]^{-1}(x,x')=\left[G_{0}^{A}\right]^{-1}(x,x')\equiv\delta(x-x')i\partial_{t}-H(x,x').
\end{equation}

The Dyson equations for the full retarded (advanced) Green functions
read

\begin{equation}
G^{R(A)}=G_{0}^{R(A)}+G_{0}^{R(A)}\circ\Sigma^{R(A)}\circ G^{R(A)},\label{eq:full green r-a}
\end{equation}
where $\Sigma^{R(A)}$ is the retarded (advanced) self-energy due
to the part $H'$ of the Hamiltonian.

The information about the state of the system is contained in the
Keldysh Green function, which is parametrized by

\begin{equation}
G^{K}=G^{R}\circ F-F\circ G^{A},\label{eq:G_K def}
\end{equation}
where $\circ$ denotes the convolution operation. From the Dyson equations
for $G^{R,A,K}$, one obtains \cite{Kamenev2011}

\begin{equation}
i\left(F\circ\left[G_{0}^{A}\right]^{-1}-\left[G_{0}^{R}\right]^{-1}\circ F\right)=i\left[\Sigma^{K}-\left(\Sigma^{R}\circ F-F\circ\Sigma^{A}\right)\right].\label{eq:Dyson eq F}
\end{equation}
Let us introduce the Wigner-transform (WT), which transforms two-point
functions to functions of the center of mass and momentum coordinates,

\begin{equation}
O(x_{1},x_{2})\overset{\textrm{WT}}{\longrightarrow}{\cal O}(x,k)\equiv\intop dx_{-}e^{-ikx_{-}}O(x+x_{-}/2,x-x_{-}/2),
\end{equation}
where $x\equiv\left(\boldsymbol{R},T\right)$ and $k\equiv\left(\boldsymbol{k},\epsilon\right)$
represent the central point and momentum coordinates, respectively.
Under the Wigner transformation, convolutions $C\equiv A\circ B$
transform according to the following formula (up to linear order in
gradients of the central coordinate $x$):

\begin{equation}
{\cal C}(x,k)={\cal A}(x,k){\cal B}(x,k)+\frac{i}{2}\left(\partial_{x}{\cal A}\partial_{k}{\cal B}-\partial_{k}{\cal A}\partial_{x}{\cal B}\right).
\end{equation}
Performing the Wigner transform on the Dyson equation (\ref{eq:Dyson eq F})
results in the quantum kinetic equation

\begin{equation}
\frac{\partial}{\partial t}F-i\left[F,H\right]_{-}+\frac{1}{2}\left[\partial_{x}F,\partial_{p}\tilde{H}\right]_{+}-\frac{1}{2}\left[\partial_{p}F,\partial_{x}\tilde{H}\right]_{+}=I_{F}\left[F\right],\label{eq:multiband WT kinetic eq}
\end{equation}
where $[A,B]_{-(+)}$ denotes the commutator (anti-commutator), $\tilde{H}\equiv H+\mathfrak{R}[\Sigma^{R}]$
is the Hamiltonian including renormalization effects from the self-energy,
and $I_{F}\left[F\right]$ is the collision integral for $F$, given
by

\begin{equation}
I_{F}\left[F\right]\equiv i\left[\Sigma^{K}-\left(\Sigma^{R}F-F\Sigma^{A}\right)\right].\label{eq:I_F coll}
\end{equation}
Note that all functions from Eq. (\ref{eq:multiband WT kinetic eq})
onwards are in the Wigner-transform space, i.e., $F=F(x,k)$. For
a single band, evaluating Eq. (\ref{eq:multiband WT kinetic eq})
on the energy shell $\epsilon=\epsilon_{\boldsymbol{k}}+U_{\textrm{ext}}(x)+\mathfrak{R}[\Sigma^{R}]$
reduces to the Boltzmann equation for the semiclassical distribution
function $f(x,\boldsymbol{k})$. Omitting renormalization effects
(approximating $\mathfrak{R}[\Sigma^{R}]$ as constant), one obtains

\begin{equation}
\left(\frac{\partial}{\partial t}+\boldsymbol{\nabla}_{\boldsymbol{k}}\epsilon_{\boldsymbol{k}}\cdot\boldsymbol{\nabla}_{\boldsymbol{R}}-\boldsymbol{\nabla}_{\boldsymbol{R}}U_{\textrm{ext}}\cdot\boldsymbol{\nabla}_{\boldsymbol{k}}\right)f(x,\boldsymbol{k})=I_{x,\boldsymbol{k}}\left[f\right],
\end{equation}
with the collision integral given by

\begin{equation}
I_{x,\boldsymbol{k}}\left[f\right]\equiv-\frac{1}{2}\left[I_{F}\right]{}_{x,\boldsymbol{k},\epsilon=\epsilon_{\boldsymbol{k}}+U_{\textrm{ext}}(x)},\label{eq:coll integral from self-energies}
\end{equation}
and the semiclassical distribution function related to the on-shell
part of $F$ by
\begin{equation}
f(x,\boldsymbol{k})\equiv\frac{1-F\left(x,\boldsymbol{k},\epsilon=\epsilon_{\boldsymbol{k}}+U_{\textrm{ext}}(x)\right)}{2}.\label{eq:semiclassical f and F}
\end{equation}

{} Coming back to the case of interest of a multiple band kinetic equation
{[}Eq. (\ref{eq:multiband WT kinetic eq}){]}, it is convenient to
work in the eigenbasis of the band Hamiltonian, where $H_{0}$ is
a diagonal matrix with elements $\epsilon_{n\boldsymbol{k}}$ on the
diagonal. The trade-off in working in the eigenbasis is that it is
generally momentum-dependent, and therefore, derivatives in momentum
space generate Berry connections (to be defined shortly). Considering
an off-diagonal element in a general matrix $\partial O/\partial\boldsymbol{k}_{i}$,
simple calculation shows

\begin{align}
\left(\frac{\partial O}{\partial k_{i}}\right)_{nn'} & \equiv\left\langle u_{n\boldsymbol{k}}\left|\frac{\partial O}{\partial k_{i}}\right|u_{n'\boldsymbol{k}}\right\rangle =\frac{\partial}{\partial k_{i}}\left\langle u_{n\boldsymbol{k}}\left|O\right|u_{n'\boldsymbol{k}}\right\rangle -\left\langle u_{n\boldsymbol{k}}\left|O\right|\frac{\partial}{\partial k_{i}}u_{n'\boldsymbol{k}}\right\rangle -\left\langle \frac{\partial}{\partial k_{i}}u_{n\boldsymbol{k}}\left|O\right|u_{n'\boldsymbol{k}}\right\rangle \nonumber \\
 & =\frac{\partial}{\partial k_{i}}O_{nn'}(\boldsymbol{k})+i\left(O{\cal A}_{i}-{\cal A}_{i}O\right)_{nn'},\label{eq:covariant derivative}
\end{align}
with $\left|u_{n\boldsymbol{k}}\right\rangle $ being the eigenstate
of $H$ at momentum $\boldsymbol{k}$ and band\textbf{ $n$},\textbf{
}and $\boldsymbol{{\cal A}}_{nn'}(\boldsymbol{k})$ being the Berry
connection,

\begin{equation}
\boldsymbol{{\cal A}}_{nn'}(\boldsymbol{k})\equiv i\left\langle u_{n\boldsymbol{k}}\vert\boldsymbol{\nabla}_{\boldsymbol{k}}u_{n'\boldsymbol{k}}\right\rangle .\label{eq:Berry connection}
\end{equation}

In the band eigenbasis, Eq. (\ref{eq:multiband WT kinetic eq}) results
in a system of coupled equations for the matrix distribution function
$F$. One can express the off-diagonal elements $F_{nn'}$ perturbatively
in terms of the diagonal elements $F_{nn}$ in order to obtain decoupled
equations for the diagonal elements. In the presence of an external
electric field, we obtain the following expression for the off-diagonal
element of $F$ (keeping the leading order terms in the gradients):

\begin{equation}
F_{nn'}=-\left[\frac{1}{2}\boldsymbol{{\cal A}}_{nn'}(\boldsymbol{k}) \cdot \boldsymbol{\nabla}_{\boldsymbol{R}}\left( F_{n}+ F_{n'}\right)+\frac{1}{\epsilon_{n\boldsymbol{k}}-\epsilon_{n'\boldsymbol{k}}}\left[-\boldsymbol{\nabla}_{\boldsymbol{R}}U_{\textrm{ext}}(x)\cdot\boldsymbol{{\cal A}}_{nn'}(\boldsymbol{k})\left(F_{n}-F_{n'}\right)+i\left[I_{F}[F]\right]_{nn'}\right]\right]\qquad\left[n\neq n'\right].\label{eq:off-diag F}
\end{equation}
Here, we denote diagonal matrix elements as $F_{n}\equiv F_{nn}$
for brevity. In the multiband case, the semiclassical distribution
function of each band is related to the diagonal component of $F$
in the same manner as in Eq. (\ref{eq:semiclassical f and F}), $f_{n}(x,\boldsymbol{k})\equiv\left[1-F_{n}\left(x,\boldsymbol{k},\epsilon=\epsilon_{n\boldsymbol{k}}+U_{\textrm{ext}}(x)\right)\right]/2$.
We note that by substituting Eq. (\ref{eq:off-diag F}) in the diagonal
element of the kinetic equation (\ref{eq:multiband WT kinetic eq}),
one may obtain the Boltzmann equation for the semiclassical distribution
function, including corrections such as the anomalous velocity \cite{Konig2021}.
Since the purpose of this appendix is only to derive the interlayer
collision integral in terms of the semiclassical distribution functions,
Eq. (\ref{eq:off-diag F}) is all that we need from the kinetic equation.

Next, we calculate the interband elements of the Green functions.
Since we choose to include the electric field in the bare part of
the Hamiltonian $H$ {[}Eq. (\ref{eq:bare H}){]}, the bare propagators
$G_{0}^{R(A)}$ acquire interband elements. In the Wigner coordinates,
the diagonal elements of the bare Green functions are given by

\begin{equation}
G_{0,n}^{R(A)}(x,k)=\frac{1}{\epsilon-\epsilon_{n\boldsymbol{k}}-U_{\textrm{ext}}(x)\pm i0}.\label{eq:G_R bare}
\end{equation}
By requiring $G_{0}\circ G_{0}^{-1}=1$, we find the off-diagonal
correction to the bare Green functions, to the leading order in the
gradients,

\begin{align}
G_{E,nn'}^{R(A)}(x,k)\equiv & G_{0,nn'}^{R(A)}(x,k)=-\boldsymbol{{\cal A}}_{nn'}(\boldsymbol{k})\cdot[\frac{-\boldsymbol{\nabla}_{\boldsymbol{R}}U_{\textrm{ext}}}{\epsilon_{n\boldsymbol{k}}-\epsilon_{n'\boldsymbol{k}}+i0}\left(G_{0,n}^{R(A)}(x,k)-G_{0,n'}^{R(A)}(x,k)\right)\nonumber \\
 & +\frac{1}{2}\boldsymbol{\nabla}_{\boldsymbol{R}}\left(G_{0,n}^{R(A)}(x,k)+G_{0,n'}^{R(A)}(x,k)\right)]\qquad\left[n\neq n'\right].\label{eq:G_R_A_E}
\end{align}

The retarded and advanced Green functions also acquire off-diagonal
corrections due to the self-energy. To the leading order in the perturbative
Hamiltonian $H'$, the correction is given by

\begin{equation}
G_{V,nn'}^{R(A)}=G_{0,n}^{R(A)}\Sigma_{nn'}^{R(A)}G_{0,n'}^{R(A)}.\label{eq:G_R_A nn' self-energy}
\end{equation}

In total, the interband corrections to $G^{R(A)}$ are the sum of
the two terms, 
\begin{equation}
G_{nn'}^{R(A)}\equiv G_{E,nn'}^{R(A)}+G_{V,nn'}^{R(A)}.\label{eq:off-diag G_R_A}
\end{equation}
Similarly, we find the interband elements of the Keldysh Green function
and write

\begin{equation}
G_{nn'}^{K}(x,k)\equiv G_{E,nn'}^{K}(x,k)+G_{V,nn'}^{K}(x,k).\label{eq:off-diag G_K}
\end{equation}
Here, $G_{E,nn'}^{K}(x,k)$ corresponds to all the off-diagonal terms
in Eq. (\ref{eq:G_K def}) that explicitly contain spatial gradients.
These terms come from $G_{E,nn'}^{R(A)}$ {[}Eq. (\ref{eq:G_R_A_E}){]},
$F_{nn'}$ {[}Eq. (\ref{eq:off-diag F}){]}, or the gradients generated
by the Wigner transformation (e.g., $G^{R}\circ F\overset{\textrm{WT}}{\longrightarrow}\left(...\right)+i\left[\partial_{r}G^{R},\partial_{k}F\right]_{-}/2-\left[r\leftrightarrow k\right]$).
The part $G_{V,nn'}^{K}$ arises from the corrections $G_{V,nn'}^{R(A)}$
{[}Eq. (\ref{eq:G_R_A nn' self-energy}){]} and the last term in Eq.
(\ref{eq:off-diag F}) for $F_{nn'}$ (the term explicitly including
$I_{F}[F]$). Note that although the perturbative Hamiltonian $H'$
determines the non-equilibrium distribution function through the collision
integral and is thus relevant for both terms in Eq. (\ref{eq:off-diag G_K}),
the term $G_{\textrm{V},nn'}^{K}$ accounts for its effect on the
propagators themselves, generating virtual interband transitions.

The term $G_{E,nn'}^{K}(x,k)$ is given by a formula analogous to
Eq. (\ref{eq:G_R_A_E}),

\begin{equation}
G_{E,nn'}^{K}(x,k)=-\boldsymbol{{\cal A}}_{nn'}\left(\frac{-\boldsymbol{\nabla}_{\boldsymbol{R}}U_{\textrm{ext}}}{\epsilon_{n\boldsymbol{k}}-\epsilon_{n'\boldsymbol{k}}+i0}\left(G_{0,n}^{K}(\boldsymbol{k},\epsilon)-G_{0,n'}^{K}(\boldsymbol{k},\epsilon)\right)+\frac{1}{2}\boldsymbol{\nabla}_{\boldsymbol{R}}\left[G_{0,n}^{K}(\boldsymbol{k},\epsilon)+G_{0,n'}^{K}(\boldsymbol{k},\epsilon)\right]\right)\qquad\left[n\neq n'\right],\label{eq:G_K_E}
\end{equation}
where we defined $G_{0,n}^{K}\equiv\left(G_{0,n}^{R}-G_{0,n}^{A}\right)F_{n}$.
Let us note that although the expressions in Eqs. (\ref{eq:G_R_A_E})
and (\ref{eq:G_K_E}) can be simplified by explicitly calculating
$\nabla_{\boldsymbol{R}}G_{0,n}^{R(A)}$ using Eq. (\ref{eq:G_R bare}), the separation
to the two terms turns out to be convenient in the calculation of
the drag later on, with the first term giving no contribution. 

The terms $G_{V,nn'}^{R,A,K}$ depend on the perturbating term $H'$.
From now on we focus on the case studied in this work, where the dominant
scattering in the WSM is due to Gaussian disorder, so that $H'$ is
the disorder potential. The Green functions and self-energies of interest
are those averaged over the random disorder configurations. Modeling
the disorder by short-ranged dilute scalar impurities at concentration
$n_{\textrm{imp}}$ and strength $u_{0}$ (in units of energy times
volume), the self-energy is given by, to the leading order in the
impurity concentration,

\begin{equation}
\Sigma_{nn'}^{R(A)}(\boldsymbol{k},\epsilon)=n_{\textrm{imp}}\sum_{m}\intop_{\boldsymbol{k}_{1}}V_{nm}^{\boldsymbol{k}\boldsymbol{k}_{1}}G_{0,m}^{R}(\boldsymbol{k}_{1},\epsilon)V_{mn'}^{\boldsymbol{k}_{1}\boldsymbol{k}},
\end{equation}
with $V_{nn'}^{\boldsymbol{k}\boldsymbol{k}'}=u_{0}\left\langle u_{n\boldsymbol{k}}\vert u_{n'\boldsymbol{k}'}\right\rangle $
being the matrix element of the impurity potential in Fourier space,
whose momentum dependence is only due to the inner product of the Bloch
wavefunctions. The correlator of the disorder potential averaged over
the disorder configurations is given by $\left\langle H'(\boldsymbol{r})H'(\boldsymbol{r}')\right\rangle _{\textrm{disorder}}=\gamma\delta(\boldsymbol{r}-\boldsymbol{r}')$
with $\gamma\equiv n_{\textrm{imp}}u_{0}^{2}$. In this case, we find

\begin{align}
G_{V,nn'}^{R(A)}(\boldsymbol{k},\epsilon) & =\gamma\sum_{m}\intop_{\boldsymbol{k}_{1}}\left\langle u_{n\boldsymbol{k}}\vert u_{m\boldsymbol{k}_{1}}\right\rangle \left\langle u_{m\boldsymbol{k}_{1}}\vert u_{n'\boldsymbol{k}}\right\rangle G_{0,n}^{R(A)}G_{0,m}^{R(A)}G_{0,n'}^{R(A)},\label{eq:G_R_A_V}\\
G_{V,nn'}^{K}(\boldsymbol{k},\epsilon) & =\gamma\sum_{m}\intop_{\boldsymbol{k}_{1}}\left\langle u_{n\boldsymbol{k}}\vert u_{m\boldsymbol{k}_{1}}\right\rangle \left\langle u_{m\boldsymbol{k}_{1}}\vert u_{n'\boldsymbol{k}}\right\rangle \{-F_{n}G_{0,n}^{A}G_{0,m}^{A}G_{0,n'}^{A}\nonumber \\
 & +F_{n'}G_{0,n}^{R}G_{0,m}^{R}G_{0,n'}^{R}\nonumber \\
 & +\left(1-\delta_{nn'}\right)G_{0,n}^{R}G_{0,n'}^{R}\left[F_{n}G_{0,m}^{A}-F_{n'}G_{0,m}^{R}+F_{m}\left(G_{0,m}^{R}-G_{0,m}^{A}\right)\right]\}.\label{eq:G_K_V}
\end{align}
Here, all the functions on the RHS are evaluated at energy $\epsilon$,
and their momentum argument can be read from the products of the Bloch wavefunctions (i.e.,
momentum $\boldsymbol{k}$ for matrix elements of bands $n,n'$ and
$\boldsymbol{k}_{1}$ for $m$).

We are now ready to evaluate the interlayer collision integral {[}Eq.
(\ref{eq:p-a collision integral keldysh}){]}, substituting the full
Green functions in the polarization operators {[}Eqs. (\ref{eq:Polarization_R_A})
and (\ref{eq:Polarization_K}){]}. We separate the contributions coming
from the different corrections of the Green functions.

\subsection{Born-approximation part of interlayer e-e collision integral}

Taking the diagonal components of the bare Green functions in Eqs.
(\ref{eq:Polarization_R_A}) and (\ref{eq:Polarization_K}) gives
the familiar expressions for the polarization operators \cite{Kamenev2011},

\begin{align}
\left[\Pi^{R}(\boldsymbol{q},\omega)-\Pi^{A}(\boldsymbol{q},\omega)\right]_{0} & =\pi i\sum_{nn'}\intop_{\boldsymbol{k}}\left|\left\langle u_{n\boldsymbol{k}}\vert u_{n'\boldsymbol{k}+\boldsymbol{q}}\right\rangle \right|^{2}\delta(\epsilon_{n'\boldsymbol{k}+\boldsymbol{q}}-\epsilon_{n\boldsymbol{k}}-\omega)\left(F_{n'}(\boldsymbol{k}+\boldsymbol{q},\epsilon_{n'\boldsymbol{k}+\boldsymbol{q}})-F_{n}(\boldsymbol{k},\epsilon_{n\boldsymbol{k}})\right),\label{eq:Pi_R 0th order}\\
\left[\Pi^{K}(\boldsymbol{q},\omega)\right]_{0} & =-\pi i\sum_{nn'}\intop_{\boldsymbol{k}}\left|\left\langle u_{n\boldsymbol{k}}\vert u_{n'\boldsymbol{k}+\boldsymbol{q}}\right\rangle \right|^{2}\delta(\epsilon_{n'\boldsymbol{k}+\boldsymbol{q}}-\epsilon_{n\boldsymbol{k}}-\omega)\left(F_{n'}(\boldsymbol{k}+\boldsymbol{q},\epsilon_{n'\boldsymbol{k}+\boldsymbol{q}})F_{n}(\boldsymbol{k},\epsilon_{n\boldsymbol{k}})-1\right).\label{eq:Pi_K 0th order}
\end{align}
Substituting $\left[\Pi^{R,A,K}\right]_{0}$ in Eq. (\ref{eq:p-a collision integral keldysh})
gives rise to the leading term of the interlayer collision integral,
given by Eq. (\ref{eq:general coll integral}) with the Born-approximation
interlayer scattering rate 

\begin{equation}
w_{\boldsymbol{k},n\boldsymbol{k}_{1}\rightarrow\boldsymbol{k}',n'\boldsymbol{k}_{1'}}^{\textrm{Born}}=2\pi\delta_{\boldsymbol{k}+\boldsymbol{k}_{1}-\boldsymbol{k}'-\boldsymbol{k}_{1'}}\delta(\epsilon_{n\boldsymbol{k}_{1}}^{\textrm{a}}+\epsilon_{\boldsymbol{k}}^{\textrm{p}}-\epsilon_{\boldsymbol{k}'}^{\textrm{p}}-\epsilon_{n'\boldsymbol{k}_{1'}}^{\textrm{a}})\left|U_{\textrm{RPA}}^{R}(\boldsymbol{q},\omega)\right|^{2}\left|\left\langle u_{n\boldsymbol{k}_{1}}\vert u_{n'\boldsymbol{k}_{1'}}\right\rangle \right|^{2},\label{eq:symmetric scat rate}
\end{equation}
with $\boldsymbol{q}\equiv\boldsymbol{k}'-\boldsymbol{k}=\boldsymbol{k}_{1}-\boldsymbol{k}_{1'}$
and $\omega\equiv\epsilon_{\boldsymbol{k}'}^{\textrm{p}}-\epsilon_{\boldsymbol{k}}^{\textrm{p}}=\epsilon_{n\boldsymbol{k}_{1}}^{\textrm{a}}-\epsilon_{n'\boldsymbol{k}_{1'}}^{\textrm{a}}$
being the momentum and energy transferred in the collision, respectively.
The spinor inner product $\left|\left\langle u_{n\boldsymbol{k}_{1}}\vert u_{n'\boldsymbol{k}_{1'}}\right\rangle \right|^{2}$
in the scattering rate is due to the spinor structure of the WSM and
suppresses backscattering, similar to graphene \cite{Tse2007}.

\subsection{Skew scattering interlayer e-e collision integral}

Next, we collect all terms in the polarization matrices {[}Eqs. (\ref{eq:Polarization_R_A}),
(\ref{eq:Polarization_K}){]} that include one off-diagonal element
of the Green functions {[}Eqs. (\ref{eq:off-diag G_R_A}), (\ref{eq:off-diag G_K}){]}.
Substituting the resulting corrections of the polarization matrices
into the collision integral {[}Eq. (\ref{eq:p-a collision integral keldysh}){]}
gives rise to skew-scattering contributions. We find the following
contributions:

\textbf{1.} Intrinsic. Consider the first term in Eqs. (\ref{eq:G_R_A_E})
and (\ref{eq:G_K_E}) for the off-diagonal parts of the Green functions,

\begin{equation}
\left[G_{nn'}^{R,A,K}\right]_{\textrm{int}}(\boldsymbol{k},\epsilon)\equiv-\boldsymbol{{\cal A}}_{nn'}(\boldsymbol{k})\cdot\frac{e\boldsymbol{E}}{\epsilon_{n\boldsymbol{k}}-\epsilon_{n'\boldsymbol{k}}+i0}\left(G_{0,n}^{R,A,K}(\boldsymbol{k},\epsilon)-G_{0,n'}^{R,A,K}(\boldsymbol{k},\epsilon)\right).\label{eq:Green electric field correction}
\end{equation}
This correction is related to the intrinsic (Berry curvature) mechanism
of the AHE, and gives the intrinsic part of the electric current when
substituted into the expectation value of the current, $\boldsymbol{j}=\textrm{Tr}\left\{ \hat{\boldsymbol{j}}G^{K}\right\} $.
We find that this correction does not contribute to the interlayer
collision integral in the linear response regime. In more detail,
collecting all terms in the polarization operators {[}Eqs. (\ref{eq:Polarization_R_A}),
(\ref{eq:Polarization_K}){]} that contain one off-diagonal element
of a Green function taken as $G_{nn'}^{R,A,K}\rightarrow\left[G_{nn'}^{R,A,K}\right]_{\textrm{int}}$
gives

\begin{align}
\left[\Pi^{R}(\boldsymbol{q},\omega)-\Pi^{A}(\boldsymbol{q},\omega)\right]_{\textrm{int}} & =\pi i\sum_{n,n'}\intop_{\boldsymbol{k}}\delta\left(\epsilon_{n'\boldsymbol{k}+\boldsymbol{q}}-\epsilon_{n\boldsymbol{k}}-\omega\right)\left[F_{n}(\boldsymbol{k},\epsilon_{n\boldsymbol{k}})-F_{n'}(\boldsymbol{k}+\boldsymbol{q},\epsilon_{n'\boldsymbol{k}+\boldsymbol{q}})\right]e\boldsymbol{E}\nonumber \\
 & \cdot\left[\sum_{m\neq n}\frac{1}{\epsilon_{n\boldsymbol{k}}-\epsilon_{m\boldsymbol{k}}}\left(\boldsymbol{{\cal A}}_{nm}(\boldsymbol{k})\left\langle u_{m\boldsymbol{k}}\vert u_{n'\boldsymbol{k}+\boldsymbol{q}}\right\rangle \left\langle u_{n'\boldsymbol{k}+\boldsymbol{q}}\vert u_{n\boldsymbol{k}}\right\rangle +c.c\right)+\left(n,\boldsymbol{k}\leftrightarrow n',\boldsymbol{k}+\boldsymbol{q}\right)\right],\label{eq:pol_R_A_int}\\
\left[\Pi^{K}(\boldsymbol{q},\omega)\right]_{\textrm{int}} & =\pi i\sum_{n,n'}\intop_{\boldsymbol{k}}\delta\left(\epsilon_{n'\boldsymbol{k}+\boldsymbol{q}}-\epsilon_{n\boldsymbol{k}}-\omega\right)\left[F_{n'}(\boldsymbol{k}+\boldsymbol{q},\epsilon_{n'\boldsymbol{k}+\boldsymbol{q}})F_{n}(\boldsymbol{k},\epsilon_{nk})-1\right]e\boldsymbol{E}\nonumber \\
 & \cdot\left[\sum_{m\neq n}\frac{1}{\epsilon_{n\boldsymbol{k}}-\epsilon_{m\boldsymbol{k}}}\left(\boldsymbol{{\cal A}}_{nm}(\boldsymbol{k})\left\langle u_{m\boldsymbol{k}}\vert u_{n'\boldsymbol{k}+\boldsymbol{q}}\right\rangle \left\langle u_{n'\boldsymbol{k}+\boldsymbol{q}}\vert u_{n\boldsymbol{k}}\right\rangle +c.c\right)+\left(n,\boldsymbol{k}\leftrightarrow n',\boldsymbol{k}+\boldsymbol{q}\right)\right].\label{eq:pol_K_int}
\end{align}
The corrections $\left[\Pi^{R,A,K}(\boldsymbol{q},\omega)\right]_{\textrm{int}}$
are of the same form as the bare expressions for the polarization
matrices {[}Eqs. (\ref{eq:Pi_R 0th order}), (\ref{eq:Pi_K 0th order}){]},
and lead to a collision integral in the form of Eq. (\ref{eq:general coll integral})
with a renormalized scattering rate. However, the correction to the
scattering rate is linear in the electric field, and thus a non-vanishing
contribution from the collision integral starts only from the second
order of $\boldsymbol{E}$.

\textbf{2.} Side jump. Next, we consider the second term in Eqs. (\ref{eq:G_R_A_E})
and (\ref{eq:G_K_E}),

\begin{equation}
\left[G_{nn'}^{R,A,K}(\boldsymbol{k},\epsilon)\right]_{\textrm{s.j.}}\equiv-\frac{1}{2}\boldsymbol{{\cal A}}_{nn'}(\boldsymbol{k})\boldsymbol{\nabla}_{\boldsymbol{R}}\left[G_{0,n}^{R,A,K}(\boldsymbol{k},\epsilon)+G_{0,n'}^{R,A,K}(\boldsymbol{k},\epsilon)\right].
\end{equation}
Similarly to the previous part, we collect all the terms in the polarization
operators that include one off-diagonal element in one Green function
with $G_{nn'}^{R,A,K}\rightarrow\left[G_{nn'}^{R,A,K}\right]_{\textrm{\textrm{s.j.}}}$
and a diagonal element in the second Green function. During the algebra,
we utilize the following identity,

\begin{align}
\frac{\sum_{m\neq n'}\left\langle u_{n\boldsymbol{k}}\vert u_{n'\boldsymbol{k}'}\right\rangle \left\langle u_{m\boldsymbol{k}'}\vert u_{n\boldsymbol{k}}\right\rangle \boldsymbol{{\cal A}}_{n'm}(\boldsymbol{k}')-\sum_{m\neq n}\left\langle u_{n\boldsymbol{k}}\vert u_{n'\boldsymbol{k}'}\right\rangle \left\langle u_{n'\boldsymbol{k}'}\vert u_{m\boldsymbol{k}}\right\rangle \boldsymbol{{\cal A}}_{mn}(\boldsymbol{k})}{2\left|\left\langle u_{n\boldsymbol{k}}\vert u_{n'\boldsymbol{k}'}\right\rangle \right|^{2}}\nonumber \\
=i\left\langle u_{n'\boldsymbol{k}'}\vert\boldsymbol{\nabla}_{\boldsymbol{k}'}u_{n'\boldsymbol{k}'}\right\rangle -i\left\langle u_{n\boldsymbol{k}}\vert\boldsymbol{\nabla}_{\boldsymbol{k}}u_{n\boldsymbol{k}}\right\rangle -\left(\boldsymbol{\nabla}_{\boldsymbol{k}'}+\boldsymbol{\nabla}_{\boldsymbol{k}}\right)\arg\left(\left\langle u_{n'\boldsymbol{k}'}\vert u_{n\boldsymbol{k}}\right\rangle \right)\equiv & \delta\boldsymbol{r}_{n\boldsymbol{k},n'\boldsymbol{k}'},\label{eq:delta_r_def}
\end{align}
where $\delta\boldsymbol{r}_{n'\boldsymbol{k}',n\boldsymbol{k}}$
denotes the coordinate shift accumulated during a collision from state
$\left|u_{n\boldsymbol{k}}\right\rangle \rightarrow\left|u_{n'\boldsymbol{k}'}\right\rangle $
\cite{Sinitsyn2006}. To get from the first line to the second line
in Eq. (\ref{eq:delta_r_def}), we add and subtract $m=n'$ and $m=n$
to the summations, giving the identity operator. Assuming a spatially
uniform semiclassical distribution function, the spatial gradient
acts only on $G_{0}^{R,A}$ through their dependence on the electric
potential {[}Eq. (\ref{eq:G_R bare}){]}. To linear order in the electric
field, we find the corrections to the polarization operators,

\begin{align}
\left[\Pi^{R}(\boldsymbol{q},\omega)-\Pi^{A}(\boldsymbol{q},\omega)\right]_{\textrm{s.j.}} & =\pi i\sum_{nn'}\intop_{\boldsymbol{k}}\left|\left\langle u_{n\boldsymbol{k}}\vert u_{n'\boldsymbol{k}+\boldsymbol{q}}\right\rangle \right|^{2}e\boldsymbol{E}\cdot\delta\boldsymbol{r}_{n'\boldsymbol{k}+\boldsymbol{q},n\boldsymbol{k}}\frac{\partial}{\partial\epsilon_{n\boldsymbol{k}}}\delta\left(\epsilon_{n'\boldsymbol{k}+\boldsymbol{q}}-\epsilon_{n\boldsymbol{k}}-\omega\right)\nonumber \\
 & \times\left[F_{n'}(\boldsymbol{k}+\boldsymbol{q},\epsilon_{n'\boldsymbol{k}+\boldsymbol{q}})-F_{n}(\boldsymbol{k},\epsilon_{n\boldsymbol{k}})\right],\\
\left[\Pi^{K}(\boldsymbol{q},\omega)\right]_{\textrm{s.j.}} & =-\pi i\sum_{nn'}\intop_{\boldsymbol{k}}\left|\left\langle u_{n\boldsymbol{k}}\vert u_{n'\boldsymbol{k}+\boldsymbol{q}}\right\rangle \right|^{2}e\boldsymbol{E}\cdot\delta\boldsymbol{r}_{n'\boldsymbol{k}+\boldsymbol{q},n\boldsymbol{k}}\frac{\partial}{\partial\epsilon_{n\boldsymbol{k}}}\delta\left(\epsilon_{n'\boldsymbol{k}+\boldsymbol{q}}-\epsilon_{n\boldsymbol{k}}-\omega\right)\nonumber \\
 & \times\left[F_{n'}(\boldsymbol{k}+\boldsymbol{q},\epsilon_{n'\boldsymbol{k}+\boldsymbol{q}})F_{n}(\boldsymbol{k},\epsilon_{n\boldsymbol{k}})-1\right].
\end{align}
Comparing to the leading parts of the polarizations given in Eqs.
(\ref{eq:Pi_R 0th order}) and (\ref{eq:Pi_K 0th order}), we can
interpret these terms as linear corrections from the energy conservation
condition, replacing $\delta\left(\epsilon_{n'\boldsymbol{k}+\boldsymbol{q}}-\epsilon_{n\boldsymbol{k}}-\omega\right)\rightarrow\delta\left(\epsilon_{n'\boldsymbol{k}+\boldsymbol{q}}-\epsilon_{n\boldsymbol{k}}-\omega-e\boldsymbol{E}\cdot\delta\boldsymbol{r}_{n'\boldsymbol{k}+\boldsymbol{q},n\boldsymbol{k}}\right)$
\cite{Konig2021}. This can be understood as accounting for the work
done by the electric field as the WSM electron obtains a coordinate
shift due to the scattering. Substituting $\left[\Pi^{R,A,K}\right]_{\textrm{s.j.}}$
in $I^{\textrm{e-e (p,a)}}$ {[}Eq. (\ref{eq:p-a collision integral keldysh}){]}
gives the side-jump correction to the interlayer collision integral,

\begin{align}
I_{\boldsymbol{k}}^{\textrm{s.j. (p,a)}}\left[f^{\textrm{a}},f^{\textrm{p}}\right] & =-2\pi W\sum_{\xi=\pm1}\sum_{nn'}\intop_{\boldsymbol{q},\boldsymbol{k}'}\left[f_{\boldsymbol{k}}^{\textrm{p}}f_{n'\boldsymbol{k}_{1}+\boldsymbol{q}}^{\textrm{a}}\left(1-f_{\boldsymbol{k}+\boldsymbol{q}}^{\textrm{p}}\right)\left(1-f_{n\boldsymbol{k}_{1}}^{\textrm{a}}\right)-\left(1-f_{\boldsymbol{k}}^{\textrm{p}}\right)\left(1-f_{n'\boldsymbol{k}_{1}+\boldsymbol{q}}^{\textrm{a}}\right)f_{\boldsymbol{k}+\boldsymbol{q}}^{\textrm{p}}f_{n\boldsymbol{k}_{1}}^{\textrm{a}}\right]\nonumber \\
 & \times\left|U_{\textrm{RPA}}^{R}(\boldsymbol{q},\omega)\right|^{2}e\boldsymbol{E}\cdot\delta\boldsymbol{r}_{n'\boldsymbol{k}_{1}+\boldsymbol{q},n\boldsymbol{k}_{1}}\frac{\partial}{\partial\epsilon_{n\boldsymbol{k}_{1}}^{\textrm{a}}}\delta\left(\epsilon_{n'\boldsymbol{k}_{1}+\boldsymbol{q}}^{\textrm{a}}-\epsilon_{n\boldsymbol{k}_{1}}^{\textrm{a}}-\omega\right)\left|\left\langle u_{n\boldsymbol{k}_{1}}\vert u_{n'\boldsymbol{k}_{1}+\boldsymbol{q}}\right\rangle \right|^{2},\label{eq:sj coll integral}
\end{align}
where we summed over the contributions from the two Weyl nodes $\xi=\pm1$
(the node index $\xi$ is omitted from the functions in the integrand
for brevity). This correction to the interlayer collision integral
corresponds to the general form of the two-particle collision integral
{[}Eq. (\ref{eq:general coll integral}) in the main text{]} with
a scattering rate proportional to the electric field, 
\begin{equation}
w_{\boldsymbol{k},n'\boldsymbol{k}_{1}+\boldsymbol{q}\rightarrow\boldsymbol{k}+\boldsymbol{q},n\boldsymbol{k}_{1}}^{\textrm{s.j.}}=2\pi\left|U_{\textrm{RPA}}^{R}(\boldsymbol{q},\omega)\right|^{2}\left[\frac{\partial}{\partial\epsilon_{n\boldsymbol{k}_{1}}^{\textrm{a}}}\delta\left(\epsilon_{n'\boldsymbol{k}_{1}+\boldsymbol{q}}^{\textrm{a}}+\epsilon_{\boldsymbol{k}}^{\textrm{p}}-\epsilon_{n\boldsymbol{k}_{1}}^{\textrm{a}}-\epsilon_{\boldsymbol{k}+\boldsymbol{q}}^{\textrm{p}}\right)\right]e\boldsymbol{E}\cdot\delta\boldsymbol{r}_{n'\boldsymbol{k}_{1}+\boldsymbol{q},n\boldsymbol{k}_{1}}\left|\left\langle u_{n\boldsymbol{k}_{1}}\vert u_{n'\boldsymbol{k}_{1}+\boldsymbol{q}}\right\rangle \right|^{2}.\label{eq:W_k_k sj}
\end{equation}

Note that since $\epsilon_{n'\boldsymbol{k}_{1}+\boldsymbol{q}}^{\textrm{a}}-\epsilon_{n\boldsymbol{k}_{1}}^{\textrm{a}}\neq\epsilon_{\boldsymbol{k}+\boldsymbol{q}}^{\textrm{p}}-\epsilon_{\boldsymbol{k}}^{\textrm{p}}$
in the integrand of Eq. (\ref{eq:sj coll integral}), the side-jump
collision integral is not nullified by the equilibrium distribution
functions. We also note that $w^{\textrm{s.j.}}$ is symmetric in
the exchange of incoming and outgoing particles ($\boldsymbol{k},n'\boldsymbol{k}_{1}+\boldsymbol{q}\leftrightarrow\boldsymbol{k}+\boldsymbol{q},n\boldsymbol{k}_{1}$),
since both the coordinate shift $\delta\boldsymbol{r}_{n'\boldsymbol{k}_{1}+\boldsymbol{q},n\boldsymbol{k}_{1}}$
and the derivative of the delta function are odd under the exchange.

\textbf{3. }Interference with disorder. This term comes from taking
one off-diagonal Green function involving disorder scattering, $G_{nn'}^{R,A,K}\rightarrow G_{V,nn'}^{R,A,K}$
{[}Eqs. (\ref{eq:G_R_A_V}), (\ref{eq:G_K_V}){]}. This is equivalent
to dressing one bare propagator with two disorder scattering lines
and taking the correction from the last term in the expression for
$F_{nn'}$ {[}Eq. (\ref{eq:off-diag F}){]}, see Fig. \ref{fig:pol. bubble one disorder line}. 

During the calculation, we omit terms that do not contribute to skew
scattering and only lead to renormalization of the Born-approximation
scattering rate. Utilizing the symmetry of the WSM for rotations in
the x-y plane, we do so by keeping only the contributions that are
anti-symmetric in reflections of the momentum $\boldsymbol{q}$ around
the momentum arguments of the distribution functions $F$ (projected
on the x-y plane). For example, for a term of the form $\Pi^{R,A,K}(\boldsymbol{q})\sim\intop_{\boldsymbol{k}_{1}}F(\boldsymbol{k}_{1})H(\boldsymbol{k}_{1},\boldsymbol{q})$
where $H$ is an arbitrary function, we calculate $\Pi^{R,A,K}(\boldsymbol{q})=\left[\intop_{\boldsymbol{k}_{1}}F(\boldsymbol{k}_{1})H(\boldsymbol{k}_{1},\boldsymbol{q})-\intop_{\boldsymbol{k}_{1}}F(\boldsymbol{k}_{1})H(\boldsymbol{k}_{1},\boldsymbol{q}^{M(\boldsymbol{k}_{1,||})})\right]/2$,
where $\boldsymbol{q}^{M(\boldsymbol{k}_{1,||})}$ is the reflection
of $\boldsymbol{q}$ with respect to the vector $\boldsymbol{k}_{1,\parallel}$
($\boldsymbol{k}_{1}$ projected on x-y plane)\footnote{Alternatively, this is equivalent to keeping only the imaginary part
of the total product of the Bloch functions inner products {[}Eq.
(\ref{eq:Z-4-product}){]}. In our problem, this is the only object
that is odd in angles on the x-y plane and thus, can result in skew
scattering.}. The resulting corrections to the polarization matrices are

\begin{figure}[h]
\begin{centering}
\includegraphics[scale=1.7]{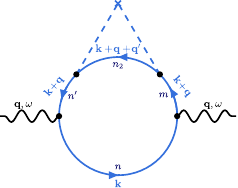}
\par\end{centering}
\caption{Polarization bubble with two disorder (dashed) lines, leading to e-e-impurity
skew scattering. Inner labels indicate the band indices. Energy arguments
of the Green functions are omitted. The upper half of the bubble corresponds
to the leading disorder-induced off-diagonal correction to the Green
functions, $G_{V,n'm}^{R,A,K}(\boldsymbol{k}+\boldsymbol{q},\epsilon+\omega)$
{[}Eqs. (\ref{eq:G_R_A_V}), (\ref{eq:G_K_V}){]}. There is an additional
diagram with the disorder lines connected to the lower half of the
bubble.\label{fig:pol. bubble one disorder line}}
\end{figure}

\begin{align}
\left[\Pi^{R}(\boldsymbol{q},\omega)-\Pi^{A}(\boldsymbol{q},\omega)\right]_{\textrm{V}} & =-\frac{1}{4}\gamma\sum_{n,n',n_{2}}\sum_{m\neq n'}\intop_{\boldsymbol{k},\boldsymbol{q}'}\intop\frac{d\epsilon}{2\pi}\{Z_{n\boldsymbol{k}\rightarrow m,\boldsymbol{k}+\boldsymbol{q}\rightarrow n_{2},\boldsymbol{k}+\boldsymbol{q}+\boldsymbol{q}'\rightarrow n',\boldsymbol{k}+\boldsymbol{q}}\nonumber \\
\times\left(F_{n_{2}}-F_{n'}\right) & \left(G_{0,n}^{R}-G_{0,n}^{A}\right)\left(G_{0,n'}^{R}-G_{0,n'}^{A}\right)\left(G_{0,n_{2}}^{R}-G_{0,n_{2}}^{A}\right)\left(G_{0,m}^{R}+G_{0,m}^{A}\right)-\left[\boldsymbol{q},\omega\rightarrow-\boldsymbol{q},-\omega\right]\},\label{eq:Pol_R_A e-e-imp}\\
\left[\Pi^{K}(\boldsymbol{q},\omega)\right]_{\textrm{V}} & =\frac{1}{4}\gamma\sum_{n,n',n_{2}}\sum_{m\neq n'}\intop_{\boldsymbol{k},\boldsymbol{q}'}\intop\frac{d\epsilon}{2\pi}\{Z_{n\boldsymbol{k}\rightarrow m,\boldsymbol{k}+\boldsymbol{q}\rightarrow n_{2},\boldsymbol{k}+\boldsymbol{q}+\boldsymbol{q}'\rightarrow n',\boldsymbol{k}+\boldsymbol{q}}\nonumber \\
\times F_{n}\left(F_{n_{2}}-F_{n'}\right) & \left(G_{0,n}^{R}-G_{0,n}^{A}\right)\left(G_{0,n'}^{R}-G_{0,n'}^{A}\right)\left(G_{0,n_{2}}^{R}-G_{0,n_{2}}^{A}\right)\left(G_{0,m}^{R}+G_{m}^{A}\right)+\left[\boldsymbol{q},\omega\rightarrow-\boldsymbol{q},-\omega\right]\},\label{eq:Pol_K e-e-imp}
\end{align}
 with $Z_{n_{1}\boldsymbol{k}_{1}\rightarrow n_{2}\boldsymbol{k}_{2}\rightarrow n_{3}\boldsymbol{k}_{3}\rightarrow n_{4}\boldsymbol{k}_{4}}$
being the imaginary part of the amplitude acquired during hopping,

\begin{equation}
Z_{n_{1}\boldsymbol{k}_{1}\rightarrow n_{2}\boldsymbol{k}_{2}\rightarrow n_{3}\boldsymbol{k}_{3}\rightarrow n_{4}\boldsymbol{k}_{4}}\equiv\textrm{Im}\left[\left\langle u_{n_{1}\boldsymbol{k}_{1}}\vert u_{n_{2}\boldsymbol{k}_{2}}\right\rangle \left\langle u_{n_{2}\boldsymbol{k}_{2}}\vert u_{n_{3}\boldsymbol{k}_{3}}\right\rangle \left\langle u_{n_{3}\boldsymbol{k}_{3}}\vert u_{n_{4}\boldsymbol{k}_{4}}\right\rangle \left\langle u_{n_{4}\boldsymbol{k}_{4}}\vert u_{n_{1}\boldsymbol{k}_{1}}\right\rangle \right].\label{eq:Z-4-product}
\end{equation}
For brevity, we have omitted the momentum and energy dependencies of
the Green functions, which can be read from the associated momentum
to the Bloch wavefunctions, e.g., $G_{0,n'}^{R}=G_{0,n'}^{R}(\boldsymbol{k}+\boldsymbol{q},\epsilon+\omega)$
(functions related to the state $n,\boldsymbol{k}$ are evaluated
on energy $\epsilon$ and all the rest are evaluated at energy $\epsilon+\omega$,
since the disorder scattering does not transfer energy).

In the two-band model, we can simplify the product of the Bloch wavefunctions appearing in
Eqs. (\ref{eq:Pol_R_A e-e-imp}) and (\ref{eq:Pol_K e-e-imp}), writing
(utilizing $m\neq n')$

\begin{align}
Z_{n\boldsymbol{k}\rightarrow m,\boldsymbol{k}+\boldsymbol{q}\rightarrow n_{2},\boldsymbol{k}+\boldsymbol{q}+\boldsymbol{q}'\rightarrow n',\boldsymbol{k}+\boldsymbol{q}} & =-\textrm{Im}\left[\left\langle u_{n\boldsymbol{k}}\vert u_{n_{2},\boldsymbol{k}+\boldsymbol{q}+\boldsymbol{q}'}\right\rangle \left\langle u_{n_{2},\boldsymbol{k}+\boldsymbol{q}+\boldsymbol{q}'}\vert u_{n',\boldsymbol{k}+\boldsymbol{q}}\right\rangle \left\langle u_{n',\boldsymbol{k}+\boldsymbol{q}}\vert u_{n\boldsymbol{k}}\right\rangle \right].\nonumber \\
\equiv & -Z_{n\boldsymbol{k}\rightarrow n_{2},\boldsymbol{k}+\boldsymbol{q}+\boldsymbol{q}'\rightarrow n',\boldsymbol{k}+\boldsymbol{q}},
\end{align}
The imaginary amplitude acquired from three hoppings is given by (measuring momentum
relative to the Weyl node) \cite{Konig2019}

\begin{equation}
Z_{n_{1}\boldsymbol{k}_{1}\rightarrow n_{2}\boldsymbol{k}_{2}\rightarrow n_{3}\boldsymbol{k}_{3}}=\frac{\xi}{4}n_{1}n_{2}n_{3}\left(\hat{\boldsymbol{k}}_{1}\times\hat{\boldsymbol{k}}_{2}\right)\cdot\hat{\boldsymbol{k}}_{3},\label{eq:Bloch phase 3 particles}
\end{equation}
where we recall that we are treating each node with chirality $\xi$
separately, assuming no internode scattering. We note that for a more
general node Hamiltonian of the form $H_{\xi}=\xi\boldsymbol{h}(\boldsymbol{k})\cdot\boldsymbol{\sigma}$,
one should replace $\hat{\boldsymbol{k}}\rightarrow\hat{\boldsymbol{h}}$
in Eq. (\ref{eq:Bloch phase 3 particles}). Substituting the corrections
$\Pi_{\textrm{V}}^{R,A,K}$ {[}Eqs. (\ref{eq:Pol_R_A e-e-imp}), (\ref{eq:Pol_K e-e-imp}){]}
into the interlayer collision integral {[}Eq. (\ref{eq:p-a collision integral keldysh}){]},
we find the e-e-impurity skew-scattering part of the collision integral
(summing over the two Weyl nodes)

\begin{align}
I_{\boldsymbol{k}}^{\textrm{e-e-imp (p,a)}} & =-W\sum_{\xi=\pm1}\sum_{nn'}\intop_{\boldsymbol{q},\boldsymbol{k}_{1},\boldsymbol{k}_{1'}}\nonumber \\
\times & \left[f_{\boldsymbol{k}}^{\textrm{p}}f_{n\boldsymbol{k}_{1}}^{\textrm{a}}\left(1-f_{\boldsymbol{k}+\boldsymbol{q}}^{\textrm{p}}\right)\left(1-f_{n'\boldsymbol{k}_{1'}}^{\textrm{a}}\right)w_{\boldsymbol{k},n\boldsymbol{k}_{1}\rightarrow\boldsymbol{k}+\boldsymbol{q},n'\boldsymbol{k}_{1'}}^{\textrm{e-e-imp}}-\left(\boldsymbol{k},n\boldsymbol{k}_{1}\leftrightarrow\boldsymbol{k}+\boldsymbol{q},n'\boldsymbol{k}_{1'}\right)\right],\label{eq:full I_skew w_asym}
\end{align}
with
\begin{align}
w_{\boldsymbol{k},n\boldsymbol{k}_{1}\rightarrow\boldsymbol{k}+\boldsymbol{q},n'\boldsymbol{k}_{1'}}^{\textrm{e-e-imp}} & \equiv4\pi^{2}\gamma\left|U_{RPA}^{R}(\boldsymbol{q},\omega)\right|^{2}\delta\left(\epsilon_{n\boldsymbol{k}_{1}}^{\textrm{a}}+\epsilon_{\boldsymbol{k}}^{\textrm{p}}-\epsilon_{n'\boldsymbol{k}_{1'}}^{\textrm{a}}-\epsilon_{\boldsymbol{k}+\boldsymbol{q}}^{\textrm{p}}\right)\nonumber \\
 & \times\sum_{n_{2}}\intop_{\boldsymbol{k}_{2}}\left\{ \frac{Z_{n\boldsymbol{k}_{1}\rightarrow n'\boldsymbol{k}_{1'}\rightarrow n_{2}\boldsymbol{k}_{2}}}{\epsilon_{n\boldsymbol{k}_{1}}^{\textrm{a}}-\epsilon_{\bar{n}\boldsymbol{k}_{1}}^{\textrm{a}}}\delta\left(\epsilon_{n\boldsymbol{k}_{1}}^{\textrm{a}}-\epsilon_{n_{2}\boldsymbol{k}_{2}}^{\textrm{a}}\right)\delta_{\boldsymbol{k}_{1}-\boldsymbol{q}-\boldsymbol{k}_{1'}}-\left[n\boldsymbol{k}_{1}\leftrightarrow n_{2}\boldsymbol{k}_{2}\right]\right\} \nonumber \\
 & +\left[\boldsymbol{k},n\boldsymbol{k}_{1}\leftrightarrow\boldsymbol{k}+\boldsymbol{q},n'\boldsymbol{k}_{1'}\right],\label{eq:W_skew full}
\end{align}
where $\bar{n}\equiv-n$. Note that $w^{\textrm{e-e-imp}}$ includes
parts where the total electron momentum is not conserved, but is rather
gained or lost due to the impurity scattering. In the level of a linearized
collision integral, some simplifications can be made due to the anti-symmetry
in $[n\boldsymbol{k}_{1}\leftrightarrow n_{2}\boldsymbol{k}_{2}]$
and $[n'\boldsymbol{k}_{1'}\leftrightarrow n_{2}\boldsymbol{k}_{2}]$
in Eq. (\ref{eq:W_skew full}). We find the linearized collision integral

\begin{align}
{\cal I}_{\boldsymbol{k}}^{\textrm{e-e-imp (p,a)}} & =-W\sum_{\xi=\pm1}\intop_{\boldsymbol{k}',\boldsymbol{k}_{1},\boldsymbol{k}_{1'}}f_{0}(\epsilon_{\boldsymbol{k}}^{\textrm{p}})f_{0}(\epsilon_{n\boldsymbol{k}_{1}}^{\textrm{a}})\left(1-f_{0}(\epsilon_{\boldsymbol{k}'}^{\textrm{p}})\right)\left(1-f_{0}(\epsilon_{n'\boldsymbol{k}_{1'}}^{\textrm{a}})\right)\nonumber \\
\times & \left[g_{n\boldsymbol{k}_{1}}^{\textrm{a}}{\cal W}_{\boldsymbol{k},n\boldsymbol{k}_{1}\rightarrow\boldsymbol{k}',n'\boldsymbol{k}_{1'}}-g_{n'\boldsymbol{k}_{1'}}^{\textrm{a}}{\cal W}_{\boldsymbol{k}',n'\boldsymbol{k}_{1'}\rightarrow\boldsymbol{k},n\boldsymbol{k}_{1}}\right],\label{eq:I e-e-imp linearized}
\end{align}
with

\begin{align}
{\cal W}_{\boldsymbol{k},n\boldsymbol{k}_{1}\rightarrow\boldsymbol{k}+\boldsymbol{q},n'\boldsymbol{k}_{1'}} & ={\cal W}_{\boldsymbol{k},n\boldsymbol{k}_{1}\rightarrow\boldsymbol{k}+\boldsymbol{q},n'\boldsymbol{k}_{1'}}^{(1)}+{\cal W}_{\boldsymbol{k},n\boldsymbol{k}_{1}\rightarrow\boldsymbol{k}+\boldsymbol{q},n'\boldsymbol{k}_{1'}}^{(2)},\label{eq:W_ee_imp_linearized}\\
{\cal W}_{\boldsymbol{k},n\boldsymbol{k}_{1}\rightarrow\boldsymbol{k}+\boldsymbol{q},n'\boldsymbol{k}_{1'}}^{(1)} & =\frac{\pi^{2}\gamma}{2}\left|U_{\textrm{RPA}}^{R}(\boldsymbol{q},\omega)\right|^{2}\delta(\epsilon_{\boldsymbol{k}}^{\textrm{p}}+\epsilon_{n\boldsymbol{k}_{1}}^{\textrm{a}}-\epsilon_{\boldsymbol{k}+\boldsymbol{q}}^{\textrm{p}}-\epsilon_{n'\boldsymbol{k}_{1'}}^{\textrm{a}})\delta_{\boldsymbol{k}_{1}-\boldsymbol{k}_{1}'-\boldsymbol{q}}\frac{\nu(\epsilon_{n\boldsymbol{k}_{1}}^{\textrm{a }})}{\epsilon_{n\boldsymbol{k}_{1}}^{\textrm{a }}}\left(\hat{\boldsymbol{k}}_{1}\times\hat{\boldsymbol{k}}_{1'}\right)\cdot\boldsymbol{M}(\epsilon_{n\boldsymbol{k}_{1}}^{\textrm{a }}),\label{eq:W_linearized_1}\\
{\cal W}_{\boldsymbol{k},n\boldsymbol{k}_{1}\rightarrow\boldsymbol{k}+\boldsymbol{q},n'\boldsymbol{k}_{1'}}^{(2)} & =\frac{\pi^{2}\gamma}{2}\left|U_{\textrm{RPA}}^{R}(\boldsymbol{q},\omega)\right|^{2}\delta(\epsilon_{\boldsymbol{k}}^{\textrm{p}}+\epsilon_{n\boldsymbol{k}_{1}}^{\textrm{a}}-\epsilon_{\boldsymbol{k}+\boldsymbol{q}}^{\textrm{p}}-\epsilon_{n'\boldsymbol{k}_{1'}}^{\textrm{a}})\delta(\epsilon_{n_{2}\boldsymbol{k}}^{\textrm{a}}-\epsilon_{n\boldsymbol{k}_{1}}^{\textrm{a }})\frac{q\xi}{v_{F}\left(k_{1'}\right)^{2}}\left(\hat{\boldsymbol{k}}_{1}\times\hat{\boldsymbol{k}}_{1'}\right)\cdot\hat{\boldsymbol{q}},\label{W_linearized_2}
\end{align}
where we assumed $k_{1'},k_{1}\gg q$ (as is the case for Coulomb
drag in the regime $k_{F}\gg1/d$) and defined the average spinor
on the Fermi surface of a single node

\begin{equation}
\boldsymbol{M}(\epsilon)\equiv\frac{1}{\nu(\epsilon)}\sum_{n}\intop_{\boldsymbol{k}}\delta\left(\epsilon_{n\boldsymbol{k}}^{\textrm{a}}-\epsilon\right)n\xi\hat{\boldsymbol{k}}.
\end{equation}

Let us comment on a peculiarity regarding the symmetry of the e-e-impurity
scattering rate under the exchange of incoming and outgoing particles,
$\boldsymbol{k},n\boldsymbol{k}_{1}\leftrightarrow\boldsymbol{k}',n'\boldsymbol{k}_{1'}$.
The full rate $w^{\textrm{e-e-imp}}$ is symmetric under the exchange,
as is explicitly seen from the exchanged term in Eq. (\ref{eq:W_skew full}).
However, in the linearized response level, the term ${\cal W}_{\boldsymbol{k},n\boldsymbol{k}_{1}\rightarrow\boldsymbol{k}',n'\boldsymbol{k}_{1'}}$
{[}Eq. (\ref{eq:W_ee_imp_linearized}){]} contains a significant anti-symmetric
part. This apparent contradiction is resolved by the fact that $w^{\textrm{e-e-imp}}$
contains non-momentum conserving terms {[}e.g., the second term in
the curly brackets in Eq. (\ref{eq:W_skew full}) where $n\boldsymbol{k}_{1}\leftrightarrow n_{2}\boldsymbol{k}_{2}$,
exchanging between the incoming and intermediate electrons{]}. In
the linearized collision integral {[}Eq. (\ref{eq:I e-e-imp linearized}){]},
we dropped terms that cancel under the integration over the angle
of the outgoing electron, e.g., under $\intop_{\boldsymbol{k}_{1'}}\delta(\epsilon_{\boldsymbol{k}_{1'}}-\epsilon)g_{\boldsymbol{k}_{1}}w_{\boldsymbol{k},n\boldsymbol{k}_{1}\rightarrow\boldsymbol{k}+\boldsymbol{q},n'\boldsymbol{k}_{1'}}^{\textrm{e-e-imp}}$
for a fixed $\epsilon$. For a typical momentum-conserving integral,
such integration only picks the momentum delta function and cannot
alter the symmetry of $w$. However, in our case, this integration
cancels a pair of terms anti-symmetric in $n'\boldsymbol{k}_{1'}\leftrightarrow n_{2}\boldsymbol{k}_{2}$,
resulting in a non-symmetric scattering rate ${\cal W}_{\boldsymbol{k},n\boldsymbol{k}_{1}\rightarrow\boldsymbol{k}+\boldsymbol{q},n'\boldsymbol{k}_{1'}}$.

\section{\label{subsec:Appendix-B:-Momentum}Drag force from the
e-e collision integral}

Here, we compute the drag force (momentum transfer rate) between the
layers for boosted distribution functions {[}Eqs. (\ref{eq:delta f parametrization}),
(\ref{eq:g_nk boost velocity}) and (\ref{eq:g_nk energy-dependent boost velocity})
of the main text{]} and an interlayer scattering integral given by
Eq. (\ref{eq:general coll integral}) with the scattering rates of
Eq. (\ref{eq:scattering rate detail}). We show that the drag force
can be written in the form of Eqs. (\ref{eq:transfered momentum response}),
(\ref{eq:F_drag high T}) in the main text, and calculate explicitly
the drag coefficients $\eta_{\parallel,H}^{D}$. We separate the calculation
for each term in the scattering rate.

\subsection{Born-approximation part of the scattering rate}

The Born-approximation part of $w^{\textrm{e-e}}$ is symmetric and
conserves momentum, $w_{\boldsymbol{k},n\boldsymbol{k}_{1}+\boldsymbol{q}\rightarrow\boldsymbol{k}+\boldsymbol{q},n'\boldsymbol{k}_{1}}^{\textrm{Born}}=w_{\boldsymbol{k}+\boldsymbol{q},n'\boldsymbol{k}_{1}\rightarrow\boldsymbol{k},n\boldsymbol{k}_{1}+\boldsymbol{q}}^{\textrm{Born}}$.
We substitute $w^{\textrm{e-e}}\rightarrow w^{\textrm{Born}}$ in
Eq. (\ref{eq:general coll integral}), linearize the collision integral
with respect to the non-equilibrium part of the distribution functions,
utilize the energy and momentum conservation of the collision integral,
and arrive at

\begin{align}
{\cal I}_{\boldsymbol{k}}^{\textrm{Born (p,a)}}\left[f^{\textrm{p}},f^{\textrm{a}}\right] & =-\frac{W}{4}\sum_{\xi=\pm1}\sum_{nn'}\intop_{\boldsymbol{q},\boldsymbol{k}_{1}}\frac{w_{\boldsymbol{k},n\boldsymbol{k}_{1}+\boldsymbol{q}\rightarrow\boldsymbol{k}+\boldsymbol{q},n'\boldsymbol{k}_{1}}^{\textrm{Born}}}{\sinh^{2}\frac{\omega}{2T}}\left(f_{0}(\epsilon_{\boldsymbol{k}+\boldsymbol{q}}^{\textrm{p}})-f_{0}(\epsilon_{\boldsymbol{k}}^{\textrm{p}})\right)\left(f_{0}(\epsilon_{n\boldsymbol{k}_{1}+\boldsymbol{q}}^{\textrm{a}})-f_{0}(\epsilon_{n'\boldsymbol{k}_{1}}^{\textrm{a}})\right)\nonumber \\
\times & \left(g_{\boldsymbol{k}}^{\textrm{p}}+g_{n\boldsymbol{k}_{1}+\boldsymbol{q}}^{\textrm{a}}-g_{\boldsymbol{k}+\boldsymbol{q}}^{\textrm{p}}-g_{n'\boldsymbol{k}_{1}}^{\textrm{a}}\right),\label{eq:I_ee born g_k}
\end{align}
where $\omega\equiv\epsilon_{\boldsymbol{k}+\boldsymbol{q}}^{\textrm{p}}-\epsilon_{\boldsymbol{k}}^{\textrm{p}}=\epsilon_{n\boldsymbol{k}_{1}+\boldsymbol{q}}^{\textrm{a}}-\epsilon_{n'\boldsymbol{k}_{1}}^{\textrm{a}}$.
The drag force between the layers is obtained by multiplying Eq. (\ref{eq:I_ee born g_k})
by $\boldsymbol{k}$ and integrating over $\boldsymbol{k}$,

\begin{align}
\boldsymbol{F}^{\textrm{p,a (Born)}} & \equiv\intop_{\boldsymbol{k}}\boldsymbol{k}{\cal I}_{\boldsymbol{k}}^{\textrm{Born (p,a)}}=-\frac{W}{4}\sum_{\xi=\pm1}\sum_{nn'}\intop_{\boldsymbol{k},\boldsymbol{q},\boldsymbol{k}_{1}}\boldsymbol{k}\frac{w_{\boldsymbol{k},n\boldsymbol{k}_{1}+\boldsymbol{q}\rightarrow\boldsymbol{k}+\boldsymbol{q},n'\boldsymbol{k}_{1}}^{\textrm{Born}}}{\sinh^{2}\frac{\omega}{2T}}\nonumber \\
 & \times\left(f_{0}(\epsilon_{\boldsymbol{k}+\boldsymbol{q}}^{\textrm{p}})-f_{0}(\epsilon_{\boldsymbol{k}}^{\textrm{p}})\right)\left(f_{0}(\epsilon_{n\boldsymbol{k}_{1}+\boldsymbol{q}}^{\textrm{a}})-f_{0}(\epsilon_{n'\boldsymbol{k}_{1}}^{\textrm{a}})\right)\left(g_{\boldsymbol{k}}^{\textrm{p}}+g_{n\boldsymbol{k}_{1}+\boldsymbol{q}}^{\textrm{a}}-g_{\boldsymbol{k}+\boldsymbol{q}}^{\textrm{p}}-g_{n'\boldsymbol{k}_{1}}^{\textrm{a}}\right).\label{eq:drag force sym not simplified}
\end{align}
The expression above may be simplified by adding the opposite scattering
process to the integrand. Concretely, writing the integral as $\intop_{\boldsymbol{q},\boldsymbol{k},\boldsymbol{k}'}h(\boldsymbol{k},\boldsymbol{k}',\boldsymbol{q})$,
we rename the integration variables $\boldsymbol{k},\boldsymbol{k}',\boldsymbol{q}\rightarrow\boldsymbol{k}+\boldsymbol{q},\boldsymbol{k}'+\boldsymbol{q},-\boldsymbol{q}$
and rewrite the integral as $\intop_{\boldsymbol{q},\boldsymbol{k},\boldsymbol{k}'}h(\boldsymbol{k},\boldsymbol{k}',\boldsymbol{q})=\frac{1}{2}\intop_{\boldsymbol{q},\boldsymbol{k},\boldsymbol{k}'}\left(h(\boldsymbol{k},\boldsymbol{k}',\boldsymbol{q})+h(\boldsymbol{k}+\boldsymbol{q},\boldsymbol{k}'+\boldsymbol{q},-\boldsymbol{q})\right)$.
Doing this for the integral in Eq. (\ref{eq:drag force sym not simplified})
leads to

\begin{align}
\boldsymbol{F}^{\textrm{p,a (Born)}} & =\frac{W}{8T}\sum_{\xi=\pm1}\sum_{nn'}\intop_{\boldsymbol{q},\boldsymbol{k},\boldsymbol{k}_{1}}\boldsymbol{q}\frac{w_{\boldsymbol{k},n\boldsymbol{k}_{1}+\boldsymbol{q}\rightarrow\boldsymbol{k}+\boldsymbol{q},n'\boldsymbol{k}_{1}}^{\textrm{Born}}}{\sinh^{2}\frac{\omega}{2T}}\left(f_{0}(\epsilon_{\boldsymbol{k}+\boldsymbol{q}}^{\textrm{p}})-f_{0}(\epsilon_{\boldsymbol{k}}^{\textrm{p}})\right)\left(f_{0}(\epsilon_{n\boldsymbol{k}_{1}+\boldsymbol{q}}^{\textrm{a}})-f_{0}(\epsilon_{n'\boldsymbol{k}_{1}}^{\textrm{a}})\right)\nonumber \\
 & \times\left[g_{\boldsymbol{k}}^{\textrm{p}}+g_{n\boldsymbol{k}_{1}+\boldsymbol{q}}^{\textrm{a}}-g_{\boldsymbol{k}+\boldsymbol{q}}^{\textrm{p}}-g_{n'\boldsymbol{k}_{1}}^{\textrm{a}}\right].\label{eq:drag force sym simplified}
\end{align}

We now use Eq. (\ref{eq:drag force sym simplified}) to calculate
the drag force for the case where the distribution functions are boosted
velocity distributions, and treat the more general case of energy-dependent
boost velocities later.

\textbf{Simple case: boosted velocity distributions}

For boosted velocity distributions, $g_{n\boldsymbol{k}}^{l}=\boldsymbol{k}\cdot\boldsymbol{u}^{l}/T$.
Substituting into Eq. (\ref{eq:drag force sym simplified}) leads
to

\begin{align}
\boldsymbol{F}^{\textrm{p,a (Born)}} & =\frac{W}{8T}\sum_{\xi=\pm1}\sum_{nn'}\intop_{\boldsymbol{q},\boldsymbol{k},\boldsymbol{k}_{1}}\boldsymbol{q}\left[\boldsymbol{q}\cdot\left(\boldsymbol{u}^{\textrm{a}}-\boldsymbol{u}^{\textrm{p}}\right)\right]\nonumber \\
 & \times\frac{w_{\boldsymbol{k},n\boldsymbol{k}_{1}+\boldsymbol{q}\rightarrow\boldsymbol{k}+\boldsymbol{q},n'\boldsymbol{k}_{1}}^{\textrm{Born}}}{\sinh^{2}\frac{\omega}{2T}}\left(f_{0}(\epsilon_{\boldsymbol{k}+\boldsymbol{q}}^{\textrm{p}})-f_{0}(\epsilon_{\boldsymbol{k}}^{\textrm{p}})\right)\left(f_{0}(\epsilon_{n\boldsymbol{k}_{1}+\boldsymbol{q}}^{\textrm{a}})-f_{0}(\epsilon_{n'\boldsymbol{k}_{1}}^{\textrm{a}})\right).\label{eq:drag force boost velocities}
\end{align}
For any isotropic system, this results in the drag force

\begin{equation}
\boldsymbol{F}^{\textrm{p,a (Born)}}=\frac{\eta_{\parallel}^{D}}{d}\left(\boldsymbol{u}^{\textrm{a}}-\boldsymbol{u}^{\textrm{p}}\right),
\end{equation}
with the drag coefficient $\eta_{\parallel}^{D}$ given by

\begin{equation}
\eta_{\parallel}^{D}=\frac{Wd}{16T}\sum_{\xi=\pm1}\sum_{nn'}\intop_{\boldsymbol{q},\boldsymbol{k},\boldsymbol{k}_{1}}\frac{q^{2}}{\sinh^{2}\frac{\omega}{2T}}w_{\boldsymbol{k},n\boldsymbol{k}_{1}+\boldsymbol{q}\rightarrow\boldsymbol{k}+\boldsymbol{q},n'\boldsymbol{k}_{1}}^{\textrm{Born}}\left(f_{0}(\epsilon_{\boldsymbol{k}+\boldsymbol{q}}^{\textrm{p}})-f_{0}(\epsilon_{\boldsymbol{k}}^{\textrm{p}})\right)\left(f_{0}(\epsilon_{n\boldsymbol{k}_{1}+\boldsymbol{q}}^{\textrm{a}})-f_{0}(\epsilon_{n'\boldsymbol{k}_{1}}^{\textrm{a}})\right).\label{eq:eta_parallel general}
\end{equation}
Substituting the specific form of $w^{\textrm{Born}}$, Eq. (\ref{eq:symmetric scat rate}),
we obtain

\begin{equation}
\eta_{\parallel}^{D}=\frac{Wd}{8\pi T}\intop_{-\infty}^{\infty}d\omega\frac{1}{\sinh^{2}\frac{\omega}{2T}}\intop_{\boldsymbol{q}}q^{2}\left|U_{\textrm{RPA}}^{R}(\boldsymbol{q},\omega)\right|^{2}\textrm{Im}\Pi_{0}^{\textrm{p,}R}(\boldsymbol{q},\omega)\textrm{Im}\Pi_{0}^{\textrm{a,}R}(\boldsymbol{q},\omega),\label{eq:eta_parallel result}
\end{equation}
where $\Pi_{0}^{l,R}(\boldsymbol{q},\omega)$ are the bare polarization
operators of the layers, with their imaginary parts given by

\begin{align}
\textrm{Im}\Pi_{0}^{\textrm{p,}R}(\boldsymbol{q},\omega) & =\pi\intop_{\boldsymbol{k}}\delta(\epsilon_{\boldsymbol{k}+\boldsymbol{q}}^{\textrm{p}}-\epsilon_{\boldsymbol{k}}^{\textrm{p }}-\omega)\left(f_{0}(\epsilon_{\boldsymbol{k}+\boldsymbol{q}}^{\textrm{p}})-f_{0}(\epsilon_{\boldsymbol{k}}^{\textrm{p }})\right),\label{eq:polarizabilities p}\\
\textrm{Im}\Pi_{0}^{\textrm{a,}R}(\boldsymbol{q},\omega) & =\pi\sum_{\xi=\pm1}\sum_{nn'}\intop_{\boldsymbol{k}}\delta(\epsilon_{n\boldsymbol{k}+\boldsymbol{q}}^{\textrm{a }}-\epsilon_{n'\boldsymbol{k}}^{\textrm{a }}-\omega)\left(f_{0}(\epsilon_{n\boldsymbol{k}+\boldsymbol{q}}^{\textrm{a }})-f_{0}(\epsilon_{n'\boldsymbol{k}}^{\textrm{a }})\right)\left|\left\langle u_{n'\boldsymbol{k}}\vert u_{n\boldsymbol{k}+\boldsymbol{q}}\right\rangle \right|^{2}.\label{eq:polarizabilities a}
\end{align}
Evaluating Eq. (\ref{eq:eta_parallel result}) with the approximate
Coulomb interaction {[}Eq. (\ref{eq:u_R_RPA_sq}){]} in the limits
$T\ll T_{d}$ and $T\gg T_{d}$ leads to Eqs. (\ref{eq:eta_parallel_low_T})
and (\ref{eq:eta_parallel high T}) of the main text. To briefly explain
the calculation, in the limit $T\ll T_{d}$, the frequency integral
in Eq. (\ref{eq:eta_parallel result}) is dominated by $\omega\sim T$
and the result breaks into the product of the independent $\omega$
and $\boldsymbol{q}$ integrals. In the limit $T\gg T_{d}$, the frequency
integration is cut off by the boundary of the particle-hole spectrum
in the layers, $\omega<\min(v_{F}^{\textrm{a}},v_{F}^{\textrm{p}})q$,
and the $\omega$ and $\boldsymbol{q}$ integrals do not factorize
\cite{Narozhny2016}. The thermal factor can be approximated by $\sinh\left(\omega/2T\right)\approx\omega/\left(2T\right)$
\cite{Jauho1993}. Note that the frequency dependence of the interlayer
scattering propagator $U_{\textrm{RPA}}^{R}(\boldsymbol{q},\omega)$
is important at high temperatures (see Appendix \ref{subsec:Geometrical-factors}
for more details about the calculation).

\textbf{General case: energy-dependent boost velocities}

We now consider the more general case, parametrizing the non-equilibrium
distribution functions with an energy-dependent boost velocity $g_{n\boldsymbol{k}}^{\textrm{a}}=\boldsymbol{k}\cdot\boldsymbol{u}^{\textrm{a}}(\epsilon_{n\boldsymbol{k}}^{\textrm{a }})/T$,
as in Sec. \ref{subsec:High-temperatures drag conductivity} of the
main text. For simplicity, we take $\boldsymbol{u}^{\textrm{p}}=0$.
Substituting this form of $g_{n\boldsymbol{k}}^{\textrm{a}}$ in Eq.
(\ref{eq:drag force sym simplified}) yields

\begin{align}
\boldsymbol{F}^{\textrm{p,a (Born)}} & =\frac{W}{8T}\sum_{\xi=\pm1}\sum_{nn'}\intop_{\boldsymbol{q},\boldsymbol{k},\boldsymbol{k}_{1}}\boldsymbol{q}\frac{w_{\boldsymbol{k},n\boldsymbol{k}_{1}+\boldsymbol{q}\rightarrow\boldsymbol{k}+\boldsymbol{q},n'\boldsymbol{k}_{1}}^{\textrm{Born}}}{\sinh^{2}\frac{\omega}{2T}}\left(f_{0}(\epsilon_{\boldsymbol{k}+\boldsymbol{q}}^{\textrm{p}})-f_{0}(\epsilon_{\boldsymbol{k}}^{\textrm{p}})\right)\left(f_{0}(\epsilon_{n\boldsymbol{k}_{1}+\boldsymbol{q}}^{\textrm{a}})-f_{0}(\epsilon_{n'\boldsymbol{k}_{1}}^{\textrm{a}})\right)\nonumber \\
 & \times\left[\left(\boldsymbol{k}_{1}+\boldsymbol{q}\right)\cdot\boldsymbol{u}^{\textrm{a}}(\epsilon_{n'\boldsymbol{k}_{1}}^{\textrm{a}}+\omega)-\boldsymbol{k}_{1}\cdot\boldsymbol{u}^{\textrm{a}}(\epsilon_{n'\boldsymbol{k}_{1}}^{\textrm{a}})\right].
\end{align}
Performing a Taylor expansion for $\boldsymbol{u}^{\textrm{a}}(\epsilon)$
up to the first-derivative term, we find

\begin{align}
\boldsymbol{F}^{\textrm{p,a (Born)}} & =\frac{W}{8T}\sum_{\xi=\pm1}\sum_{nn'}\intop_{\boldsymbol{q},\boldsymbol{k}_{1}}\boldsymbol{q}\frac{w_{\boldsymbol{k},n\boldsymbol{k}_{1}+\boldsymbol{q}\rightarrow\boldsymbol{k}+\boldsymbol{q},n'\boldsymbol{k}_{1}}^{\textrm{Born}}}{\sinh^{2}\frac{\omega}{2T}}\left(f_{0}(\epsilon_{\boldsymbol{k}+\boldsymbol{q}}^{\textrm{p}})-f_{0}(\epsilon_{\boldsymbol{k}}^{\textrm{p}})\right)\left(f_{0}(\epsilon_{n\boldsymbol{k}_{1}+\boldsymbol{q}}^{\textrm{a}})-f_{0}(\epsilon_{n'\boldsymbol{k}_{1}}^{\textrm{a}})\right)\nonumber \\
 & \left[\boldsymbol{q}\cdot\boldsymbol{u}^{\textrm{a}}(\epsilon_{n'\boldsymbol{k}_{1}}^{\textrm{a}})+\boldsymbol{k}_{1}\cdot\omega\frac{\partial\boldsymbol{u}^{\textrm{a}}(\epsilon_{n'\boldsymbol{k}_{1}}^{\textrm{a}})}{\partial\epsilon_{n'\boldsymbol{k}_{1}}^{\textrm{a}}}\right]\equiv\boldsymbol{F}^{\textrm{p,a}}[\boldsymbol{u}^{\textrm{a}}]+\boldsymbol{F}^{\textrm{p,a}}[\partial\boldsymbol{u}^{\textrm{a}}/\partial\epsilon].\label{eq:I_born energy-dependent u}
\end{align}
In the limit where $\epsilon_{F}^{\textrm{a}}\gg T,T_{d}$, one may
substitute $\boldsymbol{u}^{\textrm{a}}(\epsilon_{\boldsymbol{k}'}^{\textrm{a }})\approx\boldsymbol{u}^{\textrm{a}}(\epsilon_{F}^{\textrm{a}})$
in the first term, returning to the case of the last section. The
second term in Eq. (\ref{eq:I_born energy-dependent u}) arises from
the energy dependence of the boost velocity. We write it as

\begin{equation}
F_{\alpha}^{\textrm{p,a}}[\partial\boldsymbol{u}^{a}/\partial\epsilon]\equiv\eta_{\parallel(1)}^{D}\frac{\epsilon_{F}^{\textrm{a}}}{d}\left.\frac{\partial u_{\alpha}^{\textrm{a}}}{\partial\epsilon}\right|_{\epsilon=\epsilon_{F}^{\textrm{a}}}.
\end{equation}
This part of the force corresponds to the third term in Eq. (\ref{eq:F_drag high T})
of the main text, with the drag coefficient

\begin{equation}
\eta_{\parallel(1)}^{D}=\frac{Wd}{8\pi T}\int d\omega\frac{1}{\sinh^{2}\frac{\omega}{2T}}\intop_{\boldsymbol{q}}q^{2}\left(\frac{\omega}{v_{F}^{\textrm{a}}q}\right)^{2}\left|U_{\textrm{RPA}}^{R}(\boldsymbol{q},\omega)\right|^{2}\textrm{Im}\Pi_{0}^{\textrm{p,}R}(\boldsymbol{q},\omega)\textrm{Im}\Pi_{0}^{\textrm{a,}R}(\boldsymbol{q},\omega).\label{eq:eta_1 integral}
\end{equation}

For an e-e scattering where the WSM electron scatters from (momentum,
energy) $\left(\boldsymbol{k},\epsilon_{n\boldsymbol{k}}^{\textrm{a}}\right)\rightarrow\left(\boldsymbol{k}+\boldsymbol{q},\epsilon_{n'\boldsymbol{k}+\boldsymbol{q}}^{\textrm{a}}=\epsilon_{n\boldsymbol{k}}^{\textrm{a}}+\omega\right)$,
the factor $\omega/v_{F}^{\textrm{a}}q$ is equal to the cosine of
the angle between $\boldsymbol{v}_{n\boldsymbol{k}}^{\textrm{a}}$
and $\boldsymbol{q}$ (for $k\sim k_{F}\gg q\sim1/d$). Thus, this
factor approaches one for forward scattering (i.e., for $\boldsymbol{q}\parallel\boldsymbol{v}_{n\boldsymbol{k}}^{\textrm{a}}$)
and zero for perpendicular scattering ($\boldsymbol{q}\perp\boldsymbol{v}_{n\boldsymbol{k}}^{\textrm{a}}$).
For low temperatures ($T\ll T_{d}$), perpendicular scattering is
dominant ($\omega/v_{F}^{\textrm{a}}q\sim Td/v_{F}^{\textrm{a}}\ll1$),
and the resulting contribution to the drag from $\eta_{\parallel(1)}^{D}$
is subleading in $\left(T/T_{d}\right)^{2}$ compared to the $\eta_{\parallel(0)}^{D}$
term {[}Eq. (\ref{eq:drag force boost velocities}){]}. In the opposite
limit where $T\gg T_{d}$, the two terms are comparable. We evaluate
the integral with the approximated Coulomb interaction {[}Eq. (\ref{eq:u_R_RPA_sq}){]}
to obtain the value of $\eta_{\parallel(1)}^{D}$ presented in Eq.
(\ref{eq:eta u deriv}) of the main text. Note that in the case $v_{F}^{\textrm{a}}\gg v_{F}^{\textrm{p}}$,
interlayer collisions with forward scattering in the WSM are not possible,
and the coefficient $\eta_{\parallel(1)}^{D}$ becomes parametrically
small, as can be seen from Eq. (\ref{eq:eta u deriv}).

\subsection{Skew scattering}

Next, we calculate the drag force from the skew-scattering parts of
the e-e collision integral, corresponding to the e-e-impurity interference
and side-jump modified e-e collision integrals.

\textbf{e-e-impurity scattering}

The linearized form of the e-e-impurity part of the collision integral
is given in Eq. (\ref{eq:I e-e-imp linearized}). We calculate the
contributions from the two terms in the scattering rate ${\cal W}={\cal W}^{(1)}+{\cal W}^{(2)}$
{[}Eq. (\ref{eq:W_ee_imp_linearized}){]} separately, writing ${\cal I}_{\boldsymbol{k}}^{\textrm{e-e-imp (p,a)}}={\cal I}_{\boldsymbol{k}}^{\textrm{e-e-imp (1)}}+{\cal I}_{\boldsymbol{k}}^{\textrm{e-e-imp (2)}}$.
The term ${\cal W}^{(1)}$ is antisymmetric in incoming and outgoing
particles, and the corresponding term in the collision integral is
given by

\begin{align}
{\cal I}_{\boldsymbol{k}}^{\textrm{e-e-imp (1)}} & =-\frac{W}{4}\sum_{\xi=\pm1}\sum_{nn'}\intop_{\boldsymbol{q},\boldsymbol{k}_{1}}\frac{{\cal W}_{\boldsymbol{k},n\boldsymbol{k}_{1}+\boldsymbol{q}\rightarrow\boldsymbol{k}+\boldsymbol{q},n'\boldsymbol{k}_{1}}^{(1)}}{\sinh^{2}\frac{\omega}{2T}}\left(f_{0}(\epsilon_{\boldsymbol{k}+\boldsymbol{q}}^{\textrm{p}})-f_{0}(\epsilon_{\boldsymbol{k}}^{\textrm{p}})\right)\left(f_{0}(\epsilon_{n\boldsymbol{k}_{1}+\boldsymbol{q}}^{\textrm{a}})-f_{0}(\epsilon_{n'\boldsymbol{k}_{1}}^{\textrm{a}})\right)\nonumber \\
 & \times\left(g_{n\boldsymbol{k}_{1}+\boldsymbol{q}}^{\textrm{a}}+g_{n'\boldsymbol{k}_{1}}^{\textrm{a}}\right),\label{eq:e-e-imp1 intermediate}
\end{align}
where we utilized the momentum conservation of ${\cal W}^{(1)}$ to
eliminate one momentum integration. In the limit $k_{F}^{\textrm{a}}\gg1/d$,
the terms corresponding to $g_{n\boldsymbol{k}_{1}+\boldsymbol{q}}^{\textrm{a}}$
and $g_{n'\boldsymbol{k}_{1}}^{\textrm{a}}$ in Eq. (\ref{eq:e-e-imp1 intermediate})
give equal contributions, and we can simplify, substituting ${\cal W}^{(1)}$
from Eq. (\ref{eq:W_linearized_1}), 

\begin{align}
{\cal I}_{\boldsymbol{k}}^{\textrm{e-e-imp (1)}} & =-\frac{W}{8T}\intop_{-\infty}^{\infty}d\omega\frac{1}{\sinh^{2}\frac{\omega}{2T}}\intop_{\boldsymbol{q}}\left(f_{0}(\epsilon_{\boldsymbol{k}+\boldsymbol{q}}^{\textrm{p}})-f_{0}(\epsilon_{\boldsymbol{k}}^{\textrm{p}})\right)\delta(\epsilon_{\boldsymbol{k}+\boldsymbol{q}}^{\textrm{p}}-\epsilon_{\boldsymbol{k}}^{\textrm{p}}-\omega)\nonumber \\
 & \times\left|U_{\textrm{RPA}}^{R}(\boldsymbol{q},\omega)\right|^{2}\left[1-\left(\frac{\omega}{v_{F}^{\textrm{a}}q}\right)^{2}\right]\textrm{Im}\Pi_{0}^{\textrm{a,}R}(\boldsymbol{q},\omega)\frac{C}{\epsilon_{F}^{\textrm{a}}\tau^{\textrm{a}}}\epsilon_{\alpha\beta}q_{\alpha}u_{\beta}^{\textrm{a}}.
\end{align}
Calculating the corresponding drag force in the same manner as in
the previous subsection, we find

\begin{align}
F_{\alpha}^{\textrm{e-e-imp (1)}} & \equiv\intop_{\boldsymbol{k}}k_{\alpha}{\cal I}_{\boldsymbol{k}}^{\textrm{e-e-imp (1)}}\nonumber \\
 & =-\frac{W}{32\pi T}\frac{C}{\epsilon_{F}^{\textrm{a}}\tau^{\textrm{a}}}\intop_{-\infty}^{\infty}d\omega\frac{1}{\sinh^{2}\frac{\omega}{2T}}\intop_{\boldsymbol{q}}q^{2}\left|U_{\textrm{RPA}}^{R}(\boldsymbol{q},\omega)\right|^{2}\left[1-\left(\frac{\omega}{v_{F}^{\textrm{a}}q}\right)^{2}\right]\textrm{Im}\Pi_{0}^{\textrm{p,}R}(\boldsymbol{q},\omega)\textrm{Im}\Pi_{0}^{\textrm{a,}R}(\boldsymbol{q},\omega)\epsilon_{\alpha\beta}u_{\beta}^{\textrm{a}}.\label{eq:F_ee_imp integral}
\end{align}
This corresponds to a Hall-like drag force, $F_{\alpha}^{\textrm{e-e-imp (1)}}\sim\epsilon_{\alpha\beta}\eta_{H(\textrm{e-e-imp,1)}}^{D}u_{\beta}^{\textrm{a}}/d$
{[}second term in Eq. (\ref{eq:transfered momentum response}){]},
with

\begin{equation}
\eta_{H(\textrm{e-e-imp,1)}}^{D}=-\frac{C}{4\epsilon_{F}^{\textrm{a}}\tau^{\textrm{a}}}\eta_{D}^{\parallel}Q_{3}(\frac{v_{F}^{\textrm{p}}}{v_{F}^{\textrm{a}}}),
\end{equation}
and $Q_{3}(z)$ given in Eq. (\ref{eq:F3_factor}). The calculation
for the contribution from ${\cal I}_{\boldsymbol{k}}^{\textrm{e-e-imp (2)}}$
is similar, and leads to $\eta_{H(\textrm{e-e-imp,2)}}^{D}=\eta_{H(\textrm{e-e-imp,1)}}^{D}/3$.
In total, e-e-impurity scattering generates the drag force $F_{\alpha}^{\textrm{\textrm{p,a (\textrm{e-e-imp)}}}}=\eta_{H(\textrm{e-e-imp)}}^{D}\epsilon_{\alpha\beta}u_{\beta}^{\textrm{a}}/d$,
with 
\begin{equation}
\eta_{H(\textrm{e-e-imp)}}^{D}=\frac{4}{3}\eta_{H(\textrm{e-e-imp,1)}}^{D}.
\end{equation}

\textbf{Side-jump collision integral}

Next, we calculate the contribution from the side-jump correction
to the e-e collision integral, Eq. (\ref{eq:sj coll integral}). Since
the electric field is explicit in the scattering rate, in the linear
response level we substitute the equilibrium value of the distribution
functions, obtaining

\begin{align}
{\cal I}_{\boldsymbol{k}}^{\textrm{s.j. (p,a)}} & =\frac{\pi W}{2T}\sum_{\xi=\pm1}\sum_{nn'}\intop_{\boldsymbol{k}_{1},\boldsymbol{q}}\frac{1}{\sinh^{2}(\frac{\omega}{2T})}\left|U_{\textrm{RPA}}^{R}(\boldsymbol{q},\omega)\right|^{2}\delta(\epsilon_{\boldsymbol{k}}^{\textrm{p}}+\epsilon_{n\boldsymbol{k}_{1}+\boldsymbol{q}}^{\textrm{a}}-\epsilon_{\boldsymbol{k}+\boldsymbol{q}}^{\textrm{p}}-\epsilon_{n'\boldsymbol{k}_{1}}^{\textrm{a}})\nonumber \\
 & \times\left(f_{0}(\epsilon_{\boldsymbol{k}+\boldsymbol{q}}^{\textrm{p}})-f_{0}(\epsilon_{\boldsymbol{k}}^{\textrm{p}})\right)\left(f_{0}(\epsilon_{n\boldsymbol{k}_{1}+\boldsymbol{q}}^{\textrm{a}})-f_{0}(\epsilon_{n'\boldsymbol{k}_{1}}^{\textrm{a}})\right)\left|\left\langle n\boldsymbol{k}_{1}+\boldsymbol{q}\vert n'\boldsymbol{k}_{1}\right\rangle \right|^{2}\delta\boldsymbol{r}_{n\boldsymbol{k}_{1}+\boldsymbol{q},n'\boldsymbol{k}_{1},}\cdot e\boldsymbol{E}.
\end{align}
Substituting the value of the coordinate shift for the Weyl electrons
in a node of chirality $\xi$ \cite{Konig2019} 
\begin{equation}
\delta\boldsymbol{r}_{n\boldsymbol{k},n'\boldsymbol{k}'}=\xi\frac{\hat{\boldsymbol{k}}\times\hat{\boldsymbol{k}}'}{4\left|\left\langle \boldsymbol{k}\vert\boldsymbol{k}'\right\rangle \right|^{2}}\left(\frac{n'}{k}+\frac{n}{k'}\right),
\end{equation}
we get

\begin{align}
{\cal I}_{\boldsymbol{k}}^{\textrm{s.j. (p,a)}} & =-\frac{WC}{8T}\intop d\omega\frac{1}{4\sinh^{2}\frac{\omega}{2T}}\intop_{\boldsymbol{q}}\left|U_{\textrm{RPA}}^{R}(\boldsymbol{q},\omega)\right|^{2}\left(f_{0}(\epsilon_{\boldsymbol{k}+\boldsymbol{q}}^{\textrm{p}})-f_{0}(\epsilon_{\boldsymbol{k}}^{\textrm{p}})\right)\delta(\epsilon_{\boldsymbol{k}+\boldsymbol{q}}^{\textrm{p}}-\epsilon_{\boldsymbol{k}}^{\textrm{p}}-\omega)\nonumber \\
 & \times\textrm{Im}\Pi_{0}^{\textrm{a,}R}(\boldsymbol{q},\omega)\left[1-\left(\frac{\omega}{v_{F}^{\textrm{a}}q}\right)^{2}\right]\left(\frac{v_{F}^{\textrm{a}}}{\epsilon_{F}^{\textrm{a}}}\right)^{2}\epsilon_{\alpha\beta}q_{\alpha}eE_{\beta}.
\end{align}
The resulting drag force is given by

\begin{align}
F_{\alpha}^{\textrm{p,a (s.j.)}} & =\intop_{\boldsymbol{k}}k_{\alpha}{\cal I}_{\boldsymbol{k}}^{\textrm{s.j. (p,a)}}\nonumber \\
 & =-\frac{WC}{32\pi T}\intop d\omega\frac{1}{\sinh^{2}\frac{\omega}{2T}}\intop_{\boldsymbol{q}}\left|U_{\textrm{RPA}}^{R}(\boldsymbol{q},\omega)\right|^{2}\left[1-\left(\frac{\omega}{v_{F}^{\textrm{a}}q}\right)^{2}\right]\textrm{Im}\Pi_{0}^{\textrm{p,}R}(\boldsymbol{q},\omega)\textrm{Im}\Pi_{0}^{\textrm{a,}R}(\boldsymbol{q},\omega)\nonumber \\
 & \times\left(\frac{v_{F}^{\textrm{a}}}{\epsilon_{F}^{\textrm{a}}}\right)^{2}q^{2}\epsilon_{\alpha\beta}eE_{\beta}.\label{eq:F_sj drag E}
\end{align}
The resulting force is perpendicular to the electric field in the
active layer. Since $\boldsymbol{F}^{\textrm{p,a (s.j.)}}$ is already
subleading in $\left(1/\epsilon_{F}^{\textrm{a}}\tau^{\textrm{a}}\right)$
compared to the leading part of $\boldsymbol{F}^{\textrm{p,a}}$,
we can approximate $e\boldsymbol{E}^{\textrm{a}}\simeq\boldsymbol{u}^{\textrm{a}}\epsilon_{F}^{\textrm{a}}/\left(\left(v_{F}^{\textrm{a}}\right)^{2}\tau^{\textrm{a},\parallel}\right)=2\boldsymbol{u}^{\textrm{a}}\epsilon_{F}^{\textrm{a}}/\left(3\left(v_{F}^{\textrm{a}}\right)^{2}\tau^{\textrm{a}}\right)$
and write Eq. (\ref{eq:F_sj drag E}) as

\begin{align}
F_{\alpha}^{\textrm{p,a (s.j.)}} & =-\frac{WC}{48\pi T}\intop d\omega\frac{1}{\sinh^{2}\frac{\omega}{2T}}\intop_{\boldsymbol{q}}\left|U_{\textrm{RPA}}^{R}(\boldsymbol{q},\omega)\right|^{2}\left[1-\left(\frac{\omega}{v_{F}^{\textrm{a}}q}\right)^{2}\right]\textrm{Im}\Pi_{0}^{\textrm{p,}R}(\boldsymbol{q},\omega)\textrm{Im}\Pi_{0}^{\textrm{a,}R}(\boldsymbol{q},\omega)\nonumber \\
 & \times\frac{1}{\epsilon_{F}^{\textrm{a}}\tau^{\textrm{a}}}q^{2}\epsilon_{\alpha\beta}u_{\beta}^{\textrm{a}}\equiv\eta_{H\textrm{(s.j.)}}^{D}\epsilon_{\alpha\beta}u_{\beta}^{\textrm{a}}/d,
\end{align}
with $\eta_{H\textrm{(s.j.)}}^{D}=\eta_{H\textrm{(e-e-imp)}}^{D}/2$.
Summing the contributions from the e-e-impurity scattering and e-e-side-jump
scattering, $F_{\alpha}^{\textrm{p,a }(H)}\equiv F_{\alpha}^{\textrm{p,a (\textrm{e-e-imp)}}}+F_{\alpha}^{\textrm{p,a (\textrm{s.j.)}}}\equiv\eta_{H}^{D}\epsilon_{\alpha\beta}u_{\beta}^{\textrm{a}}/d$,
we find the total result for the Hall component of drag response $\eta_{H}^{D}=\eta_{H\textrm{(e-e-imp)}}^{D}+\eta_{H\textrm{(s.j.)}}^{D}$
given in the main text, Eqs. (\ref{eq:eta_H_low_T}) and (\ref{eq:eta_H high T}).

\subsection{Frequency integrals at high temperatures\label{subsec:Geometrical-factors}}

In the high-temperature limit $T\gg T_{d}$, the calculations of the
drag coefficients involve cumbersome integrals due to the frequency
dependence of the interlayer Coulomb interaction {[}see Eqs. (\ref{eq:u_R_RPA_sq}),
(\ref{eq:eta_parallel result}), (\ref{eq:eta_1 integral}) and (\ref{eq:F_ee_imp integral}){]}.
In the main text, we write the results for the drag coefficients {[}Eqs.
(\ref{eq:eta_parallel high T})-(\ref{eq:eta_H high T}){]} by denoting
these integrals with the functions $Q_{1}(z)$, $Q_{2}(z)$ and $Q_{3}(z)$,
with $z\equiv v_{F}^{\textrm{p}}/v_{F}^{\textrm{a}}$. Let us write
the rightmost fraction in the RHS of Eq. (\ref{eq:u_R_RPA_sq}) as

\begin{equation}
Y(\tilde{\omega},z)\equiv\frac{1-\left(\frac{\tilde{\omega}}{z}\right)^{2}}{\left(1+\frac{\tilde{\omega}}{2}\log\left(\frac{1-\tilde{\omega}}{1+\tilde{\omega}}\right)\right)^{2}+\frac{\pi^{2}}{4}\tilde{\omega}^{2}},
\end{equation}
where $\tilde{\omega}\equiv\omega/v_{F}^{\textrm{a}}q$ is a rescaled
frequency. The functions $Q_{1,2,3}(z)$ denote the integrals over
$\tilde{\omega}$ in the calculations of the drag coefficients, and
are given by

\begin{align}
Q_{1}(z) & \equiv\frac{1}{\min(1,z)}\intop_{0}^{\min(1,z)}d\tilde{\omega}\frac{Y(\tilde{\omega},z)}{\sqrt{1-\left(\frac{\tilde{\omega}}{z}\right)^{2}}},\label{eq:F1_factor}\\
Q_{2}(z) & \equiv\frac{1}{\min(1,z^{3})}\intop_{0}^{\min(1,z)}d\tilde{\omega}\frac{\tilde{\omega}^{2}Y(\tilde{\omega},z)}{\sqrt{1-\left(\frac{\tilde{\omega}}{z}\right)^{2}}},\label{eq:F2_factor}\\
Q_{3}(z) & \equiv Q_{1}(z)-\min(1,z^{2})Q_{2}(z).\label{eq:F3_factor}
\end{align}

In the limit $z\rightarrow0,$ $Q_{1}(z)=\frac{3\pi}{16},Q_{2}(z)=\frac{\pi}{32}$,
and in the limit $z\rightarrow\infty$, $Q_{1}(z)\approx0.8$00, $Q_{2}\approx0.205$.
We evaluate the integrals numerically and plot the functions in Fig.
\ref{fig:Q funs numeric}.

\begin{figure}[h]
\begin{centering}
\includegraphics[scale=0.35]{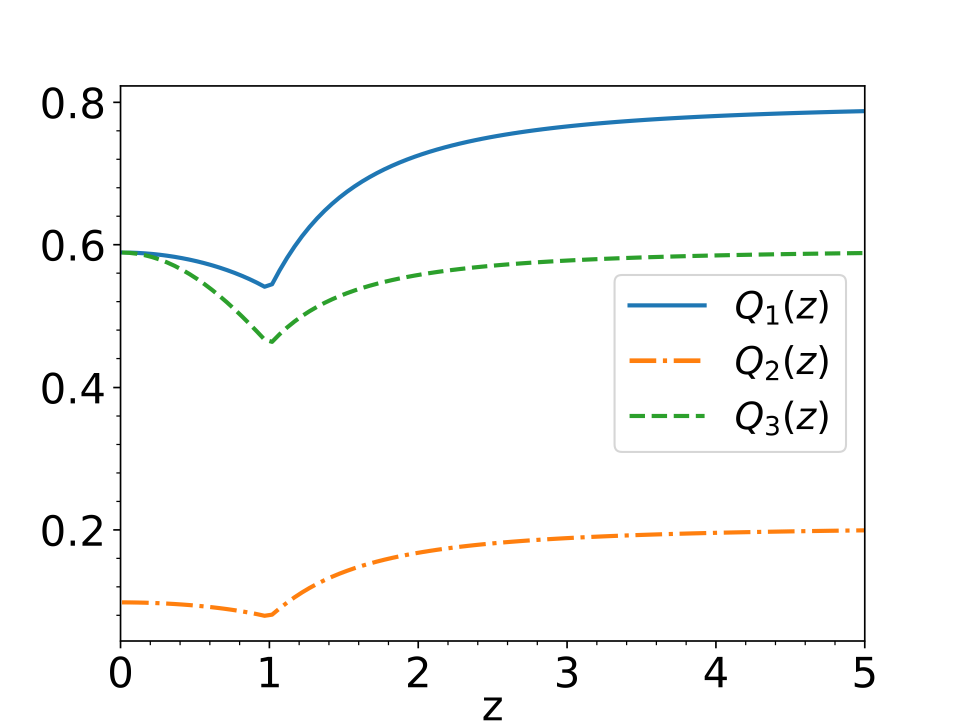}
\par\end{centering}
\caption{Numerical evaluation of the functions $Q_{1,2,3}(z)$ {[}Eqs. (\ref{eq:F1_factor})-(\ref{eq:F3_factor}){]}.\label{fig:Q funs numeric}}
\end{figure}

\section{\label{subsec:Appendix-C.-Non-interacting}Non-interacting
AHE in a WSM}

Here, we briefly summarize the calculation of the AHE conductivity
in the model of a TRS-breaking tilted WSM {[}Eq. (\ref{eq:Weyl node Hamiltonian})
of the main text{]}, following Refs. \cite{Papaj2021,Zhang2023}.
For each Weyl node of chirality $\xi$ described by the Hamiltonian
$H_{\xi}=v_{F}\left(\xi\boldsymbol{\sigma}\cdot\boldsymbol{k}+C_{\xi}k_{z}\right)$,
the energies are given by

\begin{equation}
\epsilon_{n\boldsymbol{k}}=v_{F}\left(nk+C_{\xi}k_{z}\right),
\end{equation}
where $n=\pm1$ denotes the upper and lower bands, and the momentum
is measured relative to the center of the Weyl node. It is convenient
to define the product of the node chirality and the band index, $\zeta\equiv\xi n$.
In this notation, the eigenstates of a Weyl node are written in the
spinor basis as

\begin{equation}
\left|u_{\zeta=1,\boldsymbol{k}}\right\rangle =\begin{pmatrix}\cos\frac{\theta}{2}\\
\sin\frac{\theta}{2}e^{i\varphi}
\end{pmatrix},\qquad\qquad\left|u_{\zeta=-1,\boldsymbol{k}}\right\rangle =\begin{pmatrix}-\sin\frac{\theta}{2}\\
\cos\frac{\theta}{2}e^{i\varphi}
\end{pmatrix},
\end{equation}
where $\boldsymbol{k}=k\left(\sin\theta\cos\varphi,\sin\theta\sin\varphi,\cos\theta\right)$.
We now turn to the calculation of the AHE conductivity in a single
Weyl node, multiplying the result by the number of nodes in the final
step. The corrected Boltzmann equation for the WSM in a small electric
field and in the steady state reads \cite{Sinitsyn2007}

\begin{equation}
e\boldsymbol{E}\cdot\boldsymbol{v}_{s}\frac{\partial f_{0}}{\partial\epsilon_{s}}=-\intop_{s'}w_{s,s'}\left(f_{s}-f_{s'}\right)+\intop_{s'}w_{s,s'}e\boldsymbol{E}\cdot\delta\boldsymbol{r}_{s,s'}\left(-\frac{\partial f_{0}(\epsilon_{s'})}{\partial\epsilon_{s'}}\right),\label{eq:Boltzmann non interacting AHE}
\end{equation}
where $s=\left(\boldsymbol{k},n\right)$ denotes the combined (momentum,
band) state index, $\intop_{s}\equiv\sum_{n}\intop d\boldsymbol{k}/\left(2\pi\right)^{3}$,
and $w_{s,s'}$ is the scattering rate due to disorder. The second
term in the RHS represents the side-jump correction to the collision
integral due to the coordinate shift $\delta\boldsymbol{r}_{s,s'}$
that an electron obtains when scattering from $s'\rightarrow s$ {[}Eq.
(\ref{eq:delta_r_def}){]}. The disorder scattering rate is given
by

\begin{align}
w_{s,s'} & \equiv w_{s,s'}^{\textrm{Born}}+w_{s,s'}^{\textrm{skew}},\\
w_{s,s'}^{\textrm{Born}} & =2\pi\gamma\delta(\epsilon_{s}-\epsilon_{s'})\left|\left\langle u_{s}\vert u_{s'}\right\rangle \right|^{2},\\
w_{s,s'}^{\textrm{skew}} & =\frac{4\pi^{2}\gamma^{2}\nu(\epsilon_{s})}{3\epsilon_{s}}\delta(\epsilon_{s}-\epsilon_{s'})\sin\theta_{\boldsymbol{k}}\sin\theta_{\boldsymbol{k}'}\sin(\varphi_{\boldsymbol{k}'}-\varphi_{\boldsymbol{k}}).
\end{align}
Here, $\nu(\epsilon)$ is the density of states of a single Weyl node,
given by

\begin{equation}
\nu(\epsilon)\equiv\intop_{s'}\delta(\epsilon-\epsilon_{s})=\frac{\epsilon^{2}}{2\pi^{2}v_{F}^{3}\left(1-C^{2}\right)^{2}}.
\end{equation}

To solve Eq. (\ref{eq:Boltzmann non interacting AHE}), it is convenient
to solve the side-jump collision integral separately by writing $\delta f_{s}\equiv\delta f_{s}^{\textrm{n}}+\delta f_{s}^{\textrm{anomal.}}$
and solving the two equations

\begin{align}
e\boldsymbol{E}\cdot\boldsymbol{v}_{s}\frac{\partial f_{0}}{\partial\epsilon_{s}} & =-\intop_{s'}w_{s,s'}\left(\delta f_{s}^{\textrm{n}}-\delta f_{s'}^{\textrm{n}}\right),\label{eq:Boltzmann normal dist}\\
0 & =-\intop_{s'}w_{s,s'}\left(\delta f_{s}^{\textrm{anomal.}}-\delta f_{s'}^{\textrm{anomal.}}\right)+\intop_{s'}w_{s,s'}e\boldsymbol{E}\cdot\delta\boldsymbol{r}_{s,s'}\left(-\frac{\partial f_{0}(\epsilon_{s'})}{\partial\epsilon_{s'}}\right).\label{eq:Boltzmann anomalous dist}
\end{align}
Let us define the elastic, the transport, and the skew-scattering
times

\begin{align}
\frac{1}{\tau_{s}} & \equiv\intop_{s'}w_{s,s'}=\pi\gamma\nu(\epsilon_{s}),\\
\frac{1}{\tau_{s,\parallel}}\equiv & \intop_{s'}w_{s,s'}\left(1-\frac{\sin\theta_{\boldsymbol{k}'}}{\sin\theta_{\boldsymbol{k}}}\cos\theta_{\boldsymbol{k},\boldsymbol{k}'}\right)=\frac{2}{3}\frac{1}{\tau_{s}}+O(C^{2}),\label{eq:parallel scat time}\\
\frac{1}{\tau_{s,\perp\textrm{(skew)}}} & \equiv\intop_{s'}w_{s,s'}\frac{\sin\theta_{\boldsymbol{k}'}}{\sin\theta_{\boldsymbol{k}}}\sin(\varphi_{\boldsymbol{k}'}-\varphi_{\boldsymbol{k}})=\xi\frac{2C_{\xi}}{3\epsilon_{s}\tau_{s}}\frac{1}{\tau_{s,\parallel}}+O(C^{3}).
\end{align}
In the limit $\epsilon_{F}\tau\gg1$, the skew-scattering time is
much longer than the parallel one ($\tau_{\perp\textrm{(skew)}}\gg\tau_{\parallel}$),
and the solutions to Eqs. (\ref{eq:Boltzmann normal dist}), (\ref{eq:Boltzmann anomalous dist})
are given by

\begin{align}
\delta f_{s}^{\textrm{n}} & =-\frac{\partial f_{0}}{\partial\epsilon_{s}}\boldsymbol{v}_{s}\cdot\left(e\boldsymbol{E}+\frac{\tau_{s,\parallel}}{\tau_{s,\perp\textrm{(skew)}}}e\boldsymbol{E}\times\hat{z}\right)\tau_{s,\parallel},\label{eq:normal dist part}\\
\delta f_{s}^{\textrm{anomal.}} & =\tau_{s,\parallel}\intop_{s'}w_{s,s'}e\boldsymbol{E}\cdot\delta\boldsymbol{r}_{s,s'}\left(-\frac{\partial f_{0}(\epsilon_{s'})}{\partial\epsilon_{s'}}\right)=-\frac{\partial f_{0}}{\partial\epsilon_{s}}\boldsymbol{v}_{s}\cdot\left(e\boldsymbol{E}\times\hat{z}\right)\left(\xi\frac{5C_{\xi}}{6\epsilon_{s}\tau_{s}}+O(C^{3})\right)\tau_{s,\parallel}.
\end{align}

Since the anomalous distribution $\delta f_{s}^{\textrm{anomal.}}$
is of the same form as the second term in Eq. (\ref{eq:normal dist part}),
we write the entire non-equilibrium distribution function $\delta f_{s}$
in the form of Eq. (\ref{eq:normal dist part}) {[}Eq. (\ref{eq:decomposition of delta_n})
in the main text{]}, absorbing the anomalous distribution into the
definition of the perpendicular transport time

\begin{equation}
\frac{1}{\tau_{s,\perp}}\equiv\frac{1}{\tau_{s,\perp\textrm{(skew)}}}+\xi\frac{5C_{\xi}}{6\epsilon_{s}\tau_{s}}\frac{1}{\tau_{s,\parallel}}=\xi\frac{3C_{\xi}}{2\epsilon_{s}\tau_{s}}\frac{1}{\tau_{s,\parallel}}.\label{eq:perp scattering time}
\end{equation}

\end{widetext}
The scattering times $\tau_{\parallel},\tau_{\perp}$ {[}Eqs. (\ref{eq:parallel scat time}),
(\ref{eq:perp scattering time}){]} are those used for writing the
distribution function of the WSM layer in the main text {[}Eq. (\ref{eq:decomposition of delta_n}){]}.

The velocity operator of the WSM electrons is composed of regular
and anomalous parts,

\begin{equation}
\frac{d\boldsymbol{r}}{dt}=\boldsymbol{v}_{s}+\boldsymbol{v}_{s}^{\textrm{int.}}+\boldsymbol{v}_{s}^{\textrm{ext.}},\label{eq:velocity operator breakdown}
\end{equation}
where $\boldsymbol{v}_{s}=\partial\epsilon_{s}/\partial\boldsymbol{k}$
is the regular part corresponding to the band group velocity, and
the internal and external velocities are given by \cite{Sinitsyn2007,Xiao2010,Atencia2022}

\begin{align}
\boldsymbol{v}_{s}^{\textrm{int.}} & =\boldsymbol{\Omega}_{s}\times\frac{d\boldsymbol{k}}{dt},\label{eq:intrinsic velocity}\\
\boldsymbol{v}_{s}^{\textrm{ext.}} & =\intop_{s'}w_{s',s}\delta\boldsymbol{r}_{s',s}.\label{eq:extrinsic velocity}
\end{align}
Here, $\boldsymbol{\Omega}_{s}=\boldsymbol{\nabla}_{\boldsymbol{k}}\times\boldsymbol{{\cal A}}_{s}=i\left\langle \boldsymbol{\nabla}_{\boldsymbol{k}} u_{s} \right| \times \left| \boldsymbol{\nabla}_{\boldsymbol{k}}u_{s}\right\rangle $
is the Berry curvature. We note that the intrinsic velocity gives rise to anomalous Hall current
from the filled bands, which cannot be calculated from the low-energy
Hamiltonian {[}Eq. (\ref{eq:Weyl node Hamiltonian}){]} \cite{Burkov2011}.
One may calculate this Fermi-sea contribution by regularizing the
Hamiltonian (e.g., modifying the $\sigma_{z}$ term in the Hamiltonian
to be $v_{F}\left(\sqrt{k_{x}^{2}+k_{y}^{2}+k_{z}^{2}}-k_{0}^{2}\right)\sigma_{z}$,
putting two Weyl nodes of opposite chirality at $\boldsymbol{k}=\pm k_{0}\hat{k}_{z}$
\cite{Lu2015}) or by imposing boundary conditions. When the chemical
potential is at the neutrality point ($\epsilon_{F}=0)$, integrating
the intrinsic current over the filled lower band reproduces the known
result for the AHE conductivity for a pair of Weyl nodes, $\sigma_{xy}^{\textrm{int.}}(\epsilon_{F}=0)=e^{2}\Delta_{k}/\left(4\pi^{2}\right)$
\cite{Yang2011,Burkov2011}. For non-zero $\epsilon_{F},$ we compute
the intrinsic contribution by 
\begin{align}
\sigma_{xy}^{\textrm{int.}}(\epsilon_{F})&= \sigma_{xy}^{\textrm{int.}}(\epsilon_{F}=0)
\nonumber \\
&
+2\frac{e}{E_{y}}\intop_{s}\left(f_{0}(\epsilon_{s},\mu=\epsilon_{F})-f_{0}(\epsilon_{s},\mu=0)\right)v_{s,x}^{\textrm{int.}},\label{eq:intrinsic sigma_xy calc}
\end{align}
where the factor of $2$ in the second term of the RHS accounts for the two Weyl
nodes. 

Calculating the contributions to the AHE conductivity from each part
of the velocity operator {[}Eq. (\ref{eq:velocity operator breakdown}){]},
we find (multiplying the Fermi-surface contributions by $2$ to account
for the two nodes)

\begin{align}
\sigma_{xy}^{\textrm{reg.}} & =e^{2}\frac{3\epsilon_{F}C}{4\pi^{2}v_{F}},\\
\sigma_{xy}^{\textrm{int.}} & =e^{2}\left[\frac{\Delta_{k}}{4\pi^{2}}-\frac{\epsilon_{F}C}{6\pi^{2}v_{F}}\right],\label{eq:intrinsic sigma_xy}\\
\sigma_{xy}^{\textrm{ext. velocity}} & =e^{2}\frac{5\epsilon_{F}C}{12\pi^{2}v_{F}},
\end{align}
where $\sigma_{xy}^{\textrm{reg.}},\sigma_{xy}^{\textrm{int.}},\sigma_{xy}^{\textrm{ext. velocity}}$
correspond to $\boldsymbol{v}_{s},\boldsymbol{v}_{s}^{\textrm{int.}},\boldsymbol{v}_{s}^{\textrm{ext.}}$,
respectively\footnote{We note that the sign of the second term in Eq. (\ref{eq:intrinsic sigma_xy})
appears to disagree with Ref. \cite{Zhang2023} but to agree with
Ref. \cite{Papaj2021}.}. In the main text, we combine the anomalous contributions to one
term, $\sigma_{xy}^{\textrm{a},\textrm{int.+ext.vel.}}\equiv\sigma_{xy}^{\textrm{int.}}+\sigma_{xy}^{\textrm{ext. velocity}}$.
Note that the intrinsic Hall conductivity is the only non-vanishing
term when the Fermi energy is set in the neutrality point ($\epsilon_{F}=0$).
In the notations of Refs. \cite{Papaj2021,Zhang2023}, our expression
for $\sigma_{xy}^{\textrm{reg.}}$ is equivalent to $\sigma_{xy}^{\textrm{skew }}+\sigma_{xy}^{\textrm{s.j.}}/2$,
and $\sigma_{xy}^{\textrm{ext. velocity}}$ is equivalent to $\sigma_{xy}^{\textrm{s.j. }}/2$.

\twocolumngrid

\bibliography{dragbib}

\end{document}